# Quest Complete: The Holy Grail of Gradual Security

TIANYU CHEN and JEREMY G. SIEK, Indiana University, USA

Languages with gradual information-flow control combine static and dynamic techniques to prevent security leaks. Gradual languages should satisfy the gradual guarantee: programs that only differ in the precision of their type annotations should behave the same modulo cast errors. Unfortunately, Toro et al. [2018] identify a tension between the gradual guarantee and information security; they were unable to satisfy both properties in the language $\mathrm{GSL}_{\mathsf{Ref}}$ and had to settle for only satisfying information-flow security. Azevedo de Amorim et al. [2020] show that by sacrificing type-guided classification, one obtains a language that satisfies both noninterference and the gradual guarantee. Bichhawat et al. [2021] show that both properties can be satisfied by sacrificing the no-sensitive-upgrade mechanism, replacing it with a static analysis.

In this paper we present a language design, $\lambda^{\star}_{\mathtt{IFC}}$, that satisfies both noninterference and the gradual guarantee without making any sacrifices. We keep the type-guided classification of $\mathrm{GSL}_{\mathsf{Ref}}$ and use the standard no-sensitive-upgrade mechanism to prevent implicit flows through mutable references. The key to the design of $\lambda^{\star}_{\mathtt{IFC}}$ is to walk back the decision in $\mathrm{GSL}_{\mathsf{Ref}}$ to include the unknown label $\star$ among the runtime security labels. We give a formal definition of $\lambda^{\star}_{\mathtt{IFC}}$, prove the gradual guarantee, and prove noninterference. Of technical note, the semantics of $\lambda^{\star}_{\mathtt{IFC}}$ is the first gradual information-flow control language to be specified using coercion calculi (a la Henglein), thereby expanding the coercion-based theory of gradual typing.

CCS Concepts: • **Theory of computation**; • **Security and privacy** → **Formal security models**; • **Software and its engineering** → **Formal software verification**; **Semantics**;

Additional Key Words and Phrases: gradual typing, information flow security, machine-checked proofs, Agda

## 1 INTRODUCTION

Information-flow control (IFC) ensures that information transfers within a program adhere to a security policy, e.g., by preventing high-security data from flowing to a low-security channel. This adherence can be enforced statically using a type system [Myers 1999; Myers and Liskov 1997; Volpano et al. 1996], or dynamically using runtime monitoring [Askarov and Sabelfeld 2009; Austin and Flanagan 2009; Austin et al. 2017; Devriese and Piessens 2010; Stefan et al. 2011; Xiang and Chong 2021], or using static analysis to pre-compute information that facilitates runtime monitoring [Chandra and Franz 2007; Le Guernic 2007; Le Guernic and Jensen 2005; Moore and Chong 2011; Russo and Sabelfeld 2010; Shroff et al. 2007]. The static and dynamic approaches have complementary strengths and weaknesses; the dynamic approach requires less effort from the programmer while the static approach provides stronger guarantees and less runtime overhead.

Taking inspiration from gradual typing [Siek and Taha 2006, 2007], researchers have explored how to give programmers seamless control over which parts of the program are secured statically versus dynamically. In general, gradually typed languages support the seamless transition between static and dynamic enforcement through the precision of type annotations. The programmer can choose when it is appropriate to increase the precision of the type annotations and put in the effort to pass the static checks and when it is appropriate to reduce the precision of type annotations, deferring the enforcement to runtime. The main property of gradually typed languages is the *gradual guarantee* [Siek et al. 2015], which states that removing type annotations should not change the runtime behavior. Adding type annotations should also result in the same behavior except that it may introduce more trapped errors because those new type annotations may contain mistakes.

The main challenge in the design of gradually typed languages is controlling the flow of values (and information) between the static and dynamic regions of code, which is accomplished using

Authors' address: Tianyu Chen, chen512@iu.edu; Jeremy G. Siek, jsiek@indiana.edu, Indiana University, Bloomington, Indiana, USA, 47408.

Table 1. Proposed sources of tension between security and the gradual guarantee

| Language | Security (noninterference) | Gradual Guarantee | Type-guided classification | NSU checking | Runtime security labels |
|---|---|---|---|---|---|
| $\mathrm{GSL}_{\mathsf{Ref}}$ | ✓ Yes | ✗ No | ✓ Yes | ✓ Yes | {low, high, ★} |
| GLIO | ✓ Yes | ✓ Yes | ✗ No | ✓ Yes | {low, high} |
| $\mathrm{WHILE}^{\mathrm{G}}$ | ✓ Yes | ✓ Yes | ✓ Yes | ✗ No | {low, high, ★} |
| $\lambda^{\star}_{\mathtt{IFC}}$ (this paper) | ✓ Yes | ✓ Yes | ✓ Yes | ✓ Yes | {low, high} |

runtime casts. Typically source programs are compiled to an intermediate language, called a cast calculus, that includes explicit syntax for runtime casts. Disney and Flanagan [2011a] design a cast calculus with IFC for a pure lambda calculus and prove noninterference. Fennell and Thiemann [2013] design a cast calculus named ML-GS with mutable references using the no-sensitive-upgrade (NSU) runtime checks of Austin and Flanagan [2009]. Fennell and Thiemann [2015] design a cast calculus for an imperative, object-oriented language.

Similar to gradual security-typed languages, Hybrid LIO (HLIO) [Buiras et al. 2015] also supports the choice of static or dynamic IFC in different parts of a single program. By default the checking is static, but a programmer can insert a `defer` clause to say that the security constraints should be checked at runtime. In gradual security-typed languages, the switch between static and dynamic is directed by types, so unlike in HLIO, the developer does not need to embed explicit casts or `defer` into the program. Moreover, the gradual guarantee relates the runtime behavior of programs that differ in the precision of type annotations, but there is no comparable theorem about adding and removing `defer` in HLIO.

Since the formulation of the gradual guarantee as a criterion for gradually typed languages [Siek et al. 2015], researchers have explored the feasibility of satisfying both the gradual guarantee and noninterference. Toro et al. [2018] identify a tension between the gradual guarantee and security enforcement. They analyze the semantics of runtime casts through the lens of Abstracting Gradual Typing [Garcia et al. 2016] and propose a type-driven semantics for gradual security. However, Toro et al. [2018] discover counterexamples to the gradual guarantee in the $\mathrm{GSL}_{\mathsf{Ref}}$ language. They conjecture that it is not possible to enforce noninterference and satisfy the gradual guarantee.

Azevedo de Amorim et al. [2020] conjecture one possible source of the tension: the type-guided classification performed in $\mathrm{GSL}_{\mathsf{Ref}}$ [Toro et al. 2018]. They propose a new gradually typed language, GLIO, which sacrifices type-guided classification. They prove that GLIO satisfies both noninterference and the gradual guarantee using a denotational semantics. Bichhawat et al. [2021] conjecture that *NSU checking* could be another possible source of the tension. As an alternative, they propose a hybrid approach that leverages static analysis ahead of program execution to determine the write effects in untaken branches. They study a simple imperative language with first-order stores and prove both noninterference and the gradual guarantee.

Contrary to the prior work, we show that one does not need to give up on type-guided classification or NSU checking to resolve the tension. Instead, the tension can be resolved by walking back a design choice in $\mathrm{GSL}_{\mathsf{Ref}}$, which was to allow ★ as a runtime security label. For example, in $\mathrm{GSL}_{\mathsf{Ref}}$ one can write a literal such as $\mathtt{true}_{\star}$ in a program, and at runtime the literal becomes a value of unknown security level. That design was unusual because the unknown type ★ is traditionally used in gradual languages to represent the lack of static information, not the lack of dynamic information. The design is also unusual when compared to dynamic systems for IFC, as those systems do not use an unknown security level [Askarov and Sabelfeld 2009; Austin and Flanagan 2009; Austin et al. 2017; Devriese and Piessens 2010; Stefan et al. 2011].

One might think that allowing $\star$ as a label on literals and therefore on values is necessary so that programmers can run legacy code (without any security annotations) in a gradual language, by making $\star$ the default label for literals. However, prior information-flow languages use `low` security as the default security label for literals [Myers et al. 2006] and for good reasons. The security of a literal is something that only the programmer can know. That is, the identification of high-security data in a program must be considered as an input to an information flow system, and not something that can or should be inferred. When migrating legacy code into a system that supports secure information flow, a necessary part of the process for the programmer is to identify whether there is any high-security information in the legacy code. Our choice of `low` as the default label is because most literals (if not all) in real programs are low security. In fact, it is bad practice to embed high-security literals, such as passwords, in program text.

In our design, runtime security labels do not include $\star$, only `low` and `high`.[1] On the other hand, to support gradual typing, the security labels in a type annotation may include $\star$. Surprisingly, we find that removing $\star$ from the runtime labels is sufficient to reclaim the gradual guarantee, without sacrificing type-guided classification as in GLIO or NSU checking as in WHILE$^{G}$. This finding is the primary contribution of this paper. In our design, the security level of a literal defaults to `low`, similar to systems like Jif [Myers et al. 2006] and GLIO, but different from $\mathrm{GSL}_{\mathsf{Ref}}$ and WHILE$^{G}$. We propose a new gradual, security-typed language $\lambda^{\star}_{\mathtt{IFC}}$, which (1) enforces information flow security, (2) satisfies the gradual guarantee, (3) enjoys type-guided classification, and (4) utilizes NSU checking to enforce implicit flows through the heap with no static analysis required.

The semantics of $\lambda^{\star}_{\mathtt{IFC}}$ is given by translation to a new security cast calculus $\lambda^{c}_{\mathtt{IFC}}$, for which we define a syntax, type system, and operational semantics. We compile $\lambda^{\star}_{\mathtt{IFC}}$ into $\lambda^{c}_{\mathtt{IFC}}$ in a type-preserving way. In $\lambda^{c}_{\mathtt{IFC}}$, *security coercions* serve as our runtime security monitor, in which we adapt ideas from the Coercion Calculus [Henglein 1994; Herman et al. 2010] to IFC.

Compared to prior work on gradual IFC languages, the $\lambda^{c}_{\mathtt{IFC}}$ cast calculus supports an additional feature called *blame tracking* [Findler and Felleisen 2002]. Blame tracking is important because it enables modular runtime error messages, e.g., they play an important role in production-quality languages such as Typed Racket [Tobin-Hochstadt and Felleisen 2008; Wilson et al. 2018].

Compared to prior work on gradual IFC based on abstracting gradual typing (AGT) [Toro et al. 2018], our use of a cast calculus makes it clear where in the program there is runtime overhead from dynamic checking (the casts). This is important for the programmer to know because one may wish to avoid runtime overhead in hot regions of a programs. In our translation from $\lambda^{\star}_{\mathtt{IFC}}$ to $\lambda^{c}_{\mathtt{IFC}}$, casts are only inserted where there is insufficient information during compilation to decide whether a security policy is enforced or not. In particular, casts are not inserted in statically typed regions. In contrast, the AGT mechanism for dynamic checking (called evidence) is attached to most nodes in the syntax tree.

This paper makes the following technical contributions:

- Identify the real cause of the tension between information flow security and the gradual guarantee in $\mathrm{GSL}_{\mathsf{Ref}}$: the inclusion of $\star$ in the runtime security levels (§ 2).
- A coercion calculus for security labels (§ 3) and a coercion calculus for secure values (§ 4). The two coercion calculi serve as our runtime IFC monitor. Our paper is the first work to apply the coercion calculus to an IFC setting.

[1] Of course, any lattice of security labels could be used in place of `low` and `high`.

- A cast calculus $\lambda^{c}_{\texttt{IFC}}$ with IFC that defines the dynamic semantics of $\lambda^{\star}_{\texttt{IFC}}$ (§ 5). The proofs of (1) the gradual guarantee (§ 6.4), (2) compilation from $\lambda^{\star}_{\texttt{IFC}}$ to $\lambda^{c}_{\texttt{IFC}}$ preserves types (§ 6.2), and (3) type safety of $\lambda^{c}_{\texttt{IFC}}$ (§ 5.2.1) are mechanized in Agda.
- A proof of noninterference for $\lambda^{\star}_{\texttt{IFC}}$ (§ 6.3) through the simulation between its cast calculus $\lambda^{c}_{\texttt{IFC}}$ and a dynamic IFC programming language (§ 5.2.3).
- The first design of a gradual security-typed language with type-guided classification that satisfies the gradual guarantee (§ 6).
- We mechanize the proof of the gradual guarantee in the Agda proof assistant.

The Appendix is in the supplementary material of this paper. The Agda code is available at:

https://github.com/Gradual-Typing/LambdaIFCStar at release `v2.0 (PLDI 2024)`

## 2 $\lambda^{\star}_{\texttt{IFC}}$ IN ACTION

In this section we present example programs that demonstrate how $\lambda^{\star}_{\texttt{IFC}}$ enables a gradual, smooth transition between static and dynamic information-flow control, while supporting type-based reasoning and satisfying the gradual guarantee. We briefly review the basics of gradual security typing in Section 2.1. In Section 2.2, we show that the tension between security and the gradual guarantee can be achieved by removing $\star$ from the runtime security labels. In Section 2.3, we demonstrate that $\lambda^{\star}_{\texttt{IFC}}$ enables the same type-based reasoning capabilities as $\text{GSL}_{\text{Ref}}$.

For simplicity, we use the security lattice $\langle\{\texttt{high}, \texttt{low}\}, \preccurlyeq, \curlyvee, \curlywedge\rangle$, where `high` is for private data and `low` is for publicly disclosable data. The ordering is standard: $\texttt{low} \preccurlyeq \texttt{high}$ and $\texttt{high} \not\preccurlyeq \texttt{low}$. So information is allowed to flow from public sources to private sinks but not the other way around. We refer to $\{\texttt{high}, \texttt{low}\}$ as *specific security labels*.

Types in $\lambda^{\star}_{\texttt{IFC}}$ have security labels associated with them, for example, $\texttt{Bool}_{\texttt{high}}$ is the type for booleans with high security, $\texttt{Unit}_{\texttt{low}}$ is the type for the unit value with low security, and $\texttt{Bool}_{\star}$ is the type of a boolean whose security level is unknown at compile time. We refer to $\{\texttt{high}, \texttt{low}, \star\}$ as *security labels*. We define a *precision* ordering $\sqsubseteq$ on them, where $\star \sqsubseteq g$ for any label $g$ and $\ell \sqsubseteq \ell$ for any specific security label $\ell$. The precision ordering extends to types in a natural way, so for example, $\texttt{Bool}_{\star} \sqsubseteq \texttt{Bool}_{\texttt{low}}$. Figure 18 of the Appendix gives the definition of precision on types.

To enable information-flow control, $\lambda^{\star}_{\texttt{IFC}}$ allows the programmer to annotate constants, mutable references, and $\lambda$-abstractions with a specific security label and ensures that if a value is annotated with `high`, it will not flow into a sink that is `low` security. If the programmer does not annotate a value with a label, $\lambda^{\star}_{\texttt{IFC}}$ defaults the value's label to `low`. So `true` is shorthand for $\texttt{true}_{\texttt{low}}$.

We model I/O with two functions, `user-input` and `publish`: the former returns a high-security boolean that represents sensitive input information; the latter takes a low-security boolean and publishes it into a publicly visible channel.

### 2.1 Reviewing the Basics of Gradual Information Flow Security, in $\lambda^{\star}_{\texttt{IFC}}$

In this section we review the basic concepts of gradual information flow control using $\lambda^{\star}_{\texttt{IFC}}$. We start with fully static $\lambda^{\star}_{\texttt{IFC}}$ programs and show that $\lambda^{\star}_{\texttt{IFC}}$ can behave like a static security-typed language, guarding against both illegal explicit and implicit flows at compile time. We then replace some security label annotations in types with $\star$, so that the programs become partially typed and the typing information alone is insufficient to enforce IFC. We show that coercions, our runtime security monitor, are able to capture both explicit flow and implicit flow violations at runtime, preventing information leakage and enforcing security.

*Gradual IFC includes static IFC.* For statically typed programs, $\lambda^{\star}_{\mathtt{IFC}}$ behaves just like a statically typed IFC language. Consider the following well-behaved $\lambda^{\star}_{\mathtt{IFC}}$ program that takes in a high-security user input, passes it to the function `fconst` that ignores the input and returns `false`, which is then published.

```
let fconst = λ b : Bool_high. false in
let input  = user-input () in
let result = fconst input in
  publish result
```

The program type-checks and runs without error, with no need for runtime checks to enforce security. Indeed, a malicious party cannot infer anything about the high-security input because (1) the return value of `fconst` is always the same value `false` (2) the value `false` is of low security, so the explicit flow into `publish` is allowed.

If we replace `fconst` with the identity function `fid` with parameter type $\mathtt{Bool}_{\mathtt{low}}$, the program becomes ill-typed because the type system of $\lambda^{\star}_{\mathtt{IFC}}$ does not allow the explicit flow from the high-security input to `fid`, as is usual for a statically typed IFC language.

```
let fid    = λ b : Bool_low . b in
let input  = user-input () in
let result = fid input in  // static error, input is high security but fid expects low
  publish result
```

Sometimes the observable behaviors of a program can depend on its branching structure. If some of the branch conditions have a data dependency on high-security input, a malicious party might be able to infer it from the observable behaviors, giving rise to illegal *implicit flows* [Denning 1976], which must be ruled out to guarantee security.

Consider the following program in which the function `flip` contains a conditional expression, whose condition is dependent on a high-security user input. Its two branches return different low-security booleans, creating a potential implicit flow from high to low:

```
let flip : Bool_high → Bool_low = λ b : Bool_high. if b then false else true in
let input  = user-input () in
let result = flip input in
  publish result
```

As is typical of statically typed IFC languages, the type system of $\lambda^{\star}_{\mathtt{IFC}}$ rejects this program, thereby preventing an information leak through an implicit flow. To see why, note that the branch condition is of high security, so the type of the `if` expression as a whole is $\mathtt{Bool}_{\mathtt{high}}$. In particular, the type checker computes the security level of an `if` to be the join of its branches (both `low`) and the condition (`high`), yielding $\mathtt{low} \vee \mathtt{high} = \mathtt{high}$. The `flip` function is expected to return $\mathtt{Bool}_{\mathtt{low}}$ according to its type annotation, but returns $\mathtt{Bool}_{\mathtt{high}}$ because of the conditional, $\mathtt{high} \not\preccurlyeq \mathtt{low}$, so the program is ill-typed.

To summarize, $\lambda^{\star}_{\mathtt{IFC}}$ behaves just like a static security-typed language in the above examples. When everything is statically typed, the type system of $\lambda^{\star}_{\mathtt{IFC}}$ guards against illegal information flows, whether explicit or implicit.

*Gradual IFC enables a mixture of static and dynamic IFC.* We have seen that security labels (`low` and `high`) can appear in type annotations in a program, such as $\mathtt{Bool}_{\mathtt{low}}$ and $\mathtt{Bool}_{\mathtt{high}}$. $\lambda^{\star}_{\mathtt{IFC}}$ also provides the *unknown security label*, written $\star$, for use in type annotations. We explain how the unknown security label works in the following discussion.

We return to the `fconst` example except this time the type of parameter `b` is $\mathtt{Bool}_{\star}$.

```
let fconst = λ b : Bool⋆ . false in
let input  = user-input () in
let result = fconst input in
  publish result
```

The type system of $\lambda^{\star}_{\mathtt{IFC}}$ accepts this program because, in the call `fconst input`, it allows an implicit conversion from the type of `input`, which is $\mathtt{Bool}_{\mathtt{high}}$, to the parameter type $\mathtt{Bool}_{\star}$. This program runs to completion and publishes `false`.

Now suppose we again replace `fconst` with `fid`, but keep the parameter type of $\mathtt{Bool}_{\star}$.

```
let fid    = λ b : Bool⋆ . b in
let input  = user-input () in
let result = fid input in
  publish result
```

The type system of $\lambda^{\star}_{\mathtt{IFC}}$ still accepts this program. The type of `result` changes to $\mathtt{Bool}_{\star}$ but in the call `publish result`, the type system allows an implicit conversion from $\mathtt{Bool}_{\star}$ to $\mathtt{Bool}_{\mathtt{low}}$. The security leak in this program is not caught statically; instead it is caught dynamically.

The dynamic semantics of $\lambda^{\star}_{\mathtt{IFC}}$ is defined by compilation into $\lambda^{c}_{\mathtt{IFC}}$ by inserting casts. In $\lambda^{c}_{\mathtt{IFC}}$, explicit casts are represented as *security coercions* that monitor the flow of information. We use the standard syntax for coercions [Henglein 1994] but with adaptations to handle IFC. A coercion whose target is $\star$ (and source is not $\star$) is an *injection*, and is indicated by an exclamation mark. A coercion whose source is $\star$ (and target is not $\star$) is a *projection*, and is indicated by a question mark. The projections perform runtime checks that may fail.

The translation from $\lambda^{\star}_{\mathtt{IFC}}$ to $\lambda^{c}_{\mathtt{IFC}}$ inserts a security coercion wherever an implicit cast occurred in the type checking of the $\lambda^{\star}_{\mathtt{IFC}}$ term. Here is the result of cast insertion on the above program:

```
1 let fid    = λ b. b in
2 let input  = user-input () in
3 let result = fid (input ⟨high!⟩) in
4   publish (result ⟨low?ᵖ⟩)
```

The coercion on Line 3 ($\mathtt{high}\,!$) is an injection, casting from $\mathtt{high}$ to $\star$. The coercion on line 4 ($\mathtt{low}\,?^{p}$) is a projection, casting from $\star$ to $\mathtt{low}$. At runtime, a projection checks whether the incoming value has a security level that is less than or equal to the target security level. Now suppose we run the above example with input `true`. The injection on line 3 will create an injected value $\mathtt{true}_{\mathtt{high}}\ \langle \mathtt{high}\,! \rangle$. This value is passed to and returned from `fid`, and then projected to `low`. Because `high` is greater than `low`, the projection fails.

Each projection is annotated with an identifier called a blame label ($p$). In case a projection fails, it raises a cast error, called *blame*, that contains its blame label. In this way, the programmer knows which cast is causing the problem. This feature is often referred to as *blame tracking* [Findler and Felleisen 2002; Wadler and Findler 2009]. Blame tracking is especially useful during the software development process, but in the context of IFC, one may not want blame to be observable in a production system as it could reveal information. This can be handled by causing the program to diverge whenever blame is detected, possibly sending a private error message to the software developer.

Next we return to the `flip` example to see how gradual IFC prevents implicit flows. Suppose that we change the parameter type of the $\lambda$ from $\mathtt{Bool}_{\mathtt{high}}$ to $\mathtt{Bool}_{\star}$. The return type remains $\mathtt{Bool}_{\mathtt{low}}$, to conform with the signature of `publish`. Line 1 thus becomes:

```
let flip : Bool⋆ → Bool_low = λ b : Bool⋆ . if b then false else true in
```

This change makes the program well-typed in $\lambda^{\star}_{\mathtt{IFC}}$. The IFC enforcement of the implicit flow is deferred until runtime because the branch condition now has type $\mathtt{Bool}_{\star}$, with an unknown security level. Next we discuss the runtime behavior of this program.

The result of cast insertion on this program is the following $\lambda^{c}_{\mathtt{IFC}}$ term:

```
let flip   = λ b. ((if⋆ b then (false ⟨low!⟩) else (true ⟨low!⟩)) ⟨low?ᵖ⟩) in
let input  = user-input () in
let result = flip (input ⟨high!⟩) in
  publish result
```

where the `if` is changed to `if⋆` because the condition expression has static security level $\star$. The type checking rule for `if⋆` requires the two branches to have security level $\star$ and the security level of the `if⋆` is a whole is also $\star$. If we run the program with `true` or `false` as input, the $\lambda^{c}_{\mathtt{IFC}}$ term reduces to `blame` $p$ with either input, thus capturing the illegal implicit flow. Consider running this program with input `true`. First, the $\lambda$ is bound to `flip` and the input `true` is bound to the `input` variable. We then inject the input, producing the value $(\mathtt{true}_{\mathtt{high}}\ \langle \mathtt{high}\,! \rangle)$. We call `flip` and the `if⋆` branches on this boolean, evaluating the "then" branch to the result $(\mathtt{false}_{\mathtt{low}}\ \langle \mathtt{low}\,! \rangle)$ and then, to protect against implicit flows, the `if⋆` upgrades the result to `high` to match the runtime security level of the condition $(\mathtt{true}_{\mathtt{high}}\ \langle \mathtt{high}\,! \rangle)$, producing $(\mathtt{false}_{\mathtt{low}}\ \langle \uparrow ;\ \mathtt{high}\,! \rangle)$. The subtype coercion $\uparrow$ sends the security level of `false` from `low` to `high`. The last step in the body of `flip` is to apply the coercion $\mathtt{low}\,?^{p}$ to the value $(\mathtt{false}_{\mathtt{low}}\ \langle \uparrow ;\ \mathtt{high}\,! \rangle)$, which errors because `high` is greater than `low`.

## 2.2 Implicit Flow, NSU Checks, Unknown Security, and the Gradual Guarantee

The tension between information-flow security and the gradual guarantee arises from an interaction between implicit flows and the use of no-sensitive-upgrade checks to guard writes to mutable references. In brief, when $\star$ is allowed as a runtime security label, some NSU checks have to conservatively trigger an error to preserve noninterference, even though no error would occur if the label was instead `high`. But the gradual guarantee says that if a program with a precise annotation runs without error, it should also run without error when that annotation is changed to $\star$.

In preparation to discuss this scenario in more detail, we review the no-sensitive-upgrade technique [Austin and Flanagan 2009] that protects against illegal implicit flows through writes to mutable references. We then show how allowing $\star$ as a runtime security label leads to a situation where a language designer is forced to choose between noninterference and the gradual guarantee. Finally, we show how this problem is resolved in the $\lambda^{\star}_{\mathtt{IFC}}$ language by walking back the choice of allowing $\star$ as a runtime security label.

The main idea of no-sensitive-upgrades is to prevent data leaks through the mutable references by terminating execution whenever the program attempts to modify a low-security heap cell in a high-security execution context. In $\lambda^{\star}_{\mathtt{IFC}}$, NSU checks happen at runtime when type information is insufficient to statically decide whether a heap-modifying operation is secure or not. Consider the following well-typed program in $\lambda^{\star}_{\mathtt{IFC}}$:

```
let input : Bool⋆ = user-input () in
let a      = ref low  true in
let _      = if input then a := false else a := true in
  publish (! a)
```

The assignments to `a` in the two branches try to write different low-security booleans into the cell at the address in `a`, depending on a branch condition whose security level is statically unknown.

However, at runtime the `user-input` function returns a high-security boolean, so the branch condition is actually high security, and if the writes were successful, the program would leak information via an implicit flow. Fortunately, if we run this program, it reduces to blame regardless of the input. The way NSU checking works in $\lambda^{\star}_{\texttt{IFC}}$ is that a security level is associated with the current program counter. At the point of every write that requires an NSU, the system projects the program counter's security level to the level of the memory location, making sure that the later is at least as high as the former. In the above example, the NSU check fails because the program counter's security is `high` but the write is to a `low` security location.

In $\text{GSL}_{\text{Ref}}$ [Toro et al. 2018], the dynamic enforcement of IFC through the heap is also based on NSU checks. Consider the following pair of programs adapted from Section 6.3 of their paper. The program on the left is derived from the program on the right by replacing some of the `high` annotations with the unknown label $\star$. Both variants of the program type check but the more precise variant runs to completion while the less precise variant triggers an error, thus violating the gradual guarantee. Let us examine their runtime behavior in further detail.

**Left:** less precise, more dynamic

```
1  let x = user-input () in
2  let y = ref Bool_⋆ true_⋆ in
3    if x then (y := false_high) else ()
```

**Right:** more precise, more static

```
1  let x = user-input () in
2  let y = ref Bool_high true_high in
3    if x then (y := false_high) else ()
```

The program on the right runs without error in $\text{GSL}_{\text{Ref}}$ because, at the assignment on line 3, variable `y` references a memory cell of `high` security and the PC's security level is also `high`, so the assignment is allowed.

In contrast, when the program on the left is run with input $\texttt{true}_{\texttt{high}}$, the assignment is conservatively rejected by the NSU check. This is because $\text{GSL}_{\text{Ref}}$ considers $\star$ as corresponding to the interval [`low`, `high`], and the lower bound of this interval is not greater than or equal to the `high` PC label. So we see that this more precise program (right) runs successfully while the less precise one (left) errors in $\text{GSL}_{\text{Ref}}$.

In $\lambda^{\star}_{\texttt{IFC}}$, the $\star$ security label can be used in type annotations, as one would expect of a gradually-typed language, but $\star$ is not allowed as a runtime security label and therefore also not allowed as a label on program literals and other introduction forms. The following adapts the above examples from $\text{GSL}_{\text{Ref}}$ to $\lambda^{\star}_{\texttt{IFC}}$. The fully static variant on the right is nearly identical to its $\text{GSL}_{\text{Ref}}$ counterpart. To obtain the less-precise program on the left we change the type annotation on variable `y` to model the similar loss of precision in the $\text{GSL}_{\text{Ref}}$ counterpart. We do not change the labels on the `ref` or `true` to $\star$ because that is not allowed in $\lambda^{\star}_{\texttt{IFC}}$ as we just mentioned.

```
1  let x = user-input () in
2  let y : (Ref Bool_⋆)_⋆ = ref high true_high in
3    if x then (y := false_high) else ()
```

```
1  let x = user-input () in
2  let y : (Ref Bool_high)_high = ref high true_high in
3    if x then (y := false_high) else ()
```

Branching on high-security input and assigning to a high-security memory location should be allowed. Indeed, both variants reduce to the unit value regardless of the input, thereby not violating the gradual guarantee. The fully annotated version (right) evaluates to unit without any overhead from runtime checking.

In the less-precise program (left), the type annotation $(\texttt{Ref Bool}_{\star})_{\star}$ replaces $(\texttt{Ref Bool}_{\texttt{high}})_{\texttt{high}}$, meaning that both the security level of the memory and the security of the reference itself are statically unknown and should be checked at runtime. When executing the program, at the assignment on line 3, an NSU check happens and the assignment to high-security memory under high PC is allowed. As a result, the less-precise program also evaluates successfully to unit.

One might worry that the less precise program has a heavy annotation burden. However, as we mentioned, the default security label is `low` so the programmer does not have to label constants in $\lambda^{\star}_{\mathtt{IFC}}$. So we can remove the labels on constants to obtain the following program, which also reduces successfully to unit:

```
1  let x = user-input () in
2  let y : (Ref Bool★)★ = ref high true in
3    if x then (y := false) else ()
```

The low-security `true` is classified as high security during reference creation because the security level of the cell is `high` (line 2). Similarly, during assignment the `false` is also classified as high security because the security level of the cell (line 3). Assigning to a high-security memory cell is allowed under a high PC by the NSU check, so the program evaluates successfully to unit.

One might think that requiring the programmer to annotate the reference creation with `high` (line 2) is still a burden and that $\mathrm{GSL}_{\mathsf{Ref}}$ is better in this regard. However, while $\mathrm{GSL}_{\mathsf{Ref}}$ allows the unannotated version of this program to compile, it errors during program execution. We believe that it is better to require the programmer to annotate references during program development than to have the programs compile but then fail during program execution, perhaps only detected after the program is deployed.

### 2.3 Type-based Reasoning in $\lambda^{\star}_{\mathtt{IFC}}$

Type-based reasoning in gradual IFC languages arises from two language design choices: vigilance and type-guided classification. Vigilance gives us type-based reasoning for explicit flows, while type-guided classification provides type-based reasoning about implicit flows. We show in this section that $\lambda^{\star}_{\mathtt{IFC}}$ is both vigilant and performs typed-guided classification, so it enables type-based reasoning in the sense of Toro et al. [2018].

*2.3.1 $\lambda^{\star}_{\mathtt{IFC}}$ is Vigilant.* A language with casts is *vigilant* if it checks whether all the casts that are applied to the same value are consistent with each other, and triggers an error if they are not.

Toro et al. [2018] present the following example to demonstrate how vigilance is needed for type-based reasoning and free theorems in the sense of Wadler [1989]. The example involves casts from `low` (line 3, the label annotation on $5_{\mathtt{low}}$) to `high` (line 1, the type annotation $\mathtt{Int}_{\mathtt{high}}$ in the signature of `mix`) and then back to `low` via the unknown security level $\star$ (line 2, the nested type annotations $\mathtt{Int}_{\star}$ and $\mathtt{Int}_{\mathtt{low}}$):

```
1  let mix : Int_low → Int_high → Int_low =
2      λ pub priv . if pub < (priv : Int★ : Int_low ) then 1 else 2 in
3  mix 1_low 5_low
```

The type signature of `mix` should guarantee the free theorem that either (1) the result of `mix`, which is low security, never depends on the high-security `priv` argument or (2) `mix` produces a runtime error. In this case, the output of `mix` does depend on `priv` via an implicit flow, so the free theorem says that an error should be triggered at runtime. Let us focus on the three casts applied to $5_{\mathtt{low}}$, where the first cast sends the security level from `low` to `high` because of the type annotation on line 1, the second cast is an injection due to the first type annotation on line 2, and the third cast is a projection to `low` due to the second type annotation on line 2:

$$5_{\mathtt{low}} \left\langle \mathtt{low} \Rightarrow \mathtt{high} \right\rangle \left\langle \mathtt{high} \Rightarrow \star \right\rangle \left\langle \star \Rightarrow \mathtt{low} \right\rangle$$

In $\lambda^{\star}_{\mathtt{IFC}}$, these casts produce the sequence of coercions: $5_{\mathtt{low}} \left\langle \uparrow ; \mathtt{high}\,! ; \mathtt{low}\,?^{p} \right\rangle$, which trigger an error when `high` collides with `low`, blaming label $p$.

Similarly, the interval refinement mechanism of $\mathrm{GSL}_{\mathsf{Ref}}$ detects the conflict between the intermediate cast to `high` and the final cast to `low`. Surprisingly, in GLIO and in systems prior to $\mathrm{GSL}_{\mathsf{Ref}}$ [Disney and Flanagan 2011b; Fennell and Thiemann 2013], the program runs successfully and produces $1_{\mathtt{low}}$ because they are forgetful [Greenberg 2014] regarding the intermediate cast of $5_{\mathtt{low}}$ to high and only check that $5_{\mathtt{low}}$ can be cast to low[2].

*2.3.2 $\lambda^{\star}_{\mathtt{IFC}}$ Performs Typed-Guided Classification.* A gradual IFC language employs *typed-guided classification* if the security-level of a value can be changed when the value flows through a cast.

The following example from Section 2 of Toro et al. [2018] demonstrates how type-guided classification interacts with implicit flow and type-based reasoning. Type-based reasoning tells us that the `smix` function below should either fail or return a value that does not depend on its high-security parameter `priv`. However, `smix` calls `mix` and there is an implicit flow from `priv` to the result value, so this program should fail.

```
let mix  : Int_low → Int_★    → Int_low = λ pub priv. if pub < priv then 1 else 2 in
let smix : Int_low → Int_high → Int_low = λ pub priv. mix pub priv in
smix 1_low 5_low
```

In $\lambda^{\star}_{\mathtt{IFC}}$ the program produces an error as type-based reasoning suggests. Security coercions in $\lambda^{c}_{\mathtt{IFC}}$ classify values, so when $5_{\mathtt{low}}$ is passed into `smix` and then `mix`, it is wrapped in a coercion: $5_{\mathtt{low}}\ \langle \uparrow;\ \mathtt{high}\,!\rangle$ and is classified as high-security. Consequently, the `if` reduces to its then-branch protected with high. This implicit flow affects the result value of the then-branch, by (1) inserting a subtype coercion before the injection and (2) promoting the source of the injection to high to preserve types. So the result of the `if` is $1_{\mathtt{low}}\ \langle \uparrow;\ \mathtt{high}\,!\rangle$. The injection, whose source is high, collides with a projection to low (to match the $\mathtt{Int}_{\mathtt{low}}$ return type of `mix`), causing the program to error as expected, blaming the projection.

## 3 A COERCION CALCULUS FOR SECURITY LABELS

In this section we present a coercion calculus on security labels. We show that we can use coercion composition and stamping to represent explicit flows and implicit flows respectively (Section 3.1). We define how these coercions act on security labels by defining a language of label expressions whose meaning is defined by a reduction relation (Section 3.2). Finally, we explore meta-theoretic properties of this coercion calculus (Section 3.3), establishing the intuition that the security coercion calculus can be used to enforce gradual IFC, while satisfying the gradual guarantee.

As we have seen in Section 2, gradual information flows can be modeled as casts. The cast sequence high $\Rightarrow \star \Rightarrow$ low should be statically accepted but dynamically rejected, while the sequence low $\Rightarrow \star \Rightarrow$ high should be statically and dynamically accepted, promoting the security of data to high. Such sequences of casts can be arbitrarily long (for example, low $\Rightarrow$ high $\Rightarrow \star \Rightarrow \star \Rightarrow$ low), which motivates us to model the casts on security labels as coercions. In $\lambda^{\star}_{\mathtt{IFC}}$, the source security label of a coercion sequence comes from literals, while the sink is whatever security level that the observer has: for example, the `publish` function of Section 2 is of low. Coercions can be easily sequenced and composed. Checking information flow at runtime is accomplished by reducing coercion sequences to their normal forms.

There are two noteworthy benefits of the coercion representation for IFC. First, coercions can be used to represent NSU checking while satisfying the gradual guarantee. In brief, whenever a memory location is written to, the current PC is coerced to the security level of that location. We are going to formally introduce label expressions as our representation for PC in Section 3.2 and discuss NSU in detail in Section 5.1.2. Second, the coercion representation benefits mechanization

[2]Those systems still satisfy noninterference, because the labels *on values* track their security similar to fully dynamic IFC. It is just that the security labels on *type annotations* will not trigger errors at runtime, even if they are inconsistent.

$$
\begin{array}{rccl}
\text{specific security labels} & \ell & \in & \{\texttt{low}, \texttt{high}\} \\
\text{security labels} & g & ::= & \star \mid \ell \\
\text{blame labels} & p, q & & \\
\text{security coercions} & c, d & ::= & \mathbf{id}(g) \mid \uparrow \mid \ell\,! \mid \ell\,?^p \mid \bot^p \\
\text{coercion sequences} & \bar{c}, \bar{d} & ::= & \mathbf{id}(g) \mid \bot^p\ g_1\ g_2 \mid \bar{c};\ c
\end{array}
$$

$\boxed{\vdash c : g_1 \Rightarrow g_2}$

$$
\frac{}{\vdash \mathbf{id}(g) : g \Rightarrow g} \qquad \frac{}{\vdash \uparrow : \texttt{low} \Rightarrow \texttt{high}} \qquad \frac{}{\vdash \ell\,! : \ell \Rightarrow \star} \qquad \frac{}{\vdash \ell\,?^p : \star \Rightarrow \ell} \qquad \frac{}{\vdash \bot^p : \texttt{high} \Rightarrow \texttt{low}}
$$

$\boxed{\vdash \bar{c} : g_1 \Rightarrow g_2}$

$$
\frac{}{\vdash \mathbf{id}(g) : g \Rightarrow g} \qquad \frac{\vdash \bar{c} : g_1 \Rightarrow g_2 \qquad \vdash c : g_2 \Rightarrow g_3}{\vdash \bar{c};\ c : g_1 \Rightarrow g_3} \qquad \frac{}{\vdash \bot^p\ g_1\ g_2 : g_1 \Rightarrow g_2}
$$

$\boxed{\mathbf{NF}\ \bar{c}}$

$$
\frac{}{\mathbf{NF}\ \mathbf{id}(g)} \qquad \frac{}{\mathbf{NF}\ \mathbf{id}(\star);\ \ell\,?^p} \qquad \frac{\mathbf{NF}\ \bar{c}}{\mathbf{NF}\ \bar{c};\ \ell\,!} \qquad \frac{\mathbf{NF}\ \bar{c}}{\mathbf{NF}\ \bar{c};\ \uparrow}
$$

$\boxed{c\ ;\ c \longrightarrow c}$

$$
\textit{?-id}\ \frac{}{\ell\,!;\ \ell\,?^p \longrightarrow \mathbf{id}(\ell)} \qquad \textit{?-}\uparrow\ \frac{}{\texttt{low}\,!;\ \texttt{high}\,?^p \longrightarrow \uparrow} \qquad \textit{?-}\bot\ \frac{}{\texttt{high}\,!;\ \texttt{low}\,?^p \longrightarrow \bot^p}
$$

$\boxed{\bar{c} \longrightarrow \bar{d}}$

$$
\textit{id}\ \frac{\mathbf{NF}\ \bar{c}}{\bar{c};\ \mathbf{id}(g) \longrightarrow \bar{c}} \qquad \bot\ \frac{\mathbf{NF}\ \bar{c} \qquad \vdash \bar{c} : g_1 \Rightarrow g_2}{\bar{c};\ \bot^p \longrightarrow \bot^p\ g_1\ \texttt{low}} \qquad \xi\text{-}\bot\ \frac{\vdash c : g_2 \Rightarrow g_3}{\bot^p\ g_1\ g_2;\ c \longrightarrow \bot^p\ g_1\ g_3}
$$

$$
\xi_L\ \frac{\bar{c} \longrightarrow \bar{d}}{\bar{c};\ c \longrightarrow \bar{d};\ c} \qquad \xi_R\ \frac{\mathbf{NF}\ \bar{c} \qquad c;\ d \longrightarrow c'}{\bar{c};\ c;\ d \longrightarrow \bar{c}; c'}
$$

Fig. 1. Syntax, typing, normal forms, and semantics of security coercions and coercion sequences

because it enables modular reasoning. The main simulation lemma (Lemma 13) depends on the simulation results of coercion sequences and label expressions, which are stated as separate lemmas and reasoned independently in our Agda code.

The syntax and typing for security coercions and coercion sequences is defined in Figure 1. A security label is either `low`, `high`, or statically unknown ($\star$). There are five security coercions: identity ($\mathbf{id}(g)$), subtype ($\uparrow$), injection ($\ell\,!$), and projection ($\ell\,?^p$), and blame ($\bot^p$). Projection, which corresponds to the notion of a runtime check, is the only one responsible for blame, so it carries a blame label $p$. A coercion sequence $\bar{c}$ starts with either success $\mathbf{id}(g)$ or failure ($\bot^p\ g_1\ g_2$). Each coercion has a source and target type $g_1 \Rightarrow g_2$. The $\mathbf{id}(g)$ casts the label $g$ to itself; $\uparrow$ promotes security from `low` to `high`; injection casts to $\star$ from a specific label $\ell$ and projection does the opposite. Appending a single coercion to a coercion sequence makes the target security label that of the single coercion.

Information flow is enforced in the reduction semantics of security coercions, shown in Figure 1. Injection followed by projection to the same label collapses to the identity (*?-id*). Flowing from `low` to `high` is allowed, so an injection from `low` followed by a projection to `high` collapses into the $\uparrow$ coercion (*?-* $\uparrow$). An information flow from `high` to `low` is prohibited, so an injection to `high` followed by a projection to `low` triggers an error that blames the projection (*?-*$\bot$). The predicate **NF** that specifies the normal forms of coercion sequences. The reduction rules for coercion sequences are also defined in Figure 1. Appending $\mathbf{id}(g)$ onto a coercion sequence reduces to the that sequence (*id*). The failure coercions annihilate the other coercions in the sequence ($\bot$ and

$\boxed{\bar{c} \fatsemi \bar{c} = \bar{c}}$

$$\bar{c} \fatsemi \bot^{p}\ g_2\ g_3 = \bot^{p}\ g_1\ g_3 \quad \text{,where } \vdash \bar{c} : g_1 \Rightarrow g_2$$
$$\bar{c} \fatsemi \mathbf{id}(g) = \bar{c};\ \mathbf{id}(g)$$
$$\bar{c}_1 \fatsemi (\bar{c}_2;\ c) = (\bar{c}_1 \fatsemi \bar{c}_2);\ c$$

$\boxed{stamp\ \bar{c}\ \ell = \bar{c}}$

$$stamp\ \bar{c}\ \texttt{low} = \bar{c}$$
$$stamp\ \mathbf{id}(\texttt{low})\ \texttt{high} = \mathbf{id}(\texttt{low});\ \uparrow$$
$$stamp\ \mathbf{id}(\texttt{high})\ \texttt{high} = \mathbf{id}(\texttt{high})$$
$$stamp\ (\mathbf{id}(\texttt{low});\ \texttt{low}\,!)\ \texttt{high} = \mathbf{id}(\texttt{low});\ \uparrow;\ \texttt{high}\,!$$
$$stamp\ (\mathbf{id}(\texttt{high});\ \texttt{high}\,!)\ \texttt{high} = \mathbf{id}(\texttt{high});\ \texttt{high}\,!$$
$$stamp\ (\mathbf{id}(\texttt{low});\ \uparrow;\ \texttt{high}\,!)\ \texttt{high} = \mathbf{id}(\texttt{low});\ \uparrow;\ \texttt{high}\,!$$
$$stamp\ (\mathbf{id}(\texttt{low});\ \uparrow)\ \texttt{high} = \mathbf{id}(\texttt{low});\ \uparrow$$

$\boxed{stamp!\ \bar{c}\ \ell = \bar{c}}$

$$stamp!\ \bar{c}\ \texttt{low} = \begin{cases} \bar{c} & \text{, if } \vdash \bar{c} : \ell \Rightarrow \star \\ \bar{c};\ \ell_2\,! & \text{, if } \vdash \bar{c} : \ell_1 \Rightarrow \ell_2 \end{cases}$$
$$stamp!\ \mathbf{id}(\texttt{low})\ \texttt{high} = \mathbf{id}(\texttt{low});\ \uparrow;\ \texttt{high}\,!$$
$$stamp!\ \mathbf{id}(\texttt{high})\ \texttt{high} = \mathbf{id}(\texttt{high});\ \texttt{high}\,!$$
$$stamp!\ (\mathbf{id}(\texttt{low});\ \texttt{low}\,!)\ \texttt{high} = \mathbf{id}(\texttt{low});\ \uparrow;\ \texttt{high}\,!$$
$$stamp!\ (\mathbf{id}(\texttt{high});\ \texttt{high}\,!)\ \texttt{high} = \mathbf{id}(\texttt{high});\ \texttt{high}\,!$$
$$stamp!\ (\mathbf{id}(\texttt{low});\ \uparrow;\ \texttt{high}\,!)\ \texttt{high} = \mathbf{id}(\texttt{low});\ \uparrow;\ \texttt{high}\,!$$
$$stamp!\ (\mathbf{id}(\texttt{low});\ \uparrow)\ \texttt{high} = \mathbf{id}(\texttt{low});\ \uparrow;\ \texttt{high}\,!$$

Fig. 2. Composing and stamping coercions

$\xi$-$\bot$). We choose the evaluation order in a coercion sequence to be from left to right ($\xi_L$ and $\xi_R$), because that corresponds to the direction of information flow from source to sink: in the example above, $\texttt{low} \Rightarrow \texttt{high} \Rightarrow \star \Rightarrow \star \Rightarrow \texttt{low}$, we validate that low can flow to high before we check the flow from high through $\star$ to low.

### 3.1 Monitoring Explicit and Implicit Flows

We model explicit flow using security coercions. We can compose two coercion sequences ($\bar{c} \fatsemi \bar{d}$), where $\bar{c} : g_1 \Rightarrow g_2$ and $\bar{d} : g_2 \Rightarrow g_3$, to form a flow from $g_1$ to $g_3$, which is defined in Figure 2.

The stamping operation captures the intuition of an implicit flow from the security level $\ell'$ to a coercion sequence $\bar{c}$. We define the stamping operation in Figure 2 as two functions, $stamp(\bar{c}, \ell)$ and $stamp!(\bar{c}, \ell)$. Both function require $\bar{c}$ to be in normal form and that its source label is not $\star$. The *stamp*! operator promotes the security of the coercion $\bar{c}$ to be at least $\ell$ and then injects the coercion if necessary, while *stamp* only promotes the security but does not inject. These stamping operations satisfy the gradual guarantee, because when stamping on a more precise coercion sequence and a less precise coercion sequence, stamping preserves the precision relation (Section 3.3.2) between them (Lemma 9). The stamping operations of coercion sequences are used in the stamping operations of (1) label expressions, which are our representation of PC and (2) values in the cast calculus. Those three types of stamping together formalize the notion of implicit flow in $\lambda^{\star}_{\texttt{IFC}}$.

### 3.2 Security Label Expressions

In this section we introduce security label expressions, which we use to model the security level of the PC. Security label expressions are crucial for implementing NSU checking in a way that satisfies the gradual guarantee.

A label expression is either (1) a specific security label, (2) blame (to signify an error), or (3) a coercion applied to a label expression (Figure 3). A label expression is in normal form (NF) if it is either (1) a specific security label or (2) an irreducible coercion applied to a specific security label. (A coercion is irreducible if it is a non-identity coercion in normal form). *PC* ranges over label expressions in normal form. The reduction relation for label expressions steps a label expression towards its normal form. The idea is that given a label expression of the form $e\ \langle \bar{d} \rangle$, we first reduce

$$\text{label expressions} \quad e \quad ::= \quad \ell \mid \texttt{blame}\ p \mid e\ \langle \bar{c} \rangle$$

$\boxed{\vdash e \Leftarrow g}$

$$\vdash l\ \frac{}{\vdash \ell \Leftarrow \ell} \qquad \vdash lcast\ \frac{\vdash e \Leftarrow g_1 \quad \vdash \bar{c} : g_1 \Rightarrow g_2}{\vdash e\ \langle \bar{c} \rangle \Leftarrow g_2} \qquad \vdash lblame\ \frac{}{\vdash \texttt{blame}\ p \Leftarrow g}$$

$\boxed{\textbf{Irreducible}\ \bar{c}}$ $\boxed{\textbf{NF}\ e}$

$$\frac{\textbf{NF}\ \bar{c} \quad \vdash \bar{c} : g_1 \Rightarrow g_2 \quad g_1 \neq g_2}{\textbf{Irreducible}\ \bar{c}} \qquad \frac{}{\textbf{NF}\ \ell} \qquad \frac{\textbf{Irreducible}\ \bar{c}}{\textbf{NF}\ (\ell\ \langle \bar{c} \rangle)}$$

$\boxed{e_1 \longrightarrow e_2}$

$$\xi\text{-}l\ \frac{e_1 \longrightarrow e_2}{e_1\ \langle \bar{c} \rangle \longrightarrow e_2\ \langle \bar{c} \rangle} \qquad \xi\text{-}lblame\ \frac{}{\texttt{blame}\ p\ \langle \bar{c} \rangle \longrightarrow \texttt{blame}\ p} \qquad \beta\text{-}id\ \frac{}{\ell\ \langle \mathbf{id}(\ell) \rangle \longrightarrow \ell}$$

$$lcast\ \frac{\bar{c} \longrightarrow^{+} \bar{d} \quad \textbf{NF}\ \bar{d}}{\ell\ \langle \bar{c} \rangle \longrightarrow \ell\ \langle \bar{d} \rangle} \qquad lblame\ \frac{\bar{c} \longrightarrow^{*} \bot^{p}\ \ell\ g}{\ell\ \langle \bar{c} \rangle \longrightarrow \texttt{blame}\ p} \qquad lcomp\ \frac{\textbf{Irreducible}\ \bar{c}}{\ell\ \langle \bar{c} \rangle\ \langle \bar{d} \rangle \longrightarrow \ell\ \langle \bar{c} \mathbin{;} \bar{d} \rangle}$$

Fig. 3. Syntax, typing, normal forms, and semantics of label expressions

$e$ to normal form and then apply the coercion $\bar{d}$. For example, if $e$ reduces to a label wrapped in coercion $\ell\ \langle \bar{c} \rangle$, then the *lcomp* rule says to reduce by composing the two coercions, producing $\ell\ \langle \bar{c} \mathbin{;} \bar{d} \rangle$. Furthermore, in a label expression of the form $e\ \langle \bar{d} \rangle$, the coercion $\bar{d}$ may also need to be reduced, which is accomplished by the *lcast* rule that refers to the reduction relation for coercion sequences (Figure 1). If the coercion reduces to an identity, then the coercion application goes away ($\beta$-*id*), whereas if the coercion reduces to a failure, then the label expression reduces to blame (*lblame*).

The stamping and security level operators for label expressions are defined in Figure 13 of the Appendix. They both require their input to be in normal form, which can be either (1) a specific security label $\ell$, or (2) a label wrapped with an irreducible coercion sequence $\ell\ \langle \bar{c} \rangle$. For (1), stamping low with high results in low $\langle \uparrow \rangle$, otherwise the label expression remains unchanged; for (2), we directly stamp the coercion sequence using *stamp* for coercion sequences defined in Figure 2. The definition of *stamp!* is analogous, except that it turns to the *stamp!* operator of coercion sequences. The security level operator $|{-}|$ is defined such that (1) a specific security label indicates the security level for itself and (2) the security of the coercion sequence $|\bar{c}|$ records the security level for $\ell\ \langle \bar{c} \rangle$.

We describe in Section 5.1 how label expressions are used to implement NSU checks, which enforce the heap policy for write operations.

### 3.3 Properties of Coercion Calculus on Labels

*3.3.1 Tracking Information Flow via Composition and Stamping.* We show that composing coercions tracks explicit flows, while stamping tracks implicit flows. We first define of the security level of a coercion sequence with the $|{-}|$ operator [3]:

Definition 1 (Security level of a coercion). *Given a coercion sequence $\bar{c}$ in normal form with type $\vdash \bar{c} : \ell \Rightarrow g$, then its security is given by the following $|{-}|$ operator:*

$$|\mathbf{id}(\ell)| = \ell \qquad |\mathbf{id}(\ell);\ \ell\,!| = \ell \qquad |\mathbf{id}(\texttt{low});\ \uparrow;\ \texttt{high}\,!| = \texttt{high} \qquad |\mathbf{id}(\texttt{low});\ \uparrow| = \texttt{high}$$

[3]In a larger lattice, the subtype coercion $\uparrow$ would be parameterized by two labels: $\vdash (\uparrow^{\ell_2}_{\ell_1}) : \ell_1 \Rightarrow \ell_2$ where $\ell_1 \prec \ell_2$. The security would be the greater one: $|\uparrow^{\ell_2}_{\ell_1}| = \ell_2$. Here the lattice only consists of low and high, so $|\uparrow| = |\uparrow^{\texttt{high}}_{\texttt{low}}| = \texttt{high}$.

We reason about explicit flows first. If we compose one coercion sequence with another and then reduce the result to normal form, the security of the resulting coercion sequence should be greater than or equal to that of the first sequence.

Lemma 2 (Composition models explicit flow).
*If* NF $\bar{c}$ *and* $\bar{c} \fatsemi \bar{d} \longrightarrow^* \bar{c}'$ *and* NF $\bar{c}'$, *then* $|\bar{c}| \preccurlyeq |\bar{c}'|$.

Next we show that stamping models implicit flow correctly, promoting the security of the stamped coercion by joining it with the stamped label:

Lemma 3 (Stamping models implicit flow).
*If* NF $\bar{c}$, *then* $|stamp\ \bar{c}\ \ell| = |\bar{c}| \vee \ell$ *and* $|stamp!\ \bar{c}\ \ell| = |\bar{c}| \vee \ell$.

*3.3.2 Simulation between more and less precise coercion sequences.* Our goal is to prove the gradual guarantee for $\lambda^{\star}_{\mathtt{IFC}}$. The proof depends on a simulation lemma between more and less precise terms of the cast calculus $\lambda^{c}_{\mathtt{IFC}}$. We use coercion sequences as the IFC monitor in $\lambda^{c}_{\mathtt{IFC}}$. Reducing a coercion sequence can result in a blame which errors the program. So we would like to prove that the simulation lemma holds for the coercion calculus on security labels. The precision relation on security coercions is defined in Figure 20 of the Appendix. The precision relation between two coercion sequences $\bar{c}, \bar{d}$ takes the form $\vdash \bar{c} \sqsubseteq \bar{d}$. Recall that the gradual guarantee states that replacing type annotations with $\star$ (decreasing type precision) should result in the same value for a correctly running program while adding annotations (increasing type precision) may trigger more runtime errors. The precision relation is a syntactical characterization of the runtime behaviors of programs of different type precision. We explain the intuition with two examples.

Example 4. Consider the following two programs that are related by precision because the first one has a $\star$ annotation where the second one has $\mathtt{high}$.

$$\mathtt{true}_{\mathtt{low}} : \mathtt{Bool}_{\star} : \mathtt{Bool}_{\star} \qquad \text{and} \qquad \mathtt{true}_{\mathtt{low}} : \mathtt{Bool}_{\mathtt{high}} : \mathtt{Bool}_{\star}$$

At runtime, the less precise program on the left produces value $(\mathtt{true}\ \langle \mathbf{id}(\mathtt{low});\ \mathtt{low}\,!\rangle)$ and the more precise program on the right produces $(\mathtt{true}\ \langle \mathbf{id}(\mathtt{low});\ \uparrow;\ \mathtt{high}\,!\rangle)$. The $\mathtt{true}$s are straightforwardly related; we need to show the two coercion sequences are also related:

$$\vdash \mathbf{id}(\mathtt{low});\ \mathtt{low}\,! \sqsubseteq \mathbf{id}(\mathtt{low});\ \uparrow;\ \mathtt{high}\,!$$

Starting at the beginning of the two sequences, we have $\mathbf{id}(\mathtt{low}) \sqsubseteq \mathbf{id}(\mathtt{low})$ because $\sqsubseteq$ is reflexive. Next we have $\mathtt{low}\,! \sqsubseteq \uparrow$, which makes sense because the source and targets of the two coercions are related by precision: $\mathtt{low} \sqsubseteq \mathtt{low}$ and $\star \sqsubseteq \mathtt{high}$. Finally, the coercion $\mathtt{high}\,!$ can be added to the end of the more precise sequence because both its source and target type or more precise than the target of the left-hand sequence. That is, $\star \sqsubseteq \mathtt{high}$ and $\star \sqsubseteq \star$. The injections at the ends of the two sequences, $\mathtt{low}\,!$ and $\mathtt{high}\,!$, cannot be directly related via precision because $\mathtt{low} \not\sqsubseteq \mathtt{high}$. Instead, $\mathtt{low}\,!$ is related with $\uparrow$. This underlines the indispensability of explicit subtype coercion $\uparrow$ for the purposes of proving the gradual guarantee.

Example 5. Next we consider an example where the less precise program produces a value but the more precise program encounters an error. This situation is allowed by the gradual guarantee, but the opposite one is not. We extend the example with a cast to $\mathtt{low}$ on the more precise side.

$$\mathtt{true}_{\mathtt{low}} : \mathtt{Bool}_{\star} : \mathtt{Bool}_{\star} : \mathtt{Bool}_{\star} \qquad \text{and} \qquad \mathtt{true}_{\mathtt{low}} : \mathtt{Bool}_{\mathtt{high}} : \mathtt{Bool}_{\star} : \mathtt{Bool}_{\mathtt{low}}$$

The first program again produces value $(\mathtt{true}\ \langle \mathbf{id}(\mathtt{low});\ \mathtt{low}\,!\rangle)$. The second, on the other hand, reduces to $(\mathtt{true}\ \left\langle \mathbf{id}(\mathtt{low});\ \uparrow;\ \mathtt{high}\,!;\ \mathtt{low}\,?^{p} \right\rangle) \longrightarrow^* (\mathtt{true}\ \langle \bot^{p}\ \mathtt{low}\ \mathtt{low}\rangle)$, because of the contradicting annotations $\mathtt{high}$ and $\mathtt{low}$ (note that $\mathtt{high}$ is in the middle of the sequence and both the

$$
\begin{array}{rrcl}
\text{base types} & \iota & ::= & \texttt{Unit} \mid \texttt{Bool} \\
\text{raw types} & T, S & ::= & \iota \mid A \xrightarrow{g^c} B \mid \texttt{Ref}\ (T_g) \\
\text{types} & A, B & ::= & T_g \\
\text{raw coercions} & c_r, d_r & ::= & \mathbf{id}(\iota) \mid \mathbf{Ref}\ \boldsymbol{c}\ \boldsymbol{d} \mid (\bar{d},\ \boldsymbol{c} \rightarrow \boldsymbol{d}) \\
\text{coercions} & \boldsymbol{c}, \boldsymbol{d} & ::= & c_r,\ \bar{c}
\end{array}
$$

$\boxed{V\ \langle \boldsymbol{c} \rangle \longrightarrow M}$

$$
\textit{cast}\ \frac{\bar{c} \longrightarrow^{+} \bar{d} \qquad \mathbf{NF}\ \bar{d}}{V_r\ \langle c_r,\ \bar{c}\rangle \longrightarrow V_r\ \langle c_r,\ \bar{d}\rangle}
\qquad
\textit{cast-blame}\ \frac{\bar{c} \longrightarrow^{*} \bot^{p}\ g_1\ g_2}{V_r\ \langle c_r,\ \bar{c}\rangle \longrightarrow \texttt{blame}\ p}
$$

$$
\textit{cast-id}\ \frac{}{V_r\ \langle \mathbf{id}(\iota),\ \mathbf{id}(g)\rangle \longrightarrow V_r}
\qquad
\textit{cast-comp}\ \frac{\mathbf{Irreducible}\ \boldsymbol{c}}{V_r\ \langle \boldsymbol{c}\rangle\ \langle \boldsymbol{d}\rangle \longrightarrow V_r\ \langle \boldsymbol{c}\ ⨟\ \boldsymbol{d}\rangle}
$$

Fig. 4. Syntax and semantics of coercions on values.

source and target labels of the blame coercion are low so that types are preserved). The failure is then propagated out and the term further reduces to blame $p$. The precision of coercion sequences relates $\bot$ on the right-hand side to any coercion sequence on the left so long as the respective source and target types are related via precision, in this case low $\sqsubseteq$ low and $\star \sqsubseteq$ low.

Consider the security levels (Definition 1) of both sides of Example 4, which are low on the less precise side and high on the more precise side. We observe that $\lambda^{\star}_{\texttt{IFC}}$ programs related by precision may produce values of different security: a less precise value may have lower security than a more precise value. Indeed, we prove the following for coercions in normal form:

Lemma 6 (Security is monotonic with respect to precision). *Suppose* NF $\bar{c}$ *and* NF $\bar{d}$. *If* $\vdash \bar{c} \sqsubseteq \bar{d}$, *then* $|\bar{c}| \preccurlyeq |\bar{d}|$.

Next we prove a catch-up lemma for coercion sequences, where the less precise side catches up with a more precise sequence that is in normal form. The proof is by casing on NF $\bar{d}$ first and then performing induction on the precision relation in each case.

Lemma 7 (Catching up to a more precise coercion sequence). *If* NF $\bar{d}$ *and* $\vdash \bar{c} \sqsubseteq \bar{d}$, *there exists* $\bar{c}'$ *such that* $\bar{c} \longrightarrow^{*} \bar{c}'$ *and* $\vdash \bar{c}' \sqsubseteq \bar{d}$.

Using Lemma 7, we then prove the following simulation lemma for coercion sequences:

Lemma 8 (Simulation between related coercion sequences). *If* $\vdash \bar{c} \sqsubseteq \bar{d}$ *and* $\bar{d} \longrightarrow \bar{d}'$, *there exists* $\bar{c}'$ *such that* $\bar{c} \longrightarrow^{*} \bar{c}'$ *and* $\vdash \bar{c}' \sqsubseteq \bar{d}'$.

We also prove that stamping on coercion sequences preserves precision:

Lemma 9 (Stamping preserves precision of coercion sequences). *If* $\vdash \bar{c} \sqsubseteq \bar{d}$, *then* $\vdash$ *stamp* $\bar{c}\ \ell \sqsubseteq$ *stamp* $\bar{d}\ \ell$ *and* $\vdash$ *stamp*! $\bar{c}\ \ell_1 \sqsubseteq$ *stamp*! $\bar{d}\ \ell_2$ *and* $\vdash$ *stamp*! $\bar{c}\ \ell_1 \sqsubseteq$ *stamp* $\bar{d}\ \ell_2$ *if* $\ell_1 \preccurlyeq \ell_2$.

## 4 A SECURITY COERCION CALCULUS ON VALUES

In this section we define a second coercion calculus whose purpose is to cast a program value from one type to another type. We use this coercion calculus as the representation of casts in the intermediate language $\lambda^{c}_{\texttt{IFC}}$. These coercions on values make use of the coercions on security labels that we defined in Section 3 because the types in $\lambda^{c}_{\texttt{IFC}}$ are annotated with security labels, as is usual for a static and gradually-typed IFC languages.

We begin with the definition of types in Figure 4, which is standard for gradual security type systems: each type has a security label ascription on it, which is either a specific label $\ell$ or $\star$. Function types have one extra label $g^c$, which is a static approximation of the security level of the PC while executing the function body. In a reference type $(\texttt{Ref}\ T_{\hat{g}})_g$, the label $\hat{g}$ of the referenced type also doubles as the security level of the memory cell.

The syntax and semantics for coercions on values is defined in Figure 4. Each coercion $\boldsymbol{c}$ consists of a raw coercion $c_r$ that casts the type of the value and the label coercion $\bar{c}$ that casts the security label of the value. There are three kinds of raw coercions: identity coercions $\mathbf{id}(g)$, coercions between reference types $(\mathbf{Ref}\ \boldsymbol{c}\ \boldsymbol{d})$, and coercions between function types $\left(\bar{d},\ \boldsymbol{c} \rightarrow \boldsymbol{d}\right)$. In the coercion on functions, the $\bar{d}$ casts the PC of the function. (The syntax for values is not defined until the next section, so here we remark that $V$ ranges over values, which can either be a raw value $V_r$ (constant, address, or $\lambda$) or an irreducible coercion applied to a raw value: $V_r\ \langle \boldsymbol{c} \rangle$, where there is no $M$ such that $V_r\ \langle \boldsymbol{c} \rangle \longrightarrow M$. The definition of irreducible coercion is in Figure 15 of the Appendix.)

The reduction rules in Figure 4 apply a coercion to a value, yielding a value or triggering blame. The *cast* rule normalizes the coercion $\bar{c}$ on the security label. If it normalizes to a failure coercion, the rule *cast-blame* triggers blame. We reduce identity coercions using rule *cast-id*. Finally, if the value is wrapped with an irreducible coercion, we compose the coercion with the coercion being applied (rule *cast-comp*). The composition operator $- \,⨟\, -$ is also defined in Figure 15 of the Appendix; the intuition of the composition operator is that $V_r\ \langle \boldsymbol{c} \rangle\ \langle \boldsymbol{d} \rangle$ must be contextual equivalent to $V_r\ \langle \boldsymbol{c} ⨟ \boldsymbol{d} \rangle$. There are no reduction rules specific to reference coercions $\mathbf{Ref}\ \boldsymbol{c}\ \boldsymbol{d}$ or function coercions $\left(\bar{d},\ \boldsymbol{c} \rightarrow \boldsymbol{d}\right)$ because they are irreducible coercions that wrap a value. Their action occurs when the value is used in an elimination form such as in a function call or a read or write to the reference, which we explain in the next section.

## 5 THE CAST CALCULUS $\lambda^c_{\texttt{IFC}}$

In this section we define the cast calculus $\lambda^c_{\texttt{IFC}}$ (§ 5.1), prove that $\lambda^c_{\texttt{IFC}}$ is type-safe (§ 5.2.1), and prove the main simulation lemma (§ 5.2.2) that is needed for the proof of the gradual guarantee. We conclude this section with a proof that $\lambda^c_{\texttt{IFC}}$ satisfies noninterference (§ 5.2.3).

### 5.1 Syntax, Typing, and Operational Semantics of $\lambda^c_{\texttt{IFC}}$

*5.1.1 Syntax and Typing of $\lambda^c_{\texttt{IFC}}$.* As usual, the cast calculus $\lambda^c_{\texttt{IFC}}$ is a statically-typed language that includes an explicit term for casts, written $M\ \langle \boldsymbol{c} \rangle$, where $M$ is a term and $\boldsymbol{c}$ is a coercion to be applied to the value of $M$. Furthermore, many of the operators in $\lambda^c_{\texttt{IFC}}$ have two variants, a "static" one for when the pertinent security label is statically known and the "dynamic" one for when the security label is statically unknown. The operational semantics of the "dynamic" variants involve runtime checking. The syntax and typing rules for $\lambda^c_{\texttt{IFC}}$ are shown in Figure 5 (excerpt of typing rules, full version in Figure 12 in the Appendix) and described in the following paragraphs.

A value is a raw value (constant, address or $\lambda$) or an irreducible coercion applied to a raw value.

The typing rules are syntax-directed. The typing judgment is of the form $\Gamma; \Sigma; g; \ell \vdash M \Leftarrow A$, which says that we are type-checking $\lambda^c_{\texttt{IFC}}$ term $M$ against the expected type $A$, where $\Gamma$ is the typing context, $\Sigma$ is the heap typing context, $g$ is the security label that PC is typed at, and $\ell$ is the security level of PC. Both $\Sigma$ and the security level $\ell$ play a role during runtime. The security level of the PC is constrained in rule $\vdash$*prot*, which is for the protection term that arises during reduction (we are going to discuss this rule momentarily). In the premises for sub-terms that do not immediately reduce, such as the body of a $\lambda$ and the branches of an `if`, we universally quantify the security level (as in $\forall\ell$), which helps us prove that compilation preserves types (Theorem 15). The heap context $\Sigma$ is mostly standard: looking up $\Sigma(\hat{\ell}, n)$, where $n$ is the index in part of the heap

$$
\begin{array}{rlll}
\text{variables} & x, y, z & & \\
\text{constants} & k & \in & \{\mathtt{unit}, \mathtt{true}, \mathtt{false}\} \\
\text{terms} & L, M, N & ::= & x \mid \$k \mid \mathtt{addr}\, n \mid \lambda x.\, N \mid \mathtt{app}\, L\, M\, A\, B\, \ell \mid \mathtt{app}\star\, L\, M\, A\, T \\
& & \mid & \mathtt{if}\, L\, A\, \ell\, M\, N \mid \mathtt{if}\star\, L\, T\, M\, N \mid \mathtt{let}\, x{=}M{:}A\, \mathtt{in}\, N \\
& & \mid & \mathtt{ref}\, \ell\, M \mid \mathtt{ref?}^{p}\, \ell\, M \mid\, !\, M\, A\, \ell \mid\, !\star\, M\, T \\
& & \mid & \mathtt{assign}\, L\, M\, T\, \hat{\ell}\, \ell \mid \mathtt{assign?}^{p}\, L\, M\, T\, \hat{g} \\
& & \mid & \mathtt{prot}\, PC\, \ell\, M\, A \mid M\, \langle c \rangle \mid \mathtt{blame}\; p \\
\text{raw values} & V_r, W_r & ::= & \$k \mid \mathtt{addr}\, n \mid \lambda x.\, N \\
\text{values} & V, W & ::= & V_r \mid V_r\, \langle c \rangle, \text{ where } \textbf{Irreducible}\ \boldsymbol{c}
\end{array}
$$

$$\boxed{\Gamma; \Sigma; g; \ell \vdash M \Leftarrow A}$$

$$
\vdash app\ \frac{\Gamma;\Sigma;\ell';\ell'' \vdash L \Leftarrow (A \xrightarrow{\ell' \vee \ell} B)_{\ell} \quad \Gamma;\Sigma;\ell';\ell'' \vdash M \Leftarrow A \quad C = stamp\ B\ \ell}{\Gamma;\Sigma;\ell';\ell'' \vdash \mathtt{app}\, L\, M\, A\, B\, \ell \Leftarrow C}
\qquad
\vdash app\star\ \frac{\Gamma;\Sigma;g;\ell \vdash L \Leftarrow (A \xrightarrow{\star} (T_{\star}))_{\star} \quad \Gamma;\Sigma;g;\ell \vdash M \Leftarrow A}{\Gamma;\Sigma;g;\ell \vdash \mathtt{app}\star\, L\, M\, A\, T \Leftarrow T_{\star}}
$$

$$
\vdash assign\ \frac{\Gamma;\Sigma;\ell';\ell'' \vdash L \Leftarrow (\mathtt{Ref}\ T_{\hat{\ell}})_{\ell} \quad \Gamma;\Sigma;\ell';\ell'' \vdash M \Leftarrow T_{\hat{\ell}} \quad \boxed{\ell' \vee \ell \preccurlyeq \hat{\ell}}}{\Gamma;\Sigma;\ell';\ell'' \vdash \mathtt{assign}\, L\, M\, T\, \hat{\ell}\, \ell \Leftarrow \mathtt{Unit}_{\mathtt{low}}}
\qquad
\vdash assign?\ \frac{\Gamma;\Sigma;g;\ell \vdash L \Leftarrow (\mathtt{Ref}\ T_{\hat{g}})_{\star} \quad \Gamma;\Sigma;g;\ell \vdash M \Leftarrow T_{\hat{g}}}{\Gamma;\Sigma;g;\ell \vdash \mathtt{assign?}^{p}\, L\, M\, T\, \hat{g} \Leftarrow \mathtt{Unit}_{\mathtt{low}}}
$$

$$
\vdash prot\ \frac{\Gamma;\Sigma;g';|PC| \vdash M \Leftarrow A \quad \vdash PC \Leftarrow g' \quad \ell' \vee \ell \preccurlyeq |PC| \quad B = stamp\ A\ \ell}{\Gamma;\Sigma;g;\ell' \vdash \mathtt{prot}\, PC\, \ell\, M\, A \Leftarrow B}
$$

Fig. 5. Syntax and selected typing rules of the cast calculus $\lambda^{c}_{\mathtt{IFC}}$. The side condition that enforces the heap policy statically during assignment is highlighted

with security $\hat{\ell}$, gives us a raw type. Each memory cell is associated with a specific security label $\hat{\ell}$, which is specified by the programmer when that cell is created.

As the typing rules always stay in checking mode, constants, addresses, and $\lambda$s do not carry any label. The security of these raw values is in their types: for example, $\mathtt{addr}\ n \Leftarrow (\mathtt{Ref}\ T_{\hat{\ell}})_{\ell}$ says that the security of the address $n$ itself is $\ell$ and it points to a memory cell labeled $\hat{\ell}$.

The typing rules for the static variants are similar to the typing rules in a static security type system. For example, in the static function application rule $\vdash app$, both top-level labels on the function type as well as the security label that the current PC is typed at are static and the rule mirrors one in a static system. On the other hand, in the dynamic version of application rule $\vdash app\star$, both top-level labels as well as the label of the co-domain type are $\star$ and PC is allowed to be typed at $\star$, indicating the presence of injections. As another example, in the static version of memory assignment rule $\vdash assign$, all labels to perform the heap policy check, including the security of the memory address itself ($\ell$), the security of the memory cell that the address references ($\hat{\ell}$), and the security of PC ($\ell'$) are known statically and satisfy $\ell' \vee \ell \preccurlyeq \hat{\ell}$. At runtime, this static assignment can happen directly without any runtime overhead (as is shown in the example of Section 2.2). Its dynamic counterpart rule $\vdash assign?$ does not maintain these static security invariants and thus requires runtime NSU checking, which is implemented as a projection on PC.

As we shall see in the reduction rules, the semantics of the protection term `prot` performs two things: (1) `prot` promotes the security of the value reduced from its body by level $\ell$ (2) it uses a new PC to reduce its body. However, the new PC cannot be any label expression. It has to be one with higher security than both the current PC and the security level $\ell$. We capture this invariant

$$\boxed{M \mid \mu \mid PC \longrightarrow N \mid \mu'}$$

$$\textit{prot-ctx}\ \frac{M \mid \mu \mid PC' \longrightarrow M' \mid \mu'}{\texttt{prot}\, PC'\, \ell\, M\, A \mid \mu \mid PC \longrightarrow \texttt{prot}\, PC'\, \ell\, M'\, A \mid \mu'} \qquad \textit{prot-val}\ \frac{}{\texttt{prot}\, PC'\, \ell\, V\, A \mid \mu \mid PC \longrightarrow \textit{stamp-val}\, V\, A\, \ell \mid \mu}$$

$$\textit{cast}\ \frac{V\ \langle \boldsymbol{c} \rangle \longrightarrow M}{V\ \langle \boldsymbol{c} \rangle \mid \mu \mid PC \longrightarrow M \mid \mu} \qquad \beta\ \frac{}{\texttt{app}\, (\lambda x.\, N)\, V\, A\, B\, \ell \mid \mu \mid PC \longrightarrow \texttt{prot}\, (\textit{stamp}\ PC\ \ell)\, \ell\, (N[x := V])\, B \mid \mu}$$

$$\textit{app-cast}\ \frac{\mathbf{NF}\ \bar{c} \qquad (\textit{stamp}\ PC\ \ell)\ \langle \bar{d} \rangle \longrightarrow^* PC' \qquad V\ \langle \boldsymbol{c} \rangle \longrightarrow^* W}{\texttt{app}\, (\lambda x.\, N\ \langle \bar{d},\ \boldsymbol{c} \rightarrow \boldsymbol{d},\ \bar{c} \rangle)\, V\, C\, D\, \ell \mid \mu \mid PC \longrightarrow \texttt{prot}\, PC'\, \ell\, ((N[x := W])\ \langle \boldsymbol{d} \rangle)\, D \mid \mu}$$

$$\textit{app}\star\textit{-cast}\ \frac{\mathbf{NF}\ \bar{c} \qquad (\textit{stamp!}\ PC\ |\bar{c}|)\ \langle \bar{d} \rangle \longrightarrow^* PC' \qquad V\ \langle \boldsymbol{c} \rangle \longrightarrow^* W}{\texttt{app}\star\, (\lambda x.\, N\ \langle \bar{d},\ \boldsymbol{c} \rightarrow \boldsymbol{d},\ \bar{c} \rangle)\, V\, C\, T \mid \mu \mid PC \longrightarrow \texttt{prot}\, PC'\, |\bar{c}|\, ((N[x := W])\ \langle \boldsymbol{d} \rangle)\, (T_\star) \mid \mu}$$

$$\textit{if-true}\ \frac{}{\texttt{if}\, \$\texttt{true}\, A\, \ell\, M\, N \mid \mu \mid PC \longrightarrow \texttt{prot}\, (\textit{stamp}\ PC\ \ell)\, \ell\, M\, A \mid \mu}$$

$$\textit{if}\star\textit{-true-cast}\ \frac{\mathbf{NF}\ \bar{c}}{\texttt{if}\star\, (\$\texttt{true}\ \langle \mathbf{id}(\texttt{Bool}),\ \bar{c} \rangle)\, T\, M\, N \mid \mu \mid PC \longrightarrow \texttt{prot}\, (\textit{stamp!}\ PC\ |\bar{c}|)\, |\bar{c}|\, M\, (T_\star) \mid \mu}$$

$$\textit{ref}\ \frac{n\ \mathbf{FreshIn}\ \mu(\ell)}{\texttt{ref}\, \ell\, V \mid \mu \mid PC \longrightarrow \texttt{addr}\, n \mid (\mu, \ell \mapsto n \mapsto V)} \qquad \textit{ref?}\ \frac{n\ \mathbf{FreshIn}\ \mu(\ell) \qquad PC\ \langle \star \Rightarrow^p \ell \rangle \longrightarrow^* PC'}{\texttt{ref?}^p\, \ell\, V \mid \mu \mid PC \longrightarrow \texttt{addr}\, n \mid (\mu, \ell \mapsto n \mapsto V)}$$

$$\textit{ref?-blame}\ \frac{PC\ \langle \star \Rightarrow^p \ell \rangle \longrightarrow^* \texttt{blame}\ q}{\texttt{ref?}^p\, \ell\, V \mid \mu \mid PC \longrightarrow \texttt{blame}\ q \mid \mu} \qquad \textit{assign}\ \frac{}{\texttt{assign}\, (\texttt{addr}\, n)\, V\, T\, \hat{\ell}\, \ell \mid \mu \mid PC \longrightarrow \$\texttt{unit} \mid [\hat{\ell} \mapsto n \mapsto V]\, \mu}$$

$$\textit{assign?-cast}\ \frac{\mathbf{NF}\ \bar{c} \qquad \vdash \boldsymbol{c} : T_g \Rightarrow S_{\hat{\ell}} \qquad \vdash \boldsymbol{d} : S_{\hat{\ell}} \Rightarrow T_g \qquad (\textit{stamp!}\ PC\ |\bar{c}|)\ \langle \star \Rightarrow^p \hat{\ell} \rangle \longrightarrow^* PC' \qquad V\ \langle \boldsymbol{c} \rangle \longrightarrow^* W}{\texttt{assign?}^p\ (\texttt{addr}\, n\ \langle \mathbf{Ref}\ \boldsymbol{c}\ \boldsymbol{d},\ \bar{c} \rangle)\, V\, T\, g \mid \mu \mid PC \longrightarrow \$\texttt{unit} \mid [\hat{\ell} \mapsto n \mapsto W]\, \mu}$$

$$\textit{deref}\ \frac{\mu(\hat{\ell}, n) = V}{!\ (\texttt{addr}\, n)\, T_{\hat{\ell}}\, \ell \mid \mu \mid PC \longrightarrow \texttt{prot_}\, \ell\, V\, T_{\hat{\ell}} \mid \mu}$$

$$\textit{deref}\star\textit{-cast}\ \frac{\mathbf{NF}\ \bar{c} \qquad \vdash \boldsymbol{c} : S_\star \Rightarrow T_\ell \qquad \vdash \boldsymbol{d} : T_\ell \Rightarrow S_\star \qquad \mu(\ell, n) = V}{!\star\ (\texttt{addr}\, n\ \langle \mathbf{Ref}\ \boldsymbol{c}\ \boldsymbol{d},\ \bar{c} \rangle)\, S \mid \mu \mid PC \longrightarrow \texttt{prot_}\, |\bar{c}|\, (V\ \langle \boldsymbol{d} \rangle)\, (S_\star) \mid \mu}$$

Fig. 6. Selected semantics rules for $\lambda^{c}_{\texttt{IFC}}$. NSU checks are represented using label expressions ( highlighted )

in side condition $\ell' \vee \ell \preccurlyeq |PC|$ in rule $\vdash$*prot*, where $\ell'$ is the security of the current PC and $|PC|$ is the security for the new PC. The invariant is used in the proof of noninterference.

*5.1.2 Operational Semantics for $\lambda^{c}_{\texttt{IFC}}$.* We show the interesting rules of the operational semantics of $\lambda^{c}_{\texttt{IFC}}$ in Figure 6 (full version in Figure 16 and 17 of the Appendix). The reduction relation takes the form $M \mid \mu \mid PC \longrightarrow N \mid \mu'$, which reduces the configuration of term $M$ and heap $\mu$ under the label expression $PC$ to another configuration $N$ and $\mu'$. The heap is a map $(\ell, n) \mapsto V$, where a cell is indexed by its security level $\ell$ and by index $n$ among the cells of $\ell$. The predicate ($n$ **FreshIn** $\mu(\ell)$) means that the index $n$ is fresh (not already in use) among all cells with security $\ell$; when performing a lookup, $\mu(\ell, n) = V$ retrieves the value $V$ at index $n$ whose security level is $\ell$.

***Protection terms.*** Following standard approaches to IFC, a protection term `prot` $PC'\ \ell\ M\ A$ has two functionalities: (1) it ensures that the reduction inside $M$ does not leak information through heap write operations (2) it promotes the security level of the computation result of $M$ to at least level $\ell$. The first functionality is achieved by switching to $PC$' from the current $PC$ when reducing the body $M$ (rule *prot-ctx*). Recall Section 5.1.1, the typing of `prot` makes sure that $PC$' has the correct security that is at least as secure as both $PC$ and $\ell$. The second functionality is achieved by stamping the value produced by the body of `prot` (rule *prot-val*). The stamping of values in $\lambda^c_{\mathtt{IFC}}$ (Figure 14 in the Appendix) is analogous to stamping of label expressions and turns to the stamping operation for coercions on labels (Figure 2) in a similar way. Again there are two cases, because the value is either (1) a raw value or (2) a coercion-wrapped value. Suppose the value is a raw value, if its type has `low` security and is stamped with `high`, the value becomes wrapped with a subtype coercion; otherwise the value stays unchanged. Otherwise, if the value is wrapped with an irreducible coercion, we stamp the top-level coercion sequence.

***Function Application.*** The $\beta$ rule is standard for IFC languages. It generates a `prot` term with the specific security label $\ell$ that comes from the label on the $\lambda$, preventing implicit flow from the function being applied through both the computation result and the heap. The *app-cast* rule applies a function wrapped in a function coercion to a value $V$. The application is "static", so the security level of `prot` comes from the function type just like $\beta$. The function coercion is distributed into its domain coercion $\boldsymbol{c}$, its co-domain coercion $\boldsymbol{d}$, and the coercion on PC $\bar{d}$. The coercion $\bar{c}$ is not used because the function type is fully static, so its security is already indicated by its type. The domain coercion $\boldsymbol{c}$ casts the input of the function $V$ to $W$ and $W$ is substituted into the body of the $\lambda$. The substituted body goes through the co-domain cast $\boldsymbol{d}$, and is then protected by $\ell$ using `prot`. The stamped PC casts to $PC'$ by $\bar{d}$ and $PC'$ is used as the PC for `prot`. The rule *app★-cast* is similar to *app-cast* except for two things: (1) the PC is stamped and then injected using *stamp!* to preserve types (2) the security level of the function proxy used in protection is indicated by $|\bar{c}|$ instead, because the top-level security label of the function is statically unknown ($\star$).

***If-conditional.*** The static rule *if-true* is standard; the `if` term reduces a `prot` whose security $\ell$ comes from the type of the branch condition, guarding against implicit flow. The rationale of *if★-true-cast* follows that of *app★-cast*: (1) a *stamp!* is generated to stamp and then inject the PC and (2) the security of `prot` is retrieved from the coercion in the branch condition.

***NSU and heap operations.*** Let us consider reference creation first. A "static" reference creation is secure because its typing (rule $\vdash$*-ref*, Figure 12 of the Appendix) already enforces the heap policy. Consequently, the allocation can happen directly (rule *ref*) without any runtime checking. Rule *ref?* does the same reference creation but with NSU checking, by casting the current PC to the security $\ell$ of the newly created memory cell. The coerce function ($- \Rightarrow^{-} -$) takes two security labels and a blame label to generate a coercion on labels. In this case, $\star \Rightarrow^{p} \ell$ generates $\mathbf{id}(\star)$; $\ell\,?^{p}$, which performs a projection whose target is $\ell$. If the projected PC reduces to a blame, it means that NSU checking fails so we lift the blame to $\lambda^c_{\mathtt{IFC}}$. Assignment follows the same pattern: a static assignment can happen directly, while `assign?` requires NSU checking, by stamping the current PC with the security indicated in the coercion and then projecting to $\hat{\ell}$, where $\hat{\ell}$ is the security of the heap cell. The input coercion $\boldsymbol{c}$ is applied before the value is stored into the cell.

The rule for static dereferencing *deref* looks up index $n$ in all memory cells with security level $\hat{\ell}$. The value from the lookup is protected with $\ell$, the top-level security label of the reference type. The PC of the `prot` does not matter, because $V$ is already a value and will not reduce. The rule *deref★-cast* dereferences a reference proxy. It looks up the value $V$ in the heap, applies the output coercion $\boldsymbol{d}$, and generates a `prot` with the security of the coercion $|\bar{c}|$. The PC of `prot` does not matter in this case either, because applying a coercion is pure and does not produce side effects.

## 5.2 Meta-theoretical Results of $\lambda^c_{\mathtt{IFC}}$

We prove type safety for $\lambda^c_{\mathtt{IFC}}$ (§ 5.2.1) and the main simulation lemma for $\lambda^c_{\mathtt{IFC}}$ (§ 5.2.2) used in the proof of the gradual guarantee (§ 6.4). We also prove that $\lambda^c_{\mathtt{IFC}}$ satisfies noninterference (§ 5.2.3).

*5.2.1 Type Safety.* We show that $\lambda^c_{\mathtt{IFC}}$ is type safe by proving progress and preservation. Progress says that a well-typed $\lambda^c_{\mathtt{IFC}}$ term does not get stuck. The term is either a value or a blame, which does not reduce, or the term takes one reduction step. Heap well-typedness is defined point-wise.

THEOREM 10 (PROGRESS). *Suppose PC is well-typed:* $\vdash PC \Leftarrow g$, *M is well-typed:* $\emptyset; \Sigma; g; |PC| \vdash M \Leftarrow A$, *and the heap* $\mu$ *is also well-typed:* $\Sigma \vdash \mu$. *Then either (1) M is a value or (2) M is a blame:* $M = \texttt{blame}\ p$ *or (3) M can take a reduction step:* $M \mid \mu \mid PC \longrightarrow N \mid \mu'$ *for some N and* $\mu'$.

The operation semantics of $\lambda^{c}_{\texttt{IFC}}$ preserves types and the well-typedness of heap:

THEOREM 11 (PRESERVATION). *Suppose PC is well-typed:* $\vdash PC \Leftarrow g$, *M is well-typed:* $\emptyset; \Sigma; g; |PC| \vdash M \Leftarrow A$ *and the heap* $\mu$ *is also well-typed:* $\Sigma \vdash \mu$. *If* $M \mid \mu \mid PC \longrightarrow N \mid \mu'$, *there exists* $\Sigma'$ *s.t* $\Sigma' \supseteq \Sigma$, $\emptyset; \Sigma'; g; |PC| \vdash N \Leftarrow A$, *and* $\Sigma' \vdash \mu'$.

*5.2.2 Simulation Between* $\lambda^{c}_{\texttt{IFC}}$ *Terms of Different Precision.* The main simulation lemma says that if two terms are related by *precision* and the more precise side takes one step, then the less precise side is able to multi-step and get back in sync. The precision relation is in form $\Gamma; \Gamma'; \Sigma; \Sigma'; g; g'; \ell; \ell' \vdash M \sqsubseteq M' \Leftarrow A \sqsubseteq A'$, where $\Gamma; \Sigma; g; \ell$ corresponds to the typing context, heap context, type of $PC$, and security of $PC$ of the less precise term $M$ and $\Gamma'; \Sigma'; g'; \ell'$ is for those of the more precise term $M'$. The types of the two terms, $A$ and $A'$, are related by precision between types. The intuition between this precision relation is that casts are allowed to appear in different places between the more precise and the less precise $\lambda^{c}_{\texttt{IFC}}$ terms. Moreover, the casts must be in shapes that preserve the precision of $\lambda^{\star}_{\texttt{IFC}}$ (more or fewer static type annotations provided by the programmer). According to the gradual guarantee, the more precise side is allowed to signal more blames, so there is a rule ($\sqsubseteq$-*blame*) that relates $\texttt{blame}\ p$ to any term $M$ on the less precise side as long as their types are in sync. We list selected precision rules in Figure 21 of the Appendix.

With the precision relation of $\lambda^{c}_{\texttt{IFC}}$ defined, we first state the catch-up lemma, which catches up to a more-precise value by multi-stepping on the less-precise side:

LEMMA 12 (CATCHING UP TO MORE PRECISE). *If term M and value* $V'$ *are related by precision:*

$$\Gamma; \Gamma'; \Sigma; \Sigma'; g; g'; \ell; \ell' \vdash M \sqsubseteq V' \Leftarrow A \sqsubseteq A'$$

*then there exists value V s.t* $M \mid \mu \mid PC \longrightarrow^{*} V \mid \mu$ *and* $\Gamma; \Gamma'; \Sigma; \Sigma'; g; g'; \ell; \ell' \vdash V \sqsubseteq V' \Leftarrow A \sqsubseteq A'$.

We prove the main simulation lemma using the catch-up lemma. Heap precision is defined point-wise similar to the definition of heap well-typedness.

LEMMA 13 (SIMULATION BETWEEN MORE PRECISE AND LESS PRECISE $\lambda^{c}_{\texttt{IFC}}$ TERMS). *Suppose* $PC, PC'$ *are related by precision:* $\vdash PC \sqsubseteq PC' \Leftarrow g \sqsubseteq g'$. *Moreover suppose* $M, M'$ *are related by precision:*

$$\emptyset; \emptyset; \Sigma_1; \Sigma_1'; g; g'; |PC|; |PC'| \vdash M \sqsubseteq M' \Leftarrow A \sqsubseteq A'$$

*heap contexts* $\Sigma_1$, $\Sigma_1'$ *are related by precision:* $\Sigma_1 \sqsubseteq \Sigma_1'$, *the initial heaps* $\mu_1$, $\mu_1'$ *are also related by precision:* $\Sigma_1; \Sigma_1' \vdash \mu_1 \sqsubseteq \mu_1'$.
*If* $M' \mid \mu_1' \mid PC' \longrightarrow N' \mid \mu_2'$, *there exists* $\Sigma_2, \Sigma_2', N, \mu_2$ *s.t* $\Sigma_2 \supseteq \Sigma_1$, $\Sigma_2' \supseteq \Sigma_1'$, $\Sigma_2 \sqsubseteq \Sigma_2'$,

$$M \mid \mu_1 \mid PC \longrightarrow^{*} N \mid \mu_2$$

*the resulting terms are related by precision:* $\emptyset; \emptyset; \Sigma_2; \Sigma_2'; g; g'; |PC|; |PC'| \vdash N \sqsubseteq N' \Leftarrow A \sqsubseteq A'$ *and the resulting heaps are also related by precision:* $\Sigma_2; \Sigma_2' \vdash \mu_2 \sqsubseteq \mu_2'$.

*5.2.3 Noninterference.* We prove that $\lambda^{c}_{\texttt{IFC}}$ satisfies termination-insensitive noninterference in Section 11 of the Appendix. The statement of termination-insensitive noninterference says that if we run a program with different high-security inputs in two executions, then their low-security output values should be the related (e.g the same boolean):

LEMMA 14 (NONINTERFERENCE FOR $\lambda^{c}_{\texttt{IFC}}$). *If M is well-typed:* $(x{:}\texttt{Bool}_{\texttt{high}}); \emptyset; \texttt{low}; \texttt{low} \vdash M : \texttt{Bool}_{\texttt{low}}$ *and* $M[x := \$\, b_1] \mid \emptyset \mid \texttt{low} \longrightarrow^{*} V_1 \mid \mu_1$ *and* $M[x := \$\, b_2] \mid \emptyset \mid \texttt{low} \longrightarrow^{*} V_2 \mid \mu_2$, *then* $V_1 = V_2$.

$$\begin{array}{llll} \text{terms} & L, M, N & ::= & x \mid (\$\, k)_{\ell} \mid (\lambda^{g} x{:}A.\, N)_{\ell} \mid (L\, M)^{p} \\ & & \mid & (\texttt{if } L \texttt{ then } M \texttt{ else } N)^{p} \mid \texttt{let } x = M \texttt{ in } N \\ & & \mid & (\texttt{ref } \ell\, M)^{p} \mid\, !^{p}\, M \mid (L := M)^{p} \mid (M : A)^{p} \end{array}$$

$\boxed{\Gamma; g \vdash M : A}$

$$\vdash lam\ \frac{(\Gamma, x{:}A); g \vdash N : B}{\Gamma; g' \vdash (\lambda^{g} x{:}A.\, N)_{\ell} : (A \xrightarrow{g} B)_{\ell}} \qquad \vdash assign\ \frac{\Gamma; g' \vdash L : (\texttt{Ref } T_{\hat{g}})_{g} \quad \Gamma; g' \vdash M : A \quad A \lesssim T_{\hat{g}} \quad g \precsim \hat{g} \quad g' \precsim \hat{g}}{\Gamma; g' \vdash (L := M)^{p} : \texttt{Unit}_{\texttt{low}}}$$

Fig. 7. Syntax and selected typing rules of $\lambda^{\star}_{\texttt{IFC}}$ (highlighted security labels $\ell$ default to low if omitted)

## 6 THE $\lambda^{\star}_{\texttt{IFC}}$ LANGUAGE WITH GRADUAL INFORMATION-FLOW CONTROL

In this section, we first define the gradual language $\lambda^{\star}_{\texttt{IFC}}$ in Section 6.1. It is similar to $\text{GSL}_{\text{Ref}}$ with respect to syntax and typing rules. The main syntactic difference is that in $\lambda^{\star}_{\texttt{IFC}}$, the security labels of literals and newly created memory cells default to a specific label such as low, while in $\text{GSL}_{\text{Ref}}$ they default to a runtime unknown security level $\star$. We show that $\lambda^{\star}_{\texttt{IFC}}$ can be compiled into our cast calculus $\lambda^{c}_{\texttt{IFC}}$ and the compilation preserves types in Section 6.2. As a result, the semantics of $\lambda^{\star}_{\texttt{IFC}}$ can be defined by the operational semantics of $\lambda^{c}_{\texttt{IFC}}$. In Section 6.3, we prove the noninterference of $\lambda^{\star}_{\texttt{IFC}}$ as a corollary of the noninterference lemma for $\lambda^{c}_{\texttt{IFC}}$ and compilation preserves types. Finally, we prove the gradual guarantee for $\lambda^{\star}_{\texttt{IFC}}$ as a corollary of the main simulation lemma of $\lambda^{c}_{\texttt{IFC}}$ (Lemma 13), thus solving the tension discovered by Toro et al. [2018].

### 6.1 Syntax and Typing of the Surface Language $\lambda^{\star}_{\texttt{IFC}}$

The syntax and selected typing rules of $\lambda^{\star}_{\texttt{IFC}}$ are shown in Figure 7 (full version in Figure 10 of the Appendix). They are directly adapted from those of $\text{GSL}_{\text{Ref}}$, by changing the security labels on values to disallow the $\star$ label.

The rule $\vdash assign$ includes two side conditions $g \precsim \hat{g}$ and $g' \precsim \hat{g}$. If $g$, $g'$, and $\hat{g}$ are all specific security labels, the heap policy is enforced statically, because the typing tells us that the security of the current PC as well as the memory address itself is lower than or equal to the security of the memory cell, thus no implicit flow through the heap. Indeed, we are going to see in the next section that an assignment where all three labels are statically known generates an static `assign`. The semantics of `assign` does not perform NSU because the static check on its typing rule suffices. Relating to the $\lambda^{\star}_{\texttt{IFC}}$ program at the end of Section 2.2, there is no runtime NSU check, because the program generates a static `assign` that enforces the heap policy statically.

### 6.2 Compiling $\lambda^{\star}_{\texttt{IFC}}$ to $\lambda^{c}_{\texttt{IFC}}$: Cast Insertion

The compile function takes the form $\mathcal{C}\ M = M'$, where $M$ is a $\lambda^{\star}_{\texttt{IFC}}$ program and $M'$ is a $\lambda^{c}_{\texttt{IFC}}$ term. Consider the case for assignment:

$$\mathcal{C}\ (L := M)^{p} = \begin{cases} \texttt{assign}\,(\mathcal{C}\ L)\,((\mathcal{C}\ M)\ \langle \boldsymbol{c}_2 \rangle)\,T\,\hat{g}\,g & \text{, if } g, g' \text{ and } \hat{g} \text{ are all specific} \\ \texttt{assign?}^{p}\ ((\mathcal{C}\ L)\ \langle \boldsymbol{c}_1 \rangle)\,((\mathcal{C}\ M)\ \langle \boldsymbol{c}_2 \rangle)\,T\,\hat{g} & \text{, otherwise} \end{cases}$$

$$\text{where } \boldsymbol{c}_1 = (\texttt{Ref } T_{\hat{g}})_{g} \Rightarrow^{p} (\texttt{Ref } T_{\hat{g}})_{\star}, \boldsymbol{c}_2 = A \Rightarrow^{p} T_{\hat{g}}, \Gamma; g' \vdash L : (\texttt{Ref } T_{\hat{g}})_{g}, \Gamma; g' \vdash M : A$$

If $g$, $g'$, and $\hat{g}$ are specific, the check for heap policy can be statically justified. We recursively compile both $L$ and $M$ and generate a static `assign`. We cast $M'$ using coercion $\boldsymbol{c}_2$, which casts from the type of $M'$ to the type of the memory cell. The coercion is produced by the coerce function, $(- \Rightarrow^{-} -)$, which takes two types and a blame label, returning a coercion on values by calling the coerce function of labels on each pair of security labels inside those two types. If at least one of the three security labels is $\star$, the typing information is insufficient to justify the assignment. The compilation produces an `assign?` instead, whose semantics performs NSU checking at runtime.

Compilation from $\lambda^{\star}_{\mathtt{IFC}}$ to $\lambda^{c}_{\mathtt{IFC}}$ preserves types:

Theorem 15 (Compilation preserves types). *If* $\Gamma; g \vdash M : A$, *then* $\Gamma; \emptyset; g; \mathtt{low} \vdash \mathcal{C}\ M \Leftarrow A$.

## 6.3 Noninterference for $\lambda^{\star}_{\mathtt{IFC}}$

The noninterference theorem of $\lambda^{\star}_{\mathtt{IFC}}$ is a straightforward corollary of the noninterference lemma for $\lambda^{c}_{\mathtt{IFC}}$ (Section 5.2.3) and compilation preserves types. The proof is in Section 11 of the Appendix.

Theorem 16 (Noninterference for $\lambda^{\star}_{\mathtt{IFC}}$). *Suppose a* $\lambda^{\star}_{\mathtt{IFC}}$ *program* $M$ *is well-typed:* $(x{:}\mathtt{Bool}_{\mathtt{high}}); \mathtt{low} \vdash M : \mathtt{Bool}_{\mathtt{low}}$. *If for any boolean inputs* $b_1, b_2$

$$(\mathcal{C}\ M)[x := \$\, b_1] \mid \emptyset \mid \mathtt{low} \longrightarrow^{*} V_1 \mid \mu_1 \quad \text{and} \quad (\mathcal{C}\ M)[x := \$\, b_2] \mid \emptyset \mid \mathtt{low} \longrightarrow^{*} V_2 \mid \mu_2$$

*then the resulting values* $V_1 = V_2$.

## 6.4 The Gradual Guarantee

Finally, we state the gradual guarantee of $\lambda^{\star}_{\mathtt{IFC}}$ and prove it as a corollary of Lemma 13. The definition of term precision for $\lambda^{\star}_{\mathtt{IFC}}$ is in Figure 19 of the Appendix.

Theorem 17 (Gradual guarantee). *Suppose* $M$ *and* $M'$ *are well-typed terms in* $\lambda^{\star}_{\mathtt{IFC}}$ *that are related by precision, that is* $\vdash M \sqsubseteq M'$. *So* $\emptyset; \mathtt{low} \vdash M : A$, $\emptyset; \mathtt{low} \vdash M' : A'$, *and* $A \sqsubseteq A'$. *If the compilation of* $M'$ *reduces to a value:* $\mathcal{C}\ M' \mid \emptyset \mid \mathtt{low} \longrightarrow^{*} V' \mid \mu'$
*there exists a value* $V$ *and heap* $\mu$ *s.t. the compilation of* $M$ *reduces to* $V$*:* $\mathcal{C}\ M \mid \emptyset \mid \mathtt{low} \longrightarrow^{*} V \mid \mu$
*and the resulting values are related by precision for some* $\Sigma, \Sigma'$*:*

$$\emptyset; \emptyset; \Sigma; \Sigma'; \mathtt{low}; \mathtt{low}; \mathtt{low}; \mathtt{low} \vdash V \sqsubseteq V' \Leftarrow A \sqsubseteq A'$$

Proof. Compilation preserves precision, so $\emptyset; \emptyset; \emptyset; \emptyset; \mathtt{low}; \mathtt{low}; \mathtt{low}; \mathtt{low} \vdash \mathcal{C}\ M \sqsubseteq \mathcal{C}\ M' \Leftarrow A \sqsubseteq A'$. We then proceed by induction on the reduction of $\mathcal{C}\ M'$ to a value $V'$, using Lemma 13 to show that $\mathcal{C}\ M$ reduces to a corresponding term at each step. So we have $\mathcal{C}\ M \longrightarrow^{*} N$ where $N \sqsubseteq V'$ for some $N$. We then apply Lemma 12 to show that $N$ reduces to a value $V$ where $V \sqsubseteq V'$. □

# 7 CONCLUSION

We presented the design of a gradual information-flow language $\lambda^{\star}_{\mathtt{IFC}}$ that satisfies both noninterference and the gradual guarantee while maintaining the principle of type-based reasoning. The key to the design of $\lambda^{\star}_{\mathtt{IFC}}$ is to walk back the decision in $\mathrm{GSL}_{\mathsf{Ref}}$ to include the unknown label $\star$ among the runtime security labels. So $\lambda^{\star}_{\mathtt{IFC}}$ takes a more standard approach to gradually-typed IFC: the $\star$ label can be used in type annotations but not as the security level of a runtime value. The $\lambda^{\star}_{\mathtt{IFC}}$ language is defined by translation to a cast calculus $\lambda^{c}_{\mathtt{IFC}}$. This intermediate language employs a coercion calculus to express the implicit conversions between more-or-less precise parts of the program. We proved that $\lambda^{\star}_{\mathtt{IFC}}$ satisfies termination-insensitive noninterference. We proved that $\lambda^{\star}_{\mathtt{IFC}}$ satisfies the gradual guarantee and mechanized the result in Agda.

# DATA-AVAILABILITY STATEMENT

The Agda code of this paper is in the supplementary material [Chen 2024].

# ACKNOWLEDGMENTS

This material is based upon work supported by the National Science Foundation under Grant No. 1763922.

## REFERENCES

Aslan Askarov and Andrei Sabelfeld. 2009. Tight Enforcement of Information-Release Policies for Dynamic Languages. In *2009 22nd IEEE Computer Security Foundations Symposium*. 43–59. https://doi.org/10.1109/CSF.2009.22

Thomas H Austin and Cormac Flanagan. 2009. Efficient purely-dynamic information flow analysis. In *Proceedings of the ACM SIGPLAN Fourth Workshop on Programming Languages and Analysis for Security*. 113–124.

Thomas H. Austin, Tommy Schmitz, and Cormac Flanagan. 2017. Multiple Facets for Dynamic Information Flow with Exceptions. *ACM Trans. Program. Lang. Syst.* 39, 3, Article 10 (may 2017), 56 pages. https://doi.org/10.1145/3024086

Arthur Azevedo de Amorim, Matt Fredrikson, and Limin Jia. 2020. Reconciling noninterference and gradual typing. In *Logic in Computer Science (LICS)*.

Abhishek Bichhawat, McKenna McCall, and Limin Jia. 2021. Gradual Security Types and Gradual Guarantees. In *2021 IEEE 34th Computer Security Foundations Symposium (CSF)*. IEEE, 1–16.

Pablo Buiras, Dimitrios Vytiniotis, and Alejandro Russo. 2015. HLIO: Mixing Static and Dynamic Typing for Information-Flow Control in Haskell. In *Proceedings of the 20th ACM SIGPLAN International Conference on Functional Programming* (Vancouver, BC, Canada) *(ICFP 2015)*. Association for Computing Machinery, New York, NY, USA, 289–301. https://doi.org/10.1145/2784731.2784758

Deepak Chandra and Michael Franz. 2007. Fine-Grained Information Flow Analysis and Enforcement in a Java Virtual Machine. In *Twenty-Third Annual Computer Security Applications Conference (ACSAC 2007)*. 463–475. https://doi.org/10.1109/ACSAC.2007.37

Tianyu Chen. 2024. *cty12/pldi2024-ae: PLDI Release 2.* https://doi.org/10.5281/zenodo.10933110

Tianyu Chen and Jeremy G. Siek. 2022. Mechanized Noninterference for Gradual Security. arXiv:2211.15745 [cs.PL]

Dorothy E Denning. 1976. A lattice model of secure information flow. *Commun. ACM* 19, 5 (1976), 236–243.

Dominique Devriese and Frank Piessens. 2010. Noninterference through Secure Multi-execution. In *2010 IEEE Symposium on Security and Privacy*. 109–124. https://doi.org/10.1109/SP.2010.15

Tim Disney and Cormac Flanagan. 2011a. Gradual Information Flow Typing. In *Workshop on Script to Program Evolution*.

Tim Disney and Cormac Flanagan. 2011b. Gradual information flow typing. In *International workshop on scripts to programs*.

L. Fennell and P. Thiemann. 2013. Gradual Security Typing with References. In *2013 IEEE 26th Computer Security Foundations Symposium*. 224–239. https://doi.org/10.1109/CSF.2013.22

Luminous Fennell and Peter Thiemann. 2015. LJGS: Gradual Security Types for Object-Oriented Languages. In *Workshop on Foundations of Computer Security (FCS)*.

Robert Bruce Findler and Matthias Felleisen. 2002. *Contracts for Higher-Order Functions*. Technical Report NU-CCS-02-05. Northeastern University.

Ronald Garcia, Alison M. Clark, and Éric Tanter. 2016. Abstracting Gradual Typing. In *Proceedings of the 43rd Annual ACM SIGPLAN-SIGACT Symposium on Principles of Programming Languages* (St. Petersburg, FL, USA) *(POPL 2016)*. ACM, New York, NY, USA, 429–442. https://doi.org/10.1145/2837614.2837670

Michael Greenberg. 2014. Space-Efficient Manifest Contracts. *CoRR* abs/1410.2813 (2014). http://arxiv.org/abs/1410.2813

Fritz Henglein. 1994. Dynamic typing: syntax and proof theory. *Science of Computer Programming* 22, 3 (June 1994), 197–230.

David Herman, Aaron Tomb, and Cormac Flanagan. 2010. Space-efficient gradual typing. *Higher-Order and Symbolic Computation* 23, 2 (2010), 167–189.

Gurvan Le Guernic. 2007. Automaton-based confidentiality monitoring of concurrent programs. In *20th IEEE Computer Security Foundations Symposium (CSF'07)*. IEEE, 218–232.

Gurvan Le Guernic and Thomas Jensen. 2005. Monitoring information flow. In *Proc. Workshop on Foundations of Computer Security*. 19–30.

Peng Li and Steve Zdancewic. 2010. Arrows for secure information flow. *Theoretical Computer Science* 411, 19 (2010), 1974 – 1994. https://doi.org/10.1016/j.tcs.2010.01.025 Mathematical Foundations of Programming Semantics (MFPS 2006).

Scott Moore and Stephen Chong. 2011. Static analysis for efficient hybrid information-flow control. In *2011 IEEE 24th Computer Security Foundations Symposium*. IEEE, 146–160.

Andrew C Myers. 1999. JFlow: Practical mostly-static information flow control. In *Proceedings of the 26th ACM SIGPLAN-SIGACT symposium on Principles of programming languages*. 228–241.

Andrew C. Myers and Barbara Liskov. 1997. A Decentralized Model for Information Flow Control. *SIGOPS Oper. Syst. Rev.* 31, 5 (oct 1997), 129–142. https://doi.org/10.1145/269005.266669

Andrew C. Myers, Lantian Zheng, Steve Zdancewic, Stephen Chong, and Nathaniel Nystrom. 2006. *Jif 3.0: Java information flow*. http://www.cs.cornell.edu/jif

Alejandro Russo and Andrei Sabelfeld. 2010. Dynamic vs. static flow-sensitive security analysis. In *2010 23rd IEEE Computer Security Foundations Symposium*. IEEE, 186–199.

Paritosh Shroff, Scott Smith, and Mark Thober. 2007. Dynamic Dependency Monitoring to Secure Information Flow. In *20th IEEE Computer Security Foundations Symposium (CSF'07)*. 203–217. https://doi.org/10.1109/CSF.2007.20

Jeremy G. Siek and Walid Taha. 2006. Gradual typing for functional languages. In *Scheme and Functional Programming Workshop*. 81–92.

Jeremy G. Siek and Walid Taha. 2007. Gradual Typing for Objects. In *European Conference on Object-Oriented Programming (LCNS, Vol. 4609)*. 2–27.

Jeremy G. Siek, Michael M. Vitousek, Matteo Cimini, and John Tang Boyland. 2015. Refined Criteria for Gradual Typing. In *SNAPL: Summit on Advances in Programming Languages (LIPIcs: Leibniz International Proceedings in Informatics)*.

Deian Stefan, David Mazières, John C. Mitchell, and Alejandro Russo. 2017. Flexible dynamic information flow control in the presence of exceptions. *Journal of Functional Programming* 27 (2017).

Deian Stefan, Alejandro Russo, John C Mitchell, and David Mazières. 2011. Flexible dynamic information flow control in Haskell. In *Proceedings of the 4th ACM symposium on Haskell*. 95–106.

Deian Stefan, Alejandro Russo, John C Mitchell, and David Mazières. 2012. Flexible dynamic information flow control in the presence of exceptions. *arXiv preprint arXiv:1207.1457* (2012).

Sam Tobin-Hochstadt and Matthias Felleisen. 2008. The Design and Implementation of Typed Scheme. In *Symposium on Principles of Programming Languages*.

Matías Toro, Ronald Garcia, and Éric Tanter. 2018. Type-Driven Gradual Security with References. *ACM Trans. Program. Lang. Syst.* 40, 4, Article 16 (Dec. 2018), 55 pages. https://doi.org/10.1145/3229061

Dennis Volpano, Cynthia Irvine, and Geoffrey Smith. 1996. A sound type system for secure flow analysis. *Journal of computer security* 4, 2-3 (1996), 167–187.

Philip Wadler. 1989. Theorems for free!. In *FPCA '89: Proceedings of the fourth international conference on Functional programming languages and computer architecture* (Imperial College, London, United Kingdom). ACM, 347–359.

Philip Wadler and Robert Bruce Findler. 2009. Well-typed programs can't be blamed. In *European Symposium on Programming (ESOP)*. 1–16.

Preston Tunnell Wilson, Ben Greenman, Justin Pombrio, and Shriram Krishnamurthi. 2018. The Behavior of Gradual Types: A User Study. In *Dynamic Languages Symposium*.

Jian Xiang and Stephen Chong. 2021. Co-Inflow: Coarse-grained Information Flow Control for Java-like Languages. In *Proceedings of the 2021 IEEE Symposium on Security and Privacy*. IEEE Press, Piscataway, NJ, USA.

## APPENDIX

The Appendix is organized as follows. In Section 8 we present the complete definitions of $\lambda^{\star}_{\mathtt{IFC}}$ and its cast calculus $\lambda^{c}_{\mathtt{IFC}}$ and we define the precision relations needed to state the gradual guarantee for $\lambda^{\star}_{\mathtt{IFC}}$ (Theorem 17), In Section 9, we show that vigilance *without* type-guided classification is insufficient to enable type-based reasoning by analyzing the semantics of $\lambda^{\star}_{\mathtt{SEC}}$ [Chen and Siek 2022]. On the contrary, the coercion calculi of $\lambda^{\star}_{\mathtt{IFC}}$ help us achieve both vigilance *and* type-guided classification, thus enabling type-based reasoning, while satisfying the gradual guarantee at the same time. In Section 10, we show that the normalization of coercion sequences is able to model information flow checks, because coercion sequences strongly normalize and the normalization is deterministic. Finally in Section 11, we present a proof for the noninterference result of $\lambda^{\star}_{\mathtt{IFC}}$, by establishing a simulation between its cast calculus $\lambda^{c}_{\mathtt{IFC}}$ and a dynamic IFC language $\lambda_{\mathtt{SEC}}$ derived from the calculus of Chen and Siek [2022].

## 8 SUPPLEMENTARY DEFINITIONS

We present the complete definition of the $\lambda^{\star}_{\mathtt{IFC}}$ and $\lambda^{c}_{\mathtt{IFC}}$ languages in Sections 8.2 and 8.3, respectively. There are some definitions shared between the two languages, which we present in Section 8.1. The definitions needed to state and prove the gradual guarantee are presented in Section 8.4.

### 8.1 Shared Definitions for $\lambda^{\star}_{\mathtt{IFC}}$ and $\lambda^{c}_{\mathtt{IFC}}$

Figure 8 presents auxiliary operators on security labels and types that are used by the type systems of $\lambda^{\star}_{\mathtt{IFC}}$ or its cast calculus $\lambda^{c}_{\mathtt{IFC}}$. These operators include consistent join, join w.r.t precision, consistent meet, and the stamping operation on types.

$$\ell \sqcup \ell = \ell \qquad \boxed{g \sqcup g}$$
$$\star \sqcup g = g$$
$$g \sqcup \star = g$$

$$\iota \sqcup \iota = \iota \qquad \boxed{T \sqcup T}$$
$$(\texttt{Ref}\ A) \sqcup (\texttt{Ref}\ B) = \texttt{Ref}\ (A \sqcup B)$$
$$(A \xrightarrow{g_1} B) \sqcup (C \xrightarrow{g_2} D) = (A \sqcup C) \xrightarrow{g_1 \sqcup g_2} (B \sqcup D)$$

$$S_{g_1} \sqcup T_{g_2} = (S \sqcup T)_{g_1 \sqcup g_2} \qquad \boxed{A \sqcup A}$$

$$\ell_1 \mathbin{\widetilde{\curlyvee}} \ell_2 = \ell_1 \vee \ell_2 \qquad \boxed{g \mathbin{\widetilde{\curlyvee}} g}$$
$$- \mathbin{\widetilde{\curlyvee}} \star = \star$$
$$\star \mathbin{\widetilde{\curlyvee}} - = \star$$

$$\iota \mathbin{\widetilde{\vee}} \iota = \iota \qquad \boxed{T \mathbin{\widetilde{\vee}} T}$$
$$(\texttt{Ref}\ A) \mathbin{\widetilde{\vee}} (\texttt{Ref}\ B) = \texttt{Ref}\ (A \sqcup B)$$
$$(A \xrightarrow{g_1} B) \mathbin{\widetilde{\vee}} (C \xrightarrow{g_2} D) = (A \mathbin{\widetilde{\wedge}} C) \xrightarrow{g_1 \mathbin{\widetilde{\curlywedge}} g_2} (B \mathbin{\widetilde{\vee}} D)$$

$$S_{g_1} \mathbin{\widetilde{\vee}} T_{g_2} = (S \mathbin{\widetilde{\vee}} T)_{g_1 \mathbin{\widetilde{\curlyvee}} g_2} \qquad \boxed{A \mathbin{\widetilde{\vee}} A}$$

$$\ell_1 \mathbin{\widetilde{\curlywedge}} \ell_2 = \ell_1 \wedge \ell_2 \qquad \boxed{g \mathbin{\widetilde{\curlywedge}} g}$$
$$- \mathbin{\widetilde{\curlywedge}} \star = \star$$
$$\star \mathbin{\widetilde{\curlywedge}} - = \star$$

$$\iota \mathbin{\widetilde{\wedge}} \iota = \iota \qquad \boxed{T \mathbin{\widetilde{\wedge}} T}$$
$$(\texttt{Ref}\ A) \mathbin{\widetilde{\wedge}} (\texttt{Ref}\ B) = \texttt{Ref}\ (A \sqcup B)$$
$$(A \xrightarrow{g_1} B) \mathbin{\widetilde{\wedge}} (C \xrightarrow{g_2} D) = (A \mathbin{\widetilde{\vee}} C) \xrightarrow{g_1 \mathbin{\widetilde{\curlyvee}} g_2} (B \mathbin{\widetilde{\wedge}} D)$$

$$S_{g_1} \mathbin{\widetilde{\wedge}} T_{g_2} = (S \mathbin{\widetilde{\wedge}} T)_{g_1 \mathbin{\widetilde{\curlywedge}} g_2} \qquad \boxed{A \mathbin{\widetilde{\wedge}} A}$$

$$\mathit{stamp}\ (T_{g_1})\ g_2 = T_{g_1 \mathbin{\widetilde{\curlyvee}} g_2}$$

Fig. 8. Auxiliary operators for security labels and types: join w.r.t precision (-⊔-), consistent join ($- \mathbin{\widetilde{\curlyvee}} -$ for labels and $- \mathbin{\widetilde{\vee}} -$ for types), and consistent meet ($- \mathbin{\widetilde{\curlywedge}} -$ for labels and $- \mathbin{\widetilde{\wedge}} -$ for types). Stamping for types

## 8.2 Definition of $\lambda^{\star}_{\mathtt{IFC}}$

Figure 10 contains the type system of $\lambda^{\star}_{\mathtt{IFC}}$. To suppor the type system of $\lambda^{\star}_{\mathtt{IFC}}$, we first present the definitions of consistent subtyping for security labels and types in Figure 9.

$\boxed{g_1 \precsim g_2,\ S \lesssim T, \text{ and } A \lesssim B}$

$$\precsim\star\ \frac{}{g \precsim \star} \qquad \star\precsim\ \frac{}{\star \precsim g} \qquad \precsim\text{-}\ell\ \frac{\ell_1 \preccurlyeq \ell_2}{\ell_1 \precsim \ell_2}$$

$$\lesssim\text{-}\iota\ \frac{}{\iota \lesssim \iota} \qquad \lesssim\text{-}\mathit{ref}\ \frac{A \lesssim B \quad B \lesssim A}{\mathtt{Ref}\ A \lesssim \mathtt{Ref}\ B}$$

$$\lesssim\text{-}\mathit{fun}\ \frac{g_2 \precsim g_1 \quad C \lesssim A \quad B \lesssim D}{A \xrightarrow{g_1} B \lesssim C \xrightarrow{g_2} D} \qquad \lesssim\text{-}\tau\ \frac{g_1 \precsim g_2 \quad S \lesssim T}{S_{g_1} \lesssim T_{g_2}}$$

Fig. 9. Consistent subtyping for labels and types

$\boxed{\Gamma; g \vdash M : A}$

$$\vdash\mathit{var}\ \frac{\Gamma \ni x : A}{\Gamma; g \vdash x : A} \qquad \vdash\mathit{const}\ \frac{k : \iota}{\Gamma; g \vdash (\$\ k)_\ell : \iota_\ell}$$

$$\vdash\mathit{lam}\ \frac{(\Gamma, x{:}A); g_2 \vdash N : B}{\Gamma; g_1 \vdash (\lambda^{g_2} x{:}A.\ N)_\ell : (A \xrightarrow{g_2} B)_\ell} \qquad \vdash\mathit{app}\ \frac{\Gamma; g \vdash L : (A \xrightarrow{g_2} B)_{g_1} \quad \Gamma; g \vdash M : A' \quad A' \lesssim A \quad g \precsim g_2 \quad g_1 \precsim g_2}{\Gamma; g \vdash (L\ M)^p : \mathit{stamp}\ B\ g_1}$$

$$\vdash\mathit{let}\ \frac{\Gamma; g \vdash M : A \quad (\Gamma, x{:}A); g \vdash N : B}{\Gamma; g \vdash \mathtt{let}\ x = M\ \mathtt{in}\ N : B} \qquad \vdash\mathit{if}\ \frac{\Gamma; g_2 \vdash L : \mathtt{Bool}_{g_1} \quad \Gamma; g_2 \mathbin{\tilde{\curlyvee}} g_1 \vdash M : A \quad \Gamma; g_2 \mathbin{\tilde{\curlyvee}} g_1 \vdash N : B \quad A \mathbin{\tilde{\vee}} B = C}{\Gamma; g_2 \vdash (\mathtt{if}\ L\ \mathtt{then}\ M\ \mathtt{else}\ N)^p : \mathit{stamp}\ C\ g_1}$$

$$\vdash\mathit{ref}\ \frac{\Gamma; g_2 \vdash M : T_{g_1} \quad T_{g_1} \lesssim T_\ell \quad \boxed{g_2 \precsim \ell}}{\Gamma; g_2 \vdash (\mathtt{ref}\ \ell\ M)^p : (\mathtt{Ref}\ T_\ell)_{\mathtt{low}}} \qquad \vdash\mathit{deref}\ \frac{\Gamma; g_2 \vdash M : (\mathtt{Ref}\ A)_{g_1}}{\Gamma; g_2 \vdash\ !^p\ M : \mathit{stamp}\ A\ g_1}$$

$$\vdash\mathit{assign}\ \frac{\Gamma; g_2 \vdash L : (\mathtt{Ref}\ T_{\hat{g}})_{g_1} \quad \Gamma; g_2 \vdash M : A \quad A \lesssim T_{\hat{g}} \quad \boxed{g_2 \precsim \hat{g}} \quad \boxed{g_1 \precsim \hat{g}}}{\Gamma; g_2 \vdash (L := M)^p : \mathtt{Unit}_{\mathtt{low}}} \qquad \vdash\mathit{ann}\ \frac{\Gamma; g \vdash M : A' \quad A' \lesssim A}{\Gamma; g \vdash (M : A)^p : A}$$

Fig. 10. Typing rules of the surface language $\lambda^{\star}_{\mathtt{IFC}}$. Side conditions about the heap policy are highlighted

$\boxed{\vdash \boldsymbol{c} : A \Rightarrow B}$

$$\vdash\! cbase\ \frac{\vdash \bar{c} : g_1 \Rightarrow g_2}{\vdash\ \mathbf{id}(\iota),\ \bar{c} : \iota_{g_1} \Rightarrow \iota_{g_2}} \qquad \vdash\! cref\ \frac{\vdash \boldsymbol{c} : B \Rightarrow A \qquad \vdash \boldsymbol{d} : A \Rightarrow B \qquad \vdash \bar{c} : g_1 \Rightarrow g_2}{\vdash\ \mathbf{Ref}\ \boldsymbol{c}\ \boldsymbol{d},\ \bar{c} : (\mathtt{Ref}\ A)_{g_1} \Rightarrow (\mathtt{Ref}\ B)_{g_2}}$$

$$\vdash\! cfun\ \frac{\vdash \bar{d} : g_4 \Rightarrow g_3 \qquad \vdash \boldsymbol{c} : C \Rightarrow A \qquad \vdash \boldsymbol{d} : B \Rightarrow D \qquad \vdash \bar{c} : g_1 \Rightarrow g_2}{\vdash\ \bar{d},\ \boldsymbol{c} \rightarrow \boldsymbol{d},\ \bar{c} : (A \xrightarrow{g_3} B)_{g_1} \Rightarrow (C \xrightarrow{g_4} D)_{g_2}}$$

Fig. 11. Typing rules for coercions on values.

## 8.3 Definition of $\lambda^{c}_{\mathtt{IFC}}$

*8.3.1 Type System.* To support the type system of $\lambda^{c}_{\mathtt{IFC}}$, we define the typing rules for coercions on values in Figure 11. The full typing rules of the cast calculus $\lambda^{c}_{\mathtt{IFC}}$ are listed in Figure 12.

$$\boxed{\Gamma;\Sigma;g;\ell \vdash M \Leftarrow A}$$

$$\vdash\!var\ \frac{\Gamma \ni x : A}{\Gamma;\Sigma;g;\ell \vdash x \Leftarrow A} \qquad \vdash\!const\ \frac{k : \iota}{\Gamma;\Sigma;g;\ell \vdash \$k \Leftarrow \iota_\ell}$$

$$\vdash\!addr\ \frac{\Sigma(\hat{\ell}, n) = T}{\Gamma;\Sigma;g;\ell' \vdash \mathtt{addr}\ n \Leftarrow (\mathtt{Ref}\ T_{\hat{\ell}})_\ell}$$

$$\vdash\!lam\ \frac{\forall \ell''.(\Gamma, x{:}A);\Sigma;g;\ell'' \vdash N \Leftarrow B}{\Gamma;\Sigma;g';\ell' \vdash \lambda x.\,N \Leftarrow (A \xrightarrow{g} B)_\ell} \qquad \vdash\!let\ \frac{\Gamma;\Sigma;g;\ell \vdash M \Leftarrow A \quad \forall \ell'.(\Gamma, x{:}A);\Sigma;g;\ell' \vdash N \Leftarrow B}{\Gamma;\Sigma;g;\ell \vdash \mathtt{let}\,x{=}M{:}A\,\mathtt{in}\,N \Leftarrow B}$$

$$\vdash\!app\ \frac{\Gamma;\Sigma;\ell';\ell'' \vdash L \Leftarrow (A \xrightarrow{\ell' \vee \ell} B)_\ell \quad \Gamma;\Sigma;\ell';\ell'' \vdash M \Leftarrow A \quad C = stamp\ B\ \ell}{\Gamma;\Sigma;\ell';\ell'' \vdash \mathtt{app}\,L\,M\,A\,B\,\ell \Leftarrow C} \qquad \vdash\!app\star\ \frac{\Gamma;\Sigma;g;\ell \vdash L \Leftarrow (A \xrightarrow{\star} (T_\star))_\star \quad \Gamma;\Sigma;g;\ell \vdash M \Leftarrow A}{\Gamma;\Sigma;g;\ell \vdash \mathtt{app}\star\,L\,M\,A\,T \Leftarrow T_\star}$$

$$\vdash\!if\ \frac{\Gamma;\Sigma;\ell';\ell'' \vdash L \Leftarrow \mathtt{Bool}_\ell \quad \forall \ell_1.\Gamma;\Sigma;\ell' \vee \ell;\ell_1 \vdash M \Leftarrow A \quad \forall \ell_2.\Gamma;\Sigma;\ell' \vee \ell;\ell_2 \vdash N \Leftarrow A \quad B = stamp\ A\ \ell}{\Gamma;\Sigma;\ell';\ell'' \vdash \mathtt{if}\,L\,A\,\ell\,M\,N \Leftarrow B} \qquad \vdash\!if\star\ \frac{\Gamma;\Sigma;g;\ell \vdash L \Leftarrow \mathtt{Bool}_\star \quad \forall \ell_1.\Gamma;\Sigma;\star;\ell_1 \vdash M \Leftarrow T_\star \quad \forall \ell_2.\Gamma;\Sigma;\star;\ell_2 \vdash N \Leftarrow T_\star}{\Gamma;\Sigma;g;\ell \vdash \mathtt{if}\star\,L\,T\,M\,N \Leftarrow T_\star}$$

$$\vdash\!ref\ \frac{\Gamma;\Sigma;\ell';\ell'' \vdash M \Leftarrow T_\ell \quad \boxed{\ell' \preccurlyeq \ell}}{\Gamma;\Sigma;\ell';\ell'' \vdash \mathtt{ref}\,\ell\,M \Leftarrow (\mathtt{Ref}\ T_\ell)_{\mathtt{low}}} \qquad \vdash\!ref?\ \frac{\Gamma;\Sigma;\star;\ell' \vdash M \Leftarrow T_\ell}{\Gamma;\Sigma;\star;\ell' \vdash \mathtt{ref?}^{p}\,\ell\,M \Leftarrow (\mathtt{Ref}\ T_\ell)_{\mathtt{low}}}$$

$$\vdash\!deref\ \frac{\Gamma;\Sigma;g;\ell' \vdash M \Leftarrow (\mathtt{Ref}\ A)_\ell \quad B = stamp\ A\ \ell}{\Gamma;\Sigma;g;\ell' \vdash !\,M\,A\,\ell \Leftarrow B} \qquad \vdash\!deref\star\ \frac{\Gamma;\Sigma;g;\ell \vdash M \Leftarrow (\mathtt{Ref}\ (T_\star))_\star}{\Gamma;\Sigma;g;\ell \vdash !\star\,M\,T \Leftarrow T_\star}$$

$$\vdash\!assign\ \frac{\Gamma;\Sigma;\ell';\ell'' \vdash L \Leftarrow (\mathtt{Ref}\ T_{\hat{\ell}})_\ell \qquad \Gamma;\Sigma;\ell';\ell'' \vdash M \Leftarrow T_{\hat{\ell}} \qquad \boxed{\ell' \vee \ell \preccurlyeq \hat{\ell}}}{\Gamma;\Sigma;\ell';\ell'' \vdash \mathtt{assign}\,L\,M\,T\,\hat{\ell}\,\ell \Leftarrow \mathtt{Unit}_{\mathtt{low}}}$$

$$\vdash\!assign?\ \frac{\Gamma;\Sigma;g;\ell \vdash L \Leftarrow (\mathtt{Ref}\ T_{\hat{g}})_\star \qquad \Gamma;\Sigma;g;\ell \vdash M \Leftarrow T_{\hat{g}}}{\Gamma;\Sigma;g;\ell \vdash \mathtt{assign?}^{p}\,L\,M\,T\,\hat{g} \Leftarrow \mathtt{Unit}_{\mathtt{low}}}$$

$$\vdash\!prot\ \frac{\Gamma;\Sigma;g';|PC| \vdash M \Leftarrow A \qquad \vdash PC \Leftarrow g' \qquad \ell' \vee \ell \preccurlyeq |PC| \qquad B = stamp\ A\ \ell}{\Gamma;\Sigma;g;\ell' \vdash \mathtt{prot}\,PC\,\ell\,M\,A \Leftarrow B}$$

$$\vdash\!cast\ \frac{\Gamma;\Sigma;g;\ell \vdash M \Leftarrow A \qquad \vdash \mathbf{c} : A \Rightarrow B}{\Gamma;\Sigma;g;\ell \vdash M\ \langle \mathbf{c} \rangle \Leftarrow B} \qquad \vdash\!blame\ \frac{}{\Gamma;\Sigma;g;\ell \vdash \mathtt{blame}\ p \Leftarrow A}$$

Fig. 12. Typing rules of the cast calculus $\lambda^{c}_{\mathtt{IFC}}$. The side conditions that enforce the heap policy statically during memory write operations are highlighted

$$\boxed{\mathit{stamp}\; e\; \ell = e}$$

$$\mathit{stamp}\; \ell\; \mathtt{low} = \ell$$
$$\mathit{stamp}\; \mathtt{low}\; \mathtt{high} = \mathtt{low}\; \langle \mathbf{id}(\mathtt{low});\, \uparrow \rangle$$
$$\mathit{stamp}\; \mathtt{high}\; \mathtt{high} = \mathtt{high}$$
$$\mathit{stamp}\; (\ell\; \langle \bar{c} \rangle)\; \ell' = \ell\; \left\langle \mathit{stamp}\; \bar{c}\; \ell' \right\rangle$$

$$\boxed{\mathit{stamp!}\; e\; \ell = e}$$

$$\mathit{stamp!}\; \ell\; \ell' = \ell\; \left\langle \mathit{stamp!}\; \mathbf{id}(\ell)\; \ell' \right\rangle$$
$$\mathit{stamp!}\; (\ell\; \langle \bar{c} \rangle)\; \ell' = \ell\; \left\langle \mathit{stamp!}\; \bar{c}\; \ell' \right\rangle$$

$$\boxed{|e| = \ell}$$

$$|\ell| = \ell$$
$$|\ell\; \langle \bar{c} \rangle\,| = |\bar{c}|$$

Fig. 13. Stamping and security level operators for security label expressions

$$\boxed{\mathit{coerce\text{-}id}\; T = c_r \text{ and } \mathit{coerce\text{-}id}\; A = \mathbf{c}}$$

$$\mathit{coerce\text{-}id}\; \iota = \mathbf{id}(\iota)$$
$$\mathit{coerce\text{-}id}\; (\mathtt{Ref}\; A) = \mathbf{Ref}\; (\mathit{coerce\text{-}id}\; A)\; (\mathit{coerce\text{-}id}\; A)$$
$$\mathit{coerce\text{-}id}\; (A \xrightarrow{g} B) = (\mathbf{id}(g),\; \mathit{coerce\text{-}id}\; A \to \mathit{coerce\text{-}id}\; B)$$
$$\mathit{coerce\text{-}id}\; T_g = (\mathit{coerce\text{-}id}\; T,\; \mathbf{id}(g))$$

$$\boxed{\mathit{stamp}\; V\; A\; \ell = W}$$

$$\mathit{stamp}\; V\; -\; \mathtt{low} = V$$
$$\mathit{stamp}\; V\; T_{\mathtt{low}}\; \mathtt{high} = V\; \langle (\mathit{coerce\text{-}id}\; T),\; \mathbf{id}(\mathtt{low});\, \uparrow \rangle$$
$$\mathit{stamp}\; V\; T_{\mathtt{high}}\; \mathtt{high} = V$$
$$\mathit{stamp}\; (V\; \langle c_r,\; \bar{c} \rangle)\; -\; \ell = V\; \langle c_r,\; \mathit{stamp}\; \bar{c}\; \ell \rangle$$

Fig. 14. Stamping on values of $\lambda^{c}_{\mathtt{IFC}}$

*8.3.2 Operational Semantics.* The operational semantics for $\lambda^{c}_{\mathtt{IFC}}$ is defined in Figure 16 and Figure 17. To support the operational semantics of $\lambda^{c}_{\mathtt{IFC}}$, we define the following operators for label expressions, and values of $\lambda^{c}_{\mathtt{IFC}}$:

(1) The stamping operations, *stamp* and *stamp!*, for label expressions in normal form are defined in Figure 13.
(2) The security level operator for label expressions in normal form is also defined in Figure 13.
(3) The stamping operation for values in $\lambda^{c}_{\mathtt{IFC}}$ is defined in Figure 14.
(4) The composition operator for coercions on values is defined in Figure 15.
(5) The "irreducible" predicate for coercions on values is also defined in Figure 15.

$\boxed{\boldsymbol{c} \mathbin{⨟} \boldsymbol{c} = \boldsymbol{c}}$

$$(\mathbf{id}(\iota),\ \bar{c}) \mathbin{⨟} \big(\mathbf{id}(\iota),\ \bar{d}\big) = \big(\mathbf{id}(\iota),\ \bar{c} \mathbin{⨟} \bar{d}\big)$$
$$(\mathbf{Ref}\ \boldsymbol{c_1}\ \boldsymbol{c_2},\ \bar{c}) \mathbin{⨟} \big(\mathbf{Ref}\ \boldsymbol{d_1}\ \boldsymbol{d_2},\ \bar{d}\big) = \big(\mathbf{Ref}\ (\boldsymbol{d_1} \mathbin{⨟} \boldsymbol{c_1})\ (\boldsymbol{c_2} \mathbin{⨟} \boldsymbol{d_2}),\ \bar{c} \mathbin{⨟} \bar{d}\big)$$
$$(\bar{c}_1,\ \boldsymbol{c_1} \rightarrow \boldsymbol{c_2},\ \bar{c}_2) \mathbin{⨟} \big(\bar{d}_1,\ \boldsymbol{d_1} \rightarrow \boldsymbol{d_2},\ \bar{d}_2\big) = \big(\bar{d}_1 \mathbin{⨟} \bar{c}_1,\ (\boldsymbol{d_1} \mathbin{⨟} \boldsymbol{c_1}) \rightarrow (\boldsymbol{c_2} \mathbin{⨟} \boldsymbol{d_2}),\ \bar{c}_2 \mathbin{⨟} \bar{d}_2\big)$$

$\boxed{\textbf{Irreducible}\ \boldsymbol{c}}$

$$\frac{\mathbf{NF}\ \bar{c} \qquad \ell \neq g}{\mathbf{Irreducible}\ \ (\mathbf{id}(\iota),\ \bar{c})}\ \vdash \bar{c} : \ell \Rightarrow g \qquad \frac{\mathbf{NF}\ \bar{c}}{\mathbf{Irreducible}\ \ (\mathbf{Ref}\ \boldsymbol{c}\ \boldsymbol{d},\ \bar{c})} \qquad \frac{\mathbf{NF}\ \bar{c}}{\mathbf{Irreducible}\ \ \big(\bar{d},\ \boldsymbol{c} \rightarrow \boldsymbol{d},\ \bar{c}\big)}$$

Fig. 15. Composition of coercions on values.

$$\boxed{M \mid \mu \mid PC \longrightarrow N \mid \mu'}$$

$$\xi\ \frac{M \mid \mu \mid PC \longrightarrow M' \mid \mu'}{\mathit{plug}\ M\ F \mid \mu \mid PC \longrightarrow \mathit{plug}\ M'\ F \mid \mu'} \qquad \xi\text{-}\mathit{blame}\ \frac{}{\mathit{plug}\ (\mathtt{blame}\ p)\ F \mid \mu \mid PC \longrightarrow \mathtt{blame}\ p \mid \mu}$$

$$\mathit{prot\text{-}ctx}\ \frac{M \mid \mu \mid PC' \longrightarrow M' \mid \mu'}{\mathtt{prot}\ PC'\ \ell\ M\ A \mid \mu \mid PC \longrightarrow \mathtt{prot}\ PC'\ \ell\ M'\ A \mid \mu'}$$

$$\mathit{prot\text{-}val}\ \frac{}{\mathtt{prot}\ PC'\ \ell\ V\ A \mid \mu \mid PC \longrightarrow \mathit{stamp\text{-}val}\ V\ A\ \ell \mid \mu}$$

$$\mathit{prot\text{-}blame}\ \frac{}{\mathtt{prot}\ PC'\ \ell\ (\mathtt{blame}\ p)\ A \mid \mu \mid PC \longrightarrow \mathtt{blame}\ p \mid \mu} \qquad \mathit{cast}\ \frac{V\ \langle \mathbf{c} \rangle \longrightarrow M}{V\ \langle \mathbf{c} \rangle \mid \mu \mid PC \longrightarrow M \mid \mu}$$

$$\beta\ \frac{}{\mathtt{app}\ (\lambda x.\ N)\ V\ A\ B\ \ell \mid \mu \mid PC \longrightarrow \mathtt{prot}\ (\mathit{stamp}\ PC\ \ell)\ \ell\ (N[x := V])\ B \mid \mu}$$

$$\mathit{app\text{-}cast}\ \frac{\mathbf{NF}\ \bar{c} \qquad (\mathit{stamp}\ PC\ \ell)\ \langle \bar{d} \rangle \longrightarrow^{*} PC' \qquad V\ \langle \mathbf{c} \rangle \longrightarrow^{*} W}{\mathtt{app}\ (\lambda x.\ N\ \langle \bar{d},\ \mathbf{c} \to \mathbf{d},\ \bar{c} \rangle)\ V\ C\ D\ \ell \mid \mu \mid PC \longrightarrow \mathtt{prot}\ PC'\ \ell\ ((N[x := W])\ \langle \mathbf{d} \rangle)\ D \mid \mu}$$

$$\mathit{app\text{-}blame\text{-}pc}\ \frac{\mathbf{NF}\ \bar{c} \qquad (\mathit{stamp}\ PC\ \ell)\ \langle \bar{d} \rangle \longrightarrow^{*} \mathtt{blame}\ p}{\mathtt{app}\ (\lambda x.\ N\ \langle \bar{d},\ \mathbf{c} \to \mathbf{d},\ \bar{c} \rangle)\ V\ C\ D\ \ell \mid \mu \mid PC \longrightarrow \mathtt{blame}\ p \mid \mu}$$

$$\mathit{app\text{-}blame}\ \frac{\mathbf{NF}\ \bar{c} \qquad (\mathit{stamp}\ PC\ \ell)\ \langle \bar{d} \rangle \longrightarrow^{*} PC' \qquad V\ \langle \mathbf{c} \rangle \longrightarrow^{*} \mathtt{blame}\ p}{\mathtt{app}\ (\lambda x.\ N\ \langle \bar{d},\ \mathbf{c} \to \mathbf{d},\ \bar{c} \rangle)\ V\ C\ D\ \ell \mid \mu \mid PC \longrightarrow \mathtt{blame}\ p \mid \mu}$$

$$\mathit{app{\star}\text{-}cast}\ \frac{\mathbf{NF}\ \bar{c} \qquad (\mathit{stamp!}\ PC\ |\bar{c}|)\ \langle \bar{d} \rangle \longrightarrow^{*} PC' \qquad V\ \langle \mathbf{c} \rangle \longrightarrow^{*} W}{\mathtt{app}{\star}\ (\lambda x.\ N\ \langle \bar{d},\ \mathbf{c} \to \mathbf{d},\ \bar{c} \rangle)\ V\ C\ T \mid \mu \mid PC \longrightarrow \mathtt{prot}\ PC'\ |\bar{c}|\ ((N[x := W])\ \langle \mathbf{d} \rangle)\ (T_{\star}) \mid \mu}$$

$$\mathit{app{\star}\text{-}blame\text{-}pc}\ \frac{\mathbf{NF}\ \bar{c} \qquad (\mathit{stamp!}\ PC\ |\bar{c}|)\ \langle \bar{d} \rangle \longrightarrow^{*} \mathtt{blame}\ p}{\mathtt{app}{\star}\ (\lambda x.\ N\ \langle \bar{d},\ \mathbf{c} \to \mathbf{d},\ \bar{c} \rangle)\ V\ C\ T \mid \mu \mid PC \longrightarrow \mathtt{blame}\ p \mid \mu}$$

$$\mathit{app{\star}\text{-}blame}\ \frac{\mathbf{NF}\ \bar{c} \qquad (\mathit{stamp!}\ PC\ |\bar{c}|)\ \langle \bar{d} \rangle \longrightarrow^{*} PC' \qquad V\ \langle \mathbf{c} \rangle \longrightarrow^{*} \mathtt{blame}\ p}{\mathtt{app}{\star}\ (\lambda x.\ N\ \langle \bar{d},\ \mathbf{c} \to \mathbf{d},\ \bar{c} \rangle)\ V\ C\ T \mid \mu \mid PC \longrightarrow \mathtt{blame}\ p \mid \mu}$$

$$\mathit{if\text{-}true}\ \frac{}{\mathtt{if}\ (\$\,\mathtt{true})\ A\ \ell\ M\ N \mid \mu \mid PC \longrightarrow \mathtt{prot}\ (\mathit{stamp}\ PC\ \ell)\ \ell\ M\ A \mid \mu}$$

$$\mathit{if\text{-}true\text{-}cast}\ \frac{}{\mathtt{if}\ (\$\,\mathtt{true}\ \langle \mathbf{id}(\mathtt{Bool}),\ \mathbf{id}(\mathtt{low});\ \uparrow \rangle)\ A\ \mathtt{high}\ M\ N \mid \mu \mid PC \longrightarrow \mathtt{prot}\ (\mathit{stamp}\ PC\ \mathtt{high})\ \mathtt{high}\ M\ A \mid \mu}$$

$$\mathit{if{\star}\text{-}true\text{-}cast}\ \frac{\mathbf{NF}\ \bar{c}}{\mathtt{if}{\star}\ (\$\,\mathtt{true}\ \langle \mathbf{id}(\mathtt{Bool}),\ \bar{c} \rangle)\ T\ M\ N \mid \mu \mid PC \longrightarrow \mathtt{prot}\ (\mathit{stamp!}\ PC\ |\bar{c}|)\ |\bar{c}|\ M\ (T_{\star}) \mid \mu}$$

$$\mathit{if\text{-}false}\ \frac{}{\mathtt{if}\ (\$\,\mathtt{false})\ A\ \ell\ M\ N \mid \mu \mid PC \longrightarrow \mathtt{prot}\ (\mathit{stamp}\ PC\ \ell)\ \ell\ N\ A \mid \mu}$$

$$\mathit{if\text{-}false\text{-}cast}\ \frac{}{\mathtt{if}\ (\$\,\mathtt{false}\ \langle \mathbf{id}(\mathtt{Bool}),\ \mathbf{id}(\mathtt{low});\ \uparrow \rangle)\ A\ \mathtt{high}\ M\ N \mid \mu \mid PC \longrightarrow \mathtt{prot}\ (\mathit{stamp}\ PC\ \mathtt{high})\ \mathtt{high}\ N\ A \mid \mu}$$

$$\mathit{if{\star}\text{-}false\text{-}cast}\ \frac{\mathbf{NF}\ \bar{c}}{\mathtt{if}{\star}\ (\$\,\mathtt{false}\ \langle \mathbf{id}(\mathtt{Bool}),\ \bar{c} \rangle)\ T\ M\ N \mid \mu \mid PC \longrightarrow \mathtt{prot}\ (\mathit{stamp!}\ PC\ |\bar{c}|)\ |\bar{c}|\ N\ (T_{\star}) \mid \mu}$$

$$\mathit{let}\ \frac{}{\mathtt{let}\ x{=}V{:}A\ \mathtt{in}\ N \mid \mu \mid PC \longrightarrow N[x := V] \mid \mu}$$

Fig. 16. Operational semantics of $\lambda^{c}_{\mathtt{IFC}}$ (Part I).

$$\boxed{M \mid \mu \mid PC \longrightarrow N \mid \mu'}$$

$$\textit{ref}\ \frac{n\ \textbf{FreshIn}\ \mu(\ell)}{\texttt{ref}\ \ell\ V \mid \mu \mid PC \longrightarrow \texttt{addr}\ n \mid (\mu, \ell \mapsto n \mapsto V)}$$

$$\textit{ref?}\ \frac{n\ \textbf{FreshIn}\ \mu(\ell) \qquad PC\ \langle \star \Rightarrow^{p} \ell \rangle \longrightarrow^{*} PC'}{\texttt{ref?}^{p}\ \ell\ V \mid \mu \mid PC \longrightarrow \texttt{addr}\ n \mid (\mu, \ell \mapsto n \mapsto V)}$$

$$\textit{ref?-blame}\ \frac{PC\ \langle \star \Rightarrow^{p} \ell \rangle \longrightarrow^{*} \texttt{blame}\ q}{\texttt{ref?}^{p}\ \ell\ V \mid \mu \mid PC \longrightarrow \texttt{blame}\ q \mid \mu}$$

$$\textit{assign}\ \frac{}{\texttt{assign}\ (\texttt{addr}\ n)\ V\ T\ \hat{\ell}\ \ell \mid \mu \mid PC \longrightarrow \$\texttt{unit} \mid [\hat{\ell} \mapsto n \mapsto V]\ \mu}$$

$$\textit{assign-cast}\ \frac{\textbf{NF}\ \bar{c} \qquad \vdash \boldsymbol{c} : T_{\hat{\ell}_2} \Rightarrow S_{\hat{\ell}_1} \qquad \vdash \boldsymbol{d} : S_{\hat{\ell}_1} \Rightarrow T_{\hat{\ell}_2} \qquad V\ \langle \boldsymbol{c} \rangle \longrightarrow^{*} W}{\texttt{assign}\ (\texttt{addr}\ n\ \langle \textbf{Ref}\ \boldsymbol{c}\ \boldsymbol{d},\ \bar{c} \rangle)\ V\ T\ \hat{\ell}_2\ \ell \mid \mu \mid PC \longrightarrow \$\texttt{unit} \mid [\hat{\ell}_1 \mapsto n \mapsto W]\ \mu}$$

$$\textit{assign-blame}\ \frac{\textbf{NF}\ \bar{c} \qquad \vdash \boldsymbol{c} : T_{\hat{\ell}_2} \Rightarrow S_{\hat{\ell}_1} \qquad \vdash \boldsymbol{d} : S_{\hat{\ell}_1} \Rightarrow T_{\hat{\ell}_2} \qquad V\ \langle \boldsymbol{c} \rangle \longrightarrow^{*} \texttt{blame}\ p}{\texttt{assign}\ (\texttt{addr}\ n\ \langle \textbf{Ref}\ \boldsymbol{c}\ \boldsymbol{d},\ \bar{c} \rangle)\ V\ T\ \hat{\ell}_2\ \ell \mid \mu \mid PC \longrightarrow \texttt{blame}\ p \mid \mu}$$

$$\textit{assign?-cast}\ \frac{\begin{array}{ccc} \textbf{NF}\ \bar{c} & \vdash \boldsymbol{c} : T_g \Rightarrow S_{\hat{\ell}} & \vdash \boldsymbol{d} : S_{\hat{\ell}} \Rightarrow T_g \\ \multicolumn{2}{c}{(\textit{stamp}!\ PC\ |\bar{c}|)\ \langle \star \Rightarrow^{p} \hat{\ell} \rangle \longrightarrow^{*} PC'} & V\ \langle \boldsymbol{c} \rangle \longrightarrow^{*} W \end{array}}{\texttt{assign?}^{p}\ (\texttt{addr}\ n\ \langle \textbf{Ref}\ \boldsymbol{c}\ \boldsymbol{d},\ \bar{c} \rangle)\ V\ T\ g \mid \mu \mid PC \longrightarrow \$\texttt{unit} \mid [\hat{\ell} \mapsto n \mapsto W]\ \mu}$$

$$\textit{assign?-cast-blame-pc}\ \frac{\begin{array}{c} \textbf{NF}\ \bar{c} \qquad \vdash \boldsymbol{c} : T_g \Rightarrow S_{\hat{\ell}} \qquad \vdash \boldsymbol{d} : S_{\hat{\ell}} \Rightarrow T_g \\ (\textit{stamp}!\ PC\ |\bar{c}|)\ \langle \star \Rightarrow^{p} \hat{\ell} \rangle \longrightarrow^{*} \texttt{blame}\ q \end{array}}{\texttt{assign?}^{p}\ (\texttt{addr}\ n\ \langle \textbf{Ref}\ \boldsymbol{c}\ \boldsymbol{d},\ \bar{c} \rangle)\ V\ T\ g \mid \mu \mid PC \longrightarrow \texttt{blame}\ q \mid \mu}$$

$$\textit{assign?-cast-blame}\ \frac{\begin{array}{ccc} \textbf{NF}\ \bar{c} & \vdash \boldsymbol{c} : T_g \Rightarrow S_{\hat{\ell}} & \vdash \boldsymbol{d} : S_{\hat{\ell}} \Rightarrow T_g \\ \multicolumn{2}{c}{(\textit{stamp}!\ PC\ |\bar{c}|)\ \langle \star \Rightarrow^{p} \hat{\ell} \rangle \longrightarrow^{*} PC'} & V\ \langle \boldsymbol{c} \rangle \longrightarrow^{*} \texttt{blame}\ q \end{array}}{\texttt{assign?}^{p}\ (\texttt{addr}\ n\ \langle \textbf{Ref}\ \boldsymbol{c}\ \boldsymbol{d},\ \bar{c} \rangle)\ V\ T\ g \mid \mu \mid PC \longrightarrow \texttt{blame}\ q \mid \mu}$$

$$\textit{deref}\ \frac{\mu(\hat{\ell}, n) = V}{!\ (\texttt{addr}\ n)\ T_{\hat{\ell}}\ \ell \mid \mu \mid PC \longrightarrow \texttt{prot_}\ell\ V\ T_{\hat{\ell}} \mid \mu}$$

$$\textit{deref-cast}\ \frac{\textbf{NF}\ \bar{c} \qquad \vdash \boldsymbol{c} : A \Rightarrow T_{\hat{\ell}} \qquad \vdash \boldsymbol{d} : T_{\hat{\ell}} \Rightarrow A \qquad \mu(\hat{\ell}, n) = V}{!\ (\texttt{addr}\ n\ \langle \textbf{Ref}\ \boldsymbol{c}\ \boldsymbol{d},\ \bar{c} \rangle)\ A\ \ell \mid \mu \mid PC \longrightarrow \texttt{prot_}\ell\ (V\ \langle \boldsymbol{d} \rangle)\ A \mid \mu}$$

$$\textit{deref}\star\textit{-cast}\ \frac{\textbf{NF}\ \bar{c} \qquad \vdash \boldsymbol{c} : S_{\star} \Rightarrow T_{\hat{\ell}} \qquad \vdash \boldsymbol{d} : T_{\hat{\ell}} \Rightarrow S_{\star} \qquad \mu(\hat{\ell}, n) = V}{!\star\ (\texttt{addr}\ n\ \langle \textbf{Ref}\ \boldsymbol{c}\ \boldsymbol{d},\ \bar{c} \rangle)\ S \mid \mu \mid PC \longrightarrow \texttt{prot_}|\bar{c}|\ (V\ \langle \boldsymbol{d} \rangle)\ (S_{\star}) \mid \mu}$$

Fig. 17. Operational semantics of $\lambda^{c}_{\texttt{IFC}}$ (Part II). NSU checks are represented using label expressions ( highlighted )

## 8.4 Definitions of Precision

To state and prove the gradual guarantee, we define the following precision relations:

(1) The precision relations for security labels and types are defined in Figure 18.
(2) The precision relation on terms of $\lambda^{\star}_{\mathtt{IFC}}$ is defined in Figure 19.
(3) The precision relation on the coercion calculus is defined in Figure 20.
(4) We present noteworthy term precision rules of $\lambda^{c}_{\mathtt{IFC}}$ in Figure 21, with the full definition in `/src/CC2/Precision.agda` of the Agda code in the supplementary material.

$\boxed{g_1 \sqsubseteq g_2}$

$$\star\sqsubseteq\ \frac{}{\star \sqsubseteq g} \qquad \ell\sqsubseteq\ell\ \frac{}{\ell \sqsubseteq \ell}$$

$\boxed{S \sqsubseteq T}$

$$\sqsubseteq\text{-}\iota\ \frac{}{\iota \sqsubseteq \iota} \qquad \sqsubseteq\text{-}\mathit{ref}\ \frac{A \sqsubseteq B}{\mathtt{Ref}\ A \sqsubseteq \mathtt{Ref}\ B} \qquad \sqsubseteq\text{-}\mathit{fun}\ \frac{g_1 \sqsubseteq g_2 \quad A \sqsubseteq C \quad B \sqsubseteq D}{A \xrightarrow{g_1} B \sqsubseteq C \xrightarrow{g_2} D}$$

$\boxed{A \sqsubseteq B}$

$$\sqsubseteq\text{-}\tau\ \frac{g_1 \sqsubseteq g_2 \quad S \sqsubseteq T}{S_{g_1} \sqsubseteq T_{g_2}}$$

Fig. 18. Precision of security labels and types

$\boxed{\vdash M \sqsubseteq M'}$

$$\frac{}{\vdash (\$\ k)_\ell \sqsubseteq (\$\ k)_\ell} \qquad \frac{g_1 \sqsubseteq g_2 \quad A_1 \sqsubseteq A_2 \quad \vdash N_1 \sqsubseteq N_2}{\vdash (\lambda^{g_1} x{:}A_1.\ N_1)_\ell \sqsubseteq (\lambda^{g_2} x{:}A_2.\ N_2)_\ell}$$

$$\frac{\vdash M_1 \sqsubseteq M_2}{\vdash (\mathtt{ref}\ \ell\ M_1)^p \sqsubseteq (\mathtt{ref}\ \ell\ M_2)^p} \qquad \frac{\vdash M_1 \sqsubseteq M_2 \quad A_1 \sqsubseteq A_2}{\vdash (M_1 : A_1)^p \sqsubseteq (M_2 : A_2)^p}$$

Fig. 19. Selected precision rules of $\lambda^{\star}_{\mathtt{IFC}}$

$\boxed{\vdash c \sqsubseteq d, \vdash c \sqsubseteq g \text{ , and} \vdash g \sqsubseteq d}$

$$\sqsubseteq\text{-}c\ \frac{g_1 \sqsubseteq g_1' \qquad g_2 \sqsubseteq g_2' \qquad \vdash c : g_1 \Rightarrow g_2 \qquad \vdash d : g_1' \Rightarrow g_2'}{\vdash c \sqsubseteq d}$$

$$\sqsubseteq\text{-}cl\ \frac{g_1 \sqsubseteq g \qquad g_2 \sqsubseteq g \qquad \vdash c : g_1 \Rightarrow g_2}{\vdash c \sqsubseteq g} \qquad \sqsubseteq\text{-}cr\ \frac{g \sqsubseteq g_1 \qquad g \sqsubseteq g_2 \qquad \vdash d : g_1 \Rightarrow g_2}{\vdash g \sqsubseteq d}$$

$\boxed{\vdash \bar{c} \sqsubseteq \bar{d}}$

$$\sqsubseteq\text{-}id\ \frac{g \sqsubseteq g'}{\vdash \mathbf{id}(g) \sqsubseteq \mathbf{id}(g')} \qquad \sqsubseteq\text{-}cast\ \frac{\vdash \bar{c} \sqsubseteq \bar{d} \qquad \vdash c \sqsubseteq d}{\vdash \bar{c};\, c \sqsubseteq \bar{d};\, d}$$

$$\sqsubseteq\text{-}castl\ \frac{\vdash \bar{c} \sqsubseteq \bar{d} \qquad \vdash c \sqsubseteq g_2 \qquad \vdash \bar{d} : g_1 \Rightarrow g_2}{\vdash \bar{c};\, c \sqsubseteq \bar{d}} \qquad \sqsubseteq\text{-}castr\ \frac{\vdash \bar{c} \sqsubseteq \bar{d} \qquad \vdash g_2 \sqsubseteq d \qquad \vdash \bar{c} : g_1 \Rightarrow g_2}{\vdash \bar{c} \sqsubseteq \bar{d};\, d}$$

$$\sqsubseteq\text{-}\bot\ \frac{g_1 \sqsubseteq g_3 \qquad g_2 \sqsubseteq g_4 \qquad \vdash \bar{c} : g_1 \Rightarrow g_2}{\vdash \bar{c} \sqsubseteq \mathbf{\bot}^{p}\ g_3\ g_4}$$

$\boxed{\vdash \bar{c} \sqsubseteq_l g}$

$$\sqsubseteq_l\text{-}id\ \frac{g \sqsubseteq g'}{\vdash \mathbf{id}(g) \sqsubseteq_l g'} \qquad \sqsubseteq_l\text{-}cast\ \frac{\vdash \bar{c} \sqsubseteq_l g \qquad \vdash c \sqsubseteq_l g}{\vdash \bar{c};\, c \sqsubseteq_l g}$$

$\boxed{\vdash g \sqsubseteq_r \bar{d}}$

$$\sqsubseteq_r\text{-}id\ \frac{g \sqsubseteq g'}{\vdash g \sqsubseteq_r \mathbf{id}(g')} \qquad \sqsubseteq_r\text{-}cast\ \frac{\vdash g \sqsubseteq_r \bar{d} \qquad \vdash g \sqsubseteq_r d}{\vdash g \sqsubseteq_r \bar{d};\, d} \qquad \sqsubseteq_r\text{-}\bot\ \frac{g \sqsubseteq g_1 \qquad g \sqsubseteq g_2}{\vdash g \sqsubseteq_r \mathbf{\bot}^{p}\ g_1\ g_2}$$

Fig. 20. Precision relation of the coercion calculus

$$\boxed{\Gamma;\Gamma';\Sigma;\Sigma';g;g';\ell;\ell' \vdash M \sqsubseteq M' \Leftarrow A \sqsubseteq A'}$$

$$\sqsubseteq\text{-}addr\ \frac{\Sigma(\hat{\ell},n) = T \qquad \Sigma'(\hat{\ell},n) = T'}{\Gamma;\Gamma';\Sigma;\Sigma';g;g';\ell;\ell' \vdash \texttt{addr}\ n \sqsubseteq \texttt{addr}\ n \Leftarrow (\texttt{Ref}\ T_{\hat{\ell}})_{\ell} \sqsubseteq (\texttt{Ref}\ T'_{\hat{\ell}})_{\ell}}$$

$$\sqsubseteq\text{-}ref?l\ \frac{\Gamma;\Gamma';\Sigma;\Sigma';\star;\ell_1;\ell_2;\ell_3 \vdash M \sqsubseteq M' \Leftarrow T_{\ell} \sqsubseteq T'_{\ell} \qquad \ell_1 \preccurlyeq \ell}{\Gamma;\Gamma';\Sigma;\Sigma';\star;\ell_1;\ell_2;\ell_3 \vdash \texttt{ref?}^{p}\ \ell\ M \sqsubseteq \texttt{ref}\ \ell\ M' \Leftarrow (\texttt{Ref}\ T_{\ell})_{\texttt{low}} \sqsubseteq (\texttt{Ref}\ T'_{\ell})_{\texttt{low}}}$$

$$\sqsubseteq\text{-}prot!\ \frac{\begin{array}{c}\Gamma;\Gamma';\Sigma;\Sigma';g_1;g'_1;|PC|;|PC'| \vdash M \sqsubseteq M' \Leftarrow T_{\star} \sqsubseteq T'_{\star} \qquad \vdash PC \sqsubseteq PC' \Leftarrow g_1 \sqsubseteq g'_1 \\ \ell_1 \curlyvee \ell_2 \preccurlyeq |PC| \qquad \ell'_1 \curlyvee \ell'_2 \preccurlyeq |PC'| \qquad \ell_2 \preccurlyeq \ell'_2\end{array}}{\Gamma;\Gamma';\Sigma;\Sigma';g_2;g'_2;\ell_1;\ell'_1 \vdash \texttt{prot}\ PC\ \ell_2\ M\ T_{\star} \sqsubseteq \texttt{prot}\ PC'\ \ell'_2\ M'\ T'_{\star} \Leftarrow T_{\star} \sqsubseteq T'_{\star}}$$

$$\sqsubseteq\text{-}cast\ \frac{\Gamma;\Gamma';\Sigma;\Sigma';g;g';\ell;\ell' \vdash M \sqsubseteq M' \Leftarrow A \sqsubseteq A' \qquad \boldsymbol{c} \sqsubseteq \boldsymbol{c'} \qquad \vdash \boldsymbol{c} : A \Rightarrow B \qquad \vdash \boldsymbol{c'} : A' \Rightarrow B'}{\Gamma;\Gamma';\Sigma;\Sigma';g;g';\ell;\ell' \vdash M\ \langle \boldsymbol{c} \rangle \sqsubseteq M'\ \langle \boldsymbol{c'} \rangle \Leftarrow B \sqsubseteq B'}$$

$$\sqsubseteq\text{-}castl\ \frac{\Gamma;\Gamma';\Sigma;\Sigma';g;g';\ell;\ell' \vdash M \sqsubseteq M' \Leftarrow A \sqsubseteq A' \qquad \boldsymbol{c} \sqsubseteq A' \qquad \vdash \boldsymbol{c} : A \Rightarrow B}{\Gamma;\Gamma';\Sigma;\Sigma';g;g';\ell;\ell' \vdash M\ \langle \boldsymbol{c} \rangle \sqsubseteq M' \Leftarrow B \sqsubseteq A'}$$

$$\sqsubseteq\text{-}castr\ \frac{\Gamma;\Gamma';\Sigma;\Sigma';g;g';\ell;\ell' \vdash M \sqsubseteq M' \Leftarrow A \sqsubseteq A' \qquad A \sqsubseteq \boldsymbol{c'} \qquad \vdash \boldsymbol{c'} : A' \Rightarrow B'}{\Gamma;\Gamma';\Sigma;\Sigma';g;g';\ell;\ell' \vdash M \sqsubseteq M'\ \langle \boldsymbol{c'} \rangle \Leftarrow A \sqsubseteq B'}$$

$$\sqsubseteq\text{-}blame\ \frac{\Gamma;\Sigma;g;\ell \vdash M \Leftarrow A \qquad A \sqsubseteq A'}{\Gamma;\Gamma';\Sigma;\Sigma';g;g';\ell;\ell' \vdash M \sqsubseteq \texttt{blame}\ p \Leftarrow A \sqsubseteq A'}$$

Fig. 21. Selected precision rules of $\lambda^{c}_{\texttt{IFC}}$

## 9 VIGILANCE ALONE DOES NOT ENSURE TYPE-BASED REASONING

Chen and Siek [2022] study a language $\lambda^{\star}_{\texttt{SEC}}$ with gradual security that omits type-guided classification, similar to GLIO. However, $\lambda^{\star}_{\texttt{SEC}}$ differs from GLIO in that it is vigilant like traditional gradually-typed languages. Here we show that vigilance alone is not enough to ensure type-based reasoning, as the ommision of type-guided classification in $\lambda^{\star}_{\texttt{SEC}}$ breaks type-based reasoning.

Returning to the mix/smix example by Toro et al. [2018] (§ 2.3), the cast insertion of $\lambda^{\star}_{\texttt{SEC}}$ produces:

```
let mix  =
  (λ pub priv . if (pub ⟨low ⇒ ★⟩) < priv then (1_low ⟨low ⇒ ★⟩) else (2_low ⟨low ⇒ ★⟩))
    ⟨low → ★ → ★ ⇒ low → ★ → low⟩ in
let smix = λ pub priv . mix pub (priv ⟨high ⇒ ★⟩) in
smix 1_low 5_low
```

According to type-based reasoning, this program should fail at runtime. However, the program reduces to $1_{\texttt{low}}$, violating the free theorem. Let us examine the reduction steps involving the if-expression that give rise to an implicit flow.

$$\longrightarrow^{*} \quad (\texttt{if}\ (\texttt{true}_{\texttt{low}}\ \langle \texttt{high} \Rightarrow \star\rangle)\ \texttt{then}\ 1_{\texttt{low}}\ \langle \texttt{low} \Rightarrow \star\rangle\ \texttt{else}\ \ldots)\ \langle \star \Rightarrow \texttt{low}\rangle \tag{1}$$

$$\longrightarrow^{*} \quad (\texttt{prot}\ \texttt{low}\ (1_{\texttt{low}}\ \langle \texttt{low} \Rightarrow \star\rangle))\ \langle \star \Rightarrow \texttt{low}\rangle \tag{2}$$

$$\longrightarrow^{*} \quad 1_{\texttt{low}} \tag{3}$$

The if-expression branches on the value $\texttt{true}_{\texttt{low}}\ \langle \texttt{high} \Rightarrow \star\rangle$ (1) and this value is classified as low security because in $\lambda^{\star}_{\texttt{SEC}}$, casts do not change the security classification of values. So the protection term gets security level low (2), and therefore $1_{\texttt{low}}\ \langle \texttt{low} \Rightarrow \star\rangle$ is allowed as the result of the protection term. Then the injection and projection collapse, producing $1_{\texttt{low}}$ (3).

## 10 THE NORMALIZATION OF COERCIONS CHECKS INFORMATION FLOW

We show that reducing coercion sequences to normal form models IFC checks because the normalization either succeeds or fails: if a coercion sequences successfully reduces to normal form, the IFC check succeeds and the flow is justified; if it reduces to a failure, then an illegal flow is detected and the program errors.

Lemma 18 (Strong normalization of coercion sequences). *If* $\vdash \bar{c} : g_1 \Rightarrow g_2$, *then either (1)* $\bar{c} \longrightarrow^{*} \bar{d}$ *and* NF $\bar{d}$ *or (2)* $\bar{c} \longrightarrow^{*} \bot^{p}\ g_1\ g_2$.

Normalization of coercion sequences is deterministic:

Lemma 19 (Reduction of coercion sequences is deterministic). *If* $\bar{c} \longrightarrow \bar{d}_1$ *and* $\bar{c} \longrightarrow \bar{d}_2$, *then* $\bar{d}_1 = \bar{d}_2$.

Lemma 20 (Normalization of coercion sequences is deterministic). *Suppose* $\bar{c} \longrightarrow^{*} \bar{d}_1$ *and* $\bar{c} \longrightarrow^{*} \bar{d}_2$. *If* NF $\bar{d}_i$ *or* $\bar{d}_i = \bot^{p}\ g_1\ g_2$, *then* $\bar{d}_1 = \bar{d}_2$.

## 11 PROVING NONINTERFERENCE FOR $\lambda^{\star}_{\texttt{IFC}}$

In this section we present a proof of noninterference for $\lambda^{\star}_{\texttt{IFC}}$. The proof employs a three-step approach. First, we adapt an IFC language and noninterference proof from Chen and Siek [2022]. In particular, we define a dynamic IFC programming language named $\lambda_{\texttt{SEC}}$ (Figure 22) and prove that it satisfies noninterference. This proof (Lemma 21) uses a standard erasure-based approach [Fennell and Thiemann 2013; Li and Zdancewic 2010; Stefan et al. 2017, 2011, 2012], where the high-security parts of a program are erased to an opaque value.

$$\begin{array}{llll} \text{terms} & L, M, N & ::= & x \mid (\$\, k)_\ell \mid (\texttt{addr}\ n_{\hat{\ell}})_\ell \mid (\lambda x.\, N)_\ell \\ & & \mid & L\ M \mid \texttt{if}\ L\ M\ N \\ & & \mid & \texttt{ref}^?\ \ell\ M \mid\ !\ M \mid L :=^? M \mid \texttt{prot}\ \ell\ M \\ \text{values} & V, W & ::= & (\$\, k)_\ell \mid (\texttt{addr}\ n_{\hat{\ell}})_\ell \mid (\lambda x.\, N)_\ell \\ \text{frames} & F & ::= & \square\ M \mid V\ \square \mid \texttt{if}\ \square\ M\ N \mid \texttt{ref}^?\ \ell\ \square \mid\ !\ \square \mid \square :=^? M \mid V :=^? \square \end{array}$$

$$\boxed{M \mid \mu \mid pc \longrightarrow N \mid \mu'}$$

$$\xi\ \frac{M \mid \mu \mid pc \longrightarrow M' \mid \mu'}{plug\ M\ F \mid \mu \mid pc \longrightarrow plug\ M'\ F \mid \mu'}$$

$$prot\text{-}val\ \frac{}{\texttt{prot}\ \ell\ V \mid \mu \mid pc \longrightarrow V \curlyvee \ell \mid \mu} \qquad prot\text{-}ctx\ \frac{M \mid \mu \mid pc \curlyvee \ell \longrightarrow M' \mid \mu'}{\texttt{prot}\ \ell\ M \mid \mu \mid pc \longrightarrow \texttt{prot}\ \ell\ M' \mid \mu'}$$

$$\beta\ \frac{}{(\lambda x.\, N)_\ell\ V \mid \mu \mid pc \longrightarrow \texttt{prot}\ \ell\ (N[x := V]) \mid \mu}$$

$$\beta\text{-}if\text{-}true\ \frac{}{\texttt{if}\ (\$\ \texttt{true})_\ell\ M\ N \mid \mu \mid pc \longrightarrow \texttt{prot}\ \ell\ M \mid \mu}$$

$$\beta\text{-}if\text{-}false\ \frac{}{\texttt{if}\ (\$\ \texttt{false})_\ell\ M\ N \mid \mu \mid pc \longrightarrow \texttt{prot}\ \ell\ N \mid \mu}$$

$$ref?\text{-}ok\ \frac{pc \preccurlyeq \ell \qquad n\ \textbf{FreshIn}\ \mu(\ell)}{\texttt{ref}^?\ \ell\ V \mid \mu \mid pc \longrightarrow (\texttt{addr}\ n_\ell)_{\texttt{low}} \mid (\mu, \ell \mapsto n \mapsto (V \curlyvee \ell))}$$

$$deref\ \frac{\mu(\hat{\ell}, n) = V}{!\ (\texttt{addr}\ n_{\hat{\ell}})_\ell \mid \mu \mid pc \longrightarrow \texttt{prot}\ \ell\ V \mid \mu}$$

$$assign?\text{-}ok\ \frac{pc \curlyvee \ell \preccurlyeq \hat{\ell}}{(\texttt{addr}\ n_{\hat{\ell}})_\ell :=^? V \mid \mu \mid pc \longrightarrow (\$\ \texttt{unit})_{\texttt{low}} \mid [\hat{\ell} \mapsto n \mapsto (V \curlyvee \hat{\ell})]\mu}$$

Fig. 22. Syntax and reduction semantics (successful cases) of $\lambda_{\texttt{SEC}}$

Second, we prove a simulation lemma between $\lambda^c_{\texttt{IFC}}$ and $\lambda_{\texttt{SEC}}$ (Lemma 34). The main intuition of the simulation relation (Figure 23) is that a $\lambda^c_{\texttt{IFC}}$ term always produces a value that is as secure as the one produced by its related $\lambda_{\texttt{SEC}}$ term. We translate $\lambda^c_{\texttt{IFC}}$ terms to $\lambda_{\texttt{SEC}}$ by (1) getting rid of all the casts and (2) converting static heap enforcement (`ref` and `assign`) to dynamic enforcement (NSU) (Figure 24). The noninterference property of $\lambda^c_{\texttt{IFC}}$ (Lemma 14) follows directly from the multi-step simulation lemma (Lemma 35) and the noninterference result of $\lambda_{\texttt{SEC}}$ (Lemma 21).

Third, the noninterference theorem of $\lambda^\star_{\texttt{IFC}}$ (Theorem 16) is a corollary of the noninterference property of $\lambda^c_{\texttt{IFC}}$.

Similar to $\text{GSL}_{\text{Ref}}$ and GLIO, the statements of noninterference for both $\lambda_{\texttt{SEC}}$ and $\lambda^c_{\texttt{IFC}}$ are termination-insensitive. Thus, we only consider successful executions that produce values. In other words, we do not consider reduction rules that trigger IFC monitor failures, such as NSU errors in $\lambda_{\texttt{SEC}}$ and cast errors in $\lambda^c_{\texttt{IFC}}$. As we explained in Section 2.1, the programming language runtime can force the program to diverge whenever blame is detected, possibly sending a private error message to the software developer.

Lemma 21 (Noninterference of $\lambda_{\texttt{SEC}}$). *If* $M[x := (\$\ b_1)_{\texttt{high}}] \mid \emptyset \mid \texttt{low} \longrightarrow^* (\$\ b_3)_{\texttt{low}} \mid \mu_1$ *and* $M[x := (\$\ b_2)_{\texttt{high}}] \mid \emptyset \mid \texttt{low} \longrightarrow^* (\$\ b_4)_{\texttt{low}} \mid \mu_2$ *then* $b_3 = b_4$.

Proof. The proof is fully mechanized in Agda [4] and can be found in the supplementary material. The proof follows the noninterference proof of $\lambda_{\mathsf{SEC}}^{\Rightarrow}$ [Chen and Siek 2022]. □

Definition 22 (Simulation between heaps). *$\Sigma \vdash \mu' \leq \mu \triangleq \forall \ell, n.$ if $\mu(\ell, n) = V$, then there exists $V'$ s.t $\mu'(\ell, n) = V'$ and $\mathtt{low} \vdash V' \leq V \Leftarrow T_\ell$ where $T = \Sigma(\ell, n)$.*

Lemma 23 (Casting preserves simulation). *If $g^c \vdash W' \leq V \Leftarrow A$, $\vdash \boldsymbol{c} : A \Rightarrow B$, and $V\ \langle \boldsymbol{c} \rangle \longrightarrow^* W$, then $g^c \vdash W' \leq W \Leftarrow B$.*

Proof. By induction on the multi-step cast reduction.

**Zero step** Directly proved by applying rule $\leq$-*cast*.

**One or more steps** Casing on the first reduction step yields three sub-cases:

***cast***

$$g^c \vdash W' \leq V_r \Leftarrow A \tag{4}$$

$$V_r\ \langle c_r,\ \bar{c} \rangle \longrightarrow V_r\ \left\langle c_r,\ \bar{d} \right\rangle \longrightarrow^* W \tag{5}$$

By induction hypothesis, $g^c \vdash W' \leq W \Leftarrow B$.

***cast-id***

$$g^c \vdash W' \leq V_r \Leftarrow A \tag{6}$$

$$V_r\ \langle \mathbf{id}(\iota),\ \mathbf{id}(g) \rangle \longrightarrow V_r \tag{7}$$

The goal is proved directly.

***cast-comp***

$$g^c \vdash W' \leq V_r\ \langle \boldsymbol{c} \rangle \Leftarrow A \tag{8}$$

$$V_r\ \langle \boldsymbol{c} \rangle\ \langle \boldsymbol{d} \rangle \longrightarrow V_r\ \langle \boldsymbol{c} \mathbin{⨟} \boldsymbol{d} \rangle \longrightarrow^* W \tag{9}$$

We further reason about $V_r\ \langle \boldsymbol{c} \mathbin{⨟} \boldsymbol{d} \rangle \longrightarrow^* W$. The coercion after composition $\boldsymbol{c} \mathbin{⨟} \boldsymbol{d}$ is reducible and not identity, so the reduction must take one step by *cast* and reduce the top-level coercion sequence to its normal form $\bar{c}_n$:

$$V_r\ \langle c_{r1},\ \bar{c} \rangle\ \left\langle c_{r2},\ \bar{d} \right\rangle \longrightarrow V_r\ \left\langle c_r,\ \bar{c} \mathbin{⨟} \bar{d} \right\rangle \longrightarrow V_r\ \langle c_r,\ \bar{c}_n \rangle \longrightarrow^* W$$

We know $|\bar{c}| \preccurlyeq |\bar{c}_n|$ because composition models explicit flow (Lemma 2).

- If $V_r = \$k$ and $\bar{c}_n = \mathbf{id}(\ell)$. $W' = (\$\ k)_{\ell'}$ and $\ell' \preccurlyeq |\bar{c}|$ (Lemma 25). We know $g^c \vdash (\$\ k)_{\ell'} \leq \$k \Leftarrow \iota_\ell$ by rule $\leq$-*const*, because $\ell' \preccurlyeq |\bar{c}| \preccurlyeq |\bar{c}_n| = |\mathbf{id}(\ell)| = \ell$.
- Otherwise, $V_r\ \langle c_r,\ \bar{c}_n \rangle$ is already a value. Apply rule $\leq$-*wrapped-const*, $\leq$-*wrapped-lam*, or $\leq$-*wrapped-addr* depending on whether $V_r$ is a constant, a $\lambda$, or an address.

□

Lemma 24 (Substitution preserves simulation). *If $g \vdash N' \leq N \Leftarrow A$, $(\Gamma, x : B); \Sigma; g; \ell \vdash N \Leftarrow A$, and $\forall g. g \vdash M' \leq M \Leftarrow B$, then $g \vdash N'[x := M'] \leq N[x := M] \Leftarrow A$.*

Proof. The proof is fully mechanized in Agda [5] and can be found in the supplementary material. □

Lemma 25 (Simulation with wrapped constant). *If $g^c \vdash M \leq (\$k)\ \langle \mathbf{id}(\iota),\ \bar{c} \rangle \Leftarrow \iota_g$ and $(\mathbf{id}(\iota),\ \bar{c})$ is irreducible, then there exists $\ell$ s.t $M = (\$\ k)_\ell$, and $\ell \preccurlyeq |\bar{c}|$.*

Proof. Inversion on the simulation relation:

[4] In `/src/Dyn/Noninterference.agda`.

[5] In `substitution-pres-≤` of `/src/Security/SubstPres.agda`. The simulation relation is in `/src/Security/SimRel.agda`.

$$\boxed{g^c \vdash M \leq N \Leftarrow A}$$

$$\textit{≤-var}\;\frac{}{g^c \vdash x \leq x \Leftarrow A} \qquad \textit{≤-const}\;\frac{\ell' \preccurlyeq \ell}{g^c \vdash (\$\, k)_{\ell'} \leq \$\, k \Leftarrow \iota_\ell}$$

$$\textit{≤-wrapped-const}\;\frac{\ell' \preccurlyeq |\bar{c}| \qquad \vdash \bar{c} : \ell \Rightarrow g \qquad \mathbf{NF}\ \bar{c} \qquad \ell \neq g}{g^c \vdash (\$\, k)_{\ell'} \leq \$\, k\ \langle \mathbf{id}(\iota),\ \bar{c}\rangle \Leftarrow \iota_g}$$

$$\textit{≤-lam}\;\frac{g \vdash N' \leq N \Leftarrow B \qquad \ell' \preccurlyeq \ell}{g^c \vdash (\lambda x.\, N')_{\ell'} \leq \lambda x.\, N \Leftarrow (A \xrightarrow{g} B)_\ell}$$

$$\textit{≤-wrapped-lam}\;\frac{g_3 \vdash N' \leq N \Leftarrow D \qquad \ell \preccurlyeq |\bar{c}| \qquad \vdash \bar{d} : g_1 \Rightarrow g_3 \qquad \vdash \mathbf{c} : A \Rightarrow C \qquad \vdash \mathbf{d} : D \Rightarrow B \qquad \mathbf{NF}\ \bar{c}}{g^c \vdash (\lambda x.\, N')_\ell \leq (\lambda x.\, N)\ \left\langle \bar{d},\ \mathbf{c} \rightarrow \mathbf{d},\ \bar{c}\right\rangle \Leftarrow (A \xrightarrow{g_1} B)_{g_2}}$$

$$\textit{≤-addr}\;\frac{\ell' \preccurlyeq \ell}{g^c \vdash (\mathsf{addr}\ n_{\hat{\ell}})_{\ell'} \leq \mathsf{addr}\ n \Leftarrow \mathsf{Ref}\ (T_{\hat{\ell}})_\ell}$$

$$\textit{≤-wrapped-addr}\;\frac{\ell \preccurlyeq |\bar{c}| \qquad \vdash \mathbf{c} : T_{g_1} \Rightarrow S_{\hat{\ell}} \qquad \vdash \mathbf{d} : S_{\hat{\ell}} \Rightarrow T_{g_1} \qquad \mathbf{NF}\ \bar{c}}{g^c \vdash (\mathsf{addr}\ n_{\hat{\ell}})_\ell \leq (\mathsf{addr}\ n)\ \langle \mathbf{Ref\ c\ d},\ \bar{c}\rangle \Leftarrow (\mathsf{Ref}\ (T_{g_1}))_{g_2}}$$

$$\textit{≤-app}\;\frac{\ell^c \vdash M' \leq M \Leftarrow (A \xrightarrow{\ell^c \vee \ell} B)_\ell \qquad \ell^c \vdash N' \leq N \Leftarrow A}{\ell^c \vdash M'\ N' \leq \mathsf{app}\ M\ N\ A\ B\ \ell \Leftarrow C}$$

$$\textit{≤-app}\star\;\frac{g^c \vdash M' \leq M \Leftarrow (A \xrightarrow{\star} T_\star)_\star \qquad g^c \vdash N' \leq N \Leftarrow A}{g^c \vdash M'\ N' \leq \mathsf{app}\star\ M\ N\ A\ T \Leftarrow T_\star}$$

$$\textit{≤-if}\;\frac{\ell^c \vdash L' \leq L \Leftarrow \mathsf{Bool}_\ell \qquad \ell^c \vee \ell \vdash M' \leq M \Leftarrow A \qquad \ell^c \vee \ell \vdash N' \leq N \Leftarrow A}{\ell^c \vdash \mathsf{if}\ L'\ M'\ N' \leq \mathsf{if}\ L\ A\ \ell\ M\ N \Leftarrow B}$$

$$\textit{≤-if}\star\;\frac{g^c \vdash L' \leq L \Leftarrow \mathsf{Bool}_\star \qquad \star \vdash M' \leq M \Leftarrow T_\star \qquad \star \vdash N' \leq N \Leftarrow T_\star}{g^c \vdash \mathsf{if}\ L'\ M'\ N' \leq \mathsf{if}\star\ L\ T\ M\ N \Leftarrow T_\star}$$

$$\textit{≤-ref}\;\frac{\ell^c \vdash M' \leq M \Leftarrow T_\ell}{\ell^c \vdash \mathsf{ref}^?\ \ell\ M' \leq \mathsf{ref}\ \ell\ M \Leftarrow (\mathsf{Ref}\ T_\ell)_{\mathtt{low}}} \qquad \textit{≤-ref?}\;\frac{\star \vdash M' \leq M \Leftarrow T_\ell}{\star \vdash \mathsf{ref}^?\ \ell\ M' \leq \mathsf{ref?}^p\ \ell\ M \Leftarrow (\mathsf{Ref}\ T_\ell)_{\mathtt{low}}}$$

$$\textit{≤-deref}\;\frac{g^c \vdash M' \leq M \Leftarrow (\mathsf{Ref}\ A)_\ell}{g^c \vdash\ !\ M' \leq\ !\ M\ A\ \ell \Leftarrow B} \qquad \textit{≤-deref}\star\;\frac{g^c \vdash M' \leq M \Leftarrow (\mathsf{Ref}\ T_\star)_\star}{g^c \vdash\ !\ M' \leq\ !\star\ M\ T \Leftarrow T_\star}$$

$$\textit{≤-assign}\;\frac{\ell^c \vdash L' \leq L \Leftarrow (\mathsf{Ref}\ T_{\hat{\ell}})_\ell \qquad \ell^c \vdash M' \leq M \Leftarrow T_{\hat{\ell}}}{\ell^c \vdash L' :=^? M' \leq \mathsf{assign}\ L\ M\ T\ \hat{\ell}\ \ell \Leftarrow \mathtt{Unit}_{\mathtt{low}}}$$

$$\textit{≤-assign?}\;\frac{g^c \vdash L' \leq L \Leftarrow (\mathsf{Ref}\ T_g)_\star \qquad g^c \vdash M' \leq M \Leftarrow T_g}{g^c \vdash L' :=^? M' \leq \mathsf{assign?}^p\ L\ M\ T\ g \Leftarrow \mathtt{Unit}_{\mathtt{low}}}$$

$$\textit{≤-cast}\;\frac{g^c \vdash M \leq N \Leftarrow A \qquad \vdash \mathbf{c} : A \Rightarrow B}{g^c \vdash M \leq N\ \langle \mathbf{c} \rangle \Leftarrow B}$$

$$\textit{≤-prot}\;\frac{\ell' \preccurlyeq \ell \qquad g_2 \vdash M' \leq M \Leftarrow A \qquad \vdash PC \Leftarrow g_2}{g_1 \vdash \mathsf{prot}\ \ell'\ M' \leq \mathsf{prot}\ PC\ \ell\ M\ A \Leftarrow B}$$

Fig. 23. Simulation relation between $\lambda^c_{\mathtt{IFC}}$ and $\lambda_{\mathtt{SEC}}$

**Related by $\leq$-*wrapped-const*** The goal is proved directly.

**Related by $\leq$-*cast*** We know $g^c \vdash M \leq \$\,k \Leftarrow \iota_\ell$ and $\bar{c} : \ell \Rightarrow g$. Note that $(\mathbf{id}(\iota),\ \bar{c})$ is irreducible, so $\bar{c}$ can be $\uparrow$, $\ell\,!$, or $\uparrow$; $\mathtt{high}\,!$, so $\ell \preccurlyeq |\bar{c}|$. By rule $\leq$-*const*, we know $M = (\$\ k)_{\ell'}$ for some $\ell'$ and $\ell' \preccurlyeq \ell$. Thus $\ell' \preccurlyeq \ell \preccurlyeq |\bar{c}|$.

□

Lemma 26 (Simulation with reference proxy). *If $g^c \vdash M \leq (addr\ n)\ \langle \mathbf{Ref}\ \boldsymbol{c}\ \boldsymbol{d},\ \bar{c}\rangle \Leftarrow (\mathtt{Ref}\ T_{g_1})_{g_2}$ $\vdash \boldsymbol{c} : T_{g_1} \Rightarrow S_{\hat{\ell}}, \vdash \boldsymbol{d} : S_{\hat{\ell}} \Rightarrow T_{g_1}$, and* NF *$\bar{c}$, then there exists $\ell$ s.t $M = (addr\ n_{\hat{\ell}})_\ell$, and $\ell \preccurlyeq |\bar{c}|$.*

Proof. Analogous to Lemma 25. The only extra case to consider is $\bar{c} = \mathbf{id}(\ell)$, which also satisfies $\ell \preccurlyeq |\bar{c}|$. □

Lemma 27 (Simulation with function proxy). *If $g^c \vdash M \leq (\lambda x.\ N)\ \langle \bar{d},\ \boldsymbol{c} \to \boldsymbol{d},\ \bar{c}\rangle \Leftarrow (A \xrightarrow{g_1} B)_{g_2}$, $\vdash \bar{d} : g_1 \Rightarrow g_3$, and* NF *$\bar{c}$, then there exists $N', \ell$ s.t $M = (\lambda x.\ N')_\ell$, $g_3 \vdash N' \leq N \Leftarrow B$, and $\ell \preccurlyeq |\bar{c}|$.*

Proof. Analogous to Lemma 25 and Lemma 26. □

Lemma 28 (Simulation with $\lambda^c_{\mathtt{IFC}}$ value). *If $g^c \vdash M \leq V \Leftarrow A$, then $M$ is a value.*

Proof. Inversion on the value $V$ and the simulation relation $g^c \vdash M \leq V \Leftarrow A$ yields 6 cases. We consider the two cases for constants; the cases for $\lambda$s and addresses are analogous.

**Related by $\leq$-*const*** We know $M = (\$\ k)_\ell$ for some $k, \ell$, so $M$ is a value.

**Related by $\leq$-*wrapped-const*** By Lemma 25, $M = (\$\ k)_\ell$ for some $k, \ell$, so $M$ is a value.

□

Lemma 29 (Stamping preserves simulation). *If $g^c \vdash V \leq W \Leftarrow A$ and $\ell' \preccurlyeq \ell$, then $g^c \vdash V \curlyvee \ell' \leq stamp\ W\ \ell \Leftarrow stamp\ A\ \ell$.*

Proof. Casing on $\ell' \preccurlyeq \ell$:

**Case 1** $\ell' = \mathtt{low}, \ell = \mathtt{low}$: Straightforward because stamping low returns the same value.

**Case 2** $\ell' = \mathtt{low}, \ell = \mathtt{high}$: Casing on value $W$ and the simulation relation $g^c \vdash V \leq W \Leftarrow A$ produces 6 sub-cases. We consider the two cases for constants below (corresponding to $\leq$-*const* and $\leq$-*wrapped-const*), because the cases for $\lambda$s and addresses are analogous:

- $g^c \vdash (\$\ k)_{\ell_1'} \leq \$\,k \Leftarrow \iota_{\ell_1}$
  - If $\ell_1 = \mathtt{low}$, $g^c \vdash (\$\ k)_{\ell_1'} \leq \$\,k\ \langle \mathbf{id}(\iota),\ \uparrow\rangle \Leftarrow \iota_{\mathtt{high}}$ because $\ell_1' \preccurlyeq |\uparrow| = \mathtt{high}$ (rule $\leq$-*wrapped-const*).
  - If $\ell_1 = \mathtt{high}$, $g^c \vdash (\$\ k)_{\ell_1'} \leq \$\,k \Leftarrow \iota_{\mathtt{high}}$ because $\ell_1' \preccurlyeq \mathtt{high}$ (rule $\leq$-*const*).
- $g^c \vdash (\$\ k)_{\ell_1'} \leq \$\,k\ \langle \mathbf{id}(\iota),\ \bar{c}\rangle \Leftarrow -$
  - $g^c \vdash (\$\ k)_{\ell_1'} \leq k\ \langle \mathbf{id}(\iota),\ stamp\ \bar{c}\ \mathtt{high}\rangle \Leftarrow -$ because $\ell_1' \preccurlyeq |stamp\ \bar{c}\ \mathtt{high}| = |\bar{c}| \curlyvee \mathtt{high} = \mathtt{high}$ (Lemma 3 (Stamping models implicit flow) and rule $\leq$-*wrapped-const*).

**Case 3** $\ell' = \mathtt{high}, \ell = \mathtt{high}$: Analogous to Case 2; the only difference is that instead of $(\$\ k)_{\ell_1'}$, the left side is always $(\$\ k)_{\mathtt{high}}$ because it is stamped with $\ell' = \mathtt{high}$.

□

Lemma 30 (Casting a label expression models explicit flow). *If* NF *$e_1$ and $e_1\ \langle \bar{c}\rangle \longrightarrow^* e_2$ and* NF *$e_2$, $|e_1| \preccurlyeq |e_2|$.*

Proof. The proof is fully mechanized in Agda [6] and can be found in the supplementary material. The proof is by inversion on the multi-step reduction. □

[6] In `cast-security` of `/src/LabelExpr/Security.agda`.

Lemma 31 (Stamping a label expression models implicit flow). *If* **NF** $e$, $|stamp\ e\ \ell| = |e| \vee \ell$ *and* $|stamp!\ e\ \ell| = |e| \vee \ell$.

Proof. The proof is fully mechanized in Agda [7] and can be found in the supplementary material. The proof is by casing on **NF** $e$. □

Lemma 32. *Suppose* $\Gamma; \Sigma; g; \ell \vdash V \Leftarrow S_{\ell_1}$ *and* $\vdash \boldsymbol{c} : S_{\ell_1} \Rightarrow T_{\ell_2}$. *If* $V\ \langle \boldsymbol{c} \rangle \longrightarrow^* W$, *then* $\ell_1 \preccurlyeq \ell_2$.

Proof. By induction on the multi-step reduction.

**Zero step** $\boldsymbol{c} = (c_r,\ \bar{c})$ is irreducible, so $\bar{c} : \ell_1 \Rightarrow \ell_2$ is in its normal form. Thus $\ell_1 \preccurlyeq \ell_2$.

**One or more steps** Case on the first reduction step:

- Rule *cast*: $V_r\ \langle c_r,\ \bar{c} \rangle \longrightarrow V_r\ \left\langle c_r,\ \bar{d} \right\rangle \longrightarrow^* W$. By preservation of coercion sequences, $\bar{d} : \ell_1 \Rightarrow \ell_2$. By induction hypothesis, $\ell_1 \preccurlyeq \ell_2$.
- Rule *cast-id*: $V_r\ \langle \mathbf{id}(\iota),\ \mathbf{id}(g) \rangle \longrightarrow V_r$. We directly know $\ell_1 = g = \ell_2$.
- Rule *cast-comp*: $V_r\ \langle \boldsymbol{c} \rangle\ \langle \boldsymbol{d} \rangle \longrightarrow V_r\ \langle \boldsymbol{c} \mathbin{⨟} \boldsymbol{d} \rangle \longrightarrow^* W$. Suppose $\boldsymbol{c} = (-,\ \bar{c}), \bar{c} : \ell_0 \Rightarrow \ell_1$ and $\boldsymbol{d} = (-,\ \bar{d}), \bar{d} : \ell_1 \Rightarrow \ell_2$. We know $V_r\ \left\langle -,\ \bar{c} \mathbin{⨟} \bar{d} \right\rangle \longrightarrow V_r\ \langle \bar{c}_n \rangle \longrightarrow^* W$. By Lemma 2 (Composition models explicit flow), $\ell_1 = |\bar{c}| \preccurlyeq |\bar{c}_n| = \ell_2$.

□

Lemma 33. *If* $g_1 \vdash M \le V \Leftarrow A$ *then* $g_2 \vdash M \le V \Leftarrow A$.

Proof. The proof is fully mechanized in Agda and can be found in the supplementary material. The proof is by casing on value and the simulation relation. □

Lemma 34 (Simulation between $\lambda^c_{\texttt{IFC}}$ and $\lambda_{\texttt{SEC}}$). *Suppose* $M_1$ *is a well-typed* $\lambda^c_{\texttt{IFC}}$ *term:* $\Gamma; \Sigma_1; g^c; |PC| \vdash M_1 \Leftarrow A$, *PC is a well-typed label expression:* $\vdash PC \Leftarrow g^c$, *and* $\mu_1$ *is a well-typed heap:* $\Sigma_1 \vdash \mu_1$. *Suppose* $g^c \vdash M_1' \le M_1 \Leftarrow A$, $\Sigma_1 \vdash \mu_1' \le \mu_1$, *and* $\ell \preccurlyeq |PC|$. *If* $M_1 \mid \mu_1 \mid PC \longrightarrow M_2 \mid \mu_2$, *then there exists* $M_3, \mu_3, M_2', \mu_2'$ *s.t* $M_2 \mid \mu_2 \mid PC \longrightarrow^* M_3 \mid \mu_3$, $M_1' \mid \mu_1' \mid \ell \longrightarrow^* M_2' \mid \mu_2'$, $g^c \vdash M_2' \le M_3 \Leftarrow A$, *and* $\Sigma_2 \vdash \mu_2' \le \mu_3$ *for some* $\Sigma_2$.

Proof. The proof is by induction on the reduction relation $M_1 \mid \mu_1 \mid PC \longrightarrow M_2 \mid \mu_2$ and inversion on the simulation relation $\vdash M_1' \le M_1 \Leftarrow A$. We only consider successful cases of the reduction (which do not produce errors), because the theorem statement of noninterference (Theorem 16) for $\lambda^\star_{\texttt{IFC}}$ is termination-insensitive.

**Case** $\xi$:

$$N_1 \mid \mu_1 \mid PC \longrightarrow N_2 \mid \mu_2 \tag{10}$$

$$plug\ N_1\ F = M_1 \tag{11}$$

$$plug\ N_2\ F = M_2 \tag{12}$$

We case on the frame $F$. We consider $F = \mathsf{app}\,\square\,M\,A\,B\,\ell$ below; the cases for other frames all follow the same pattern. By inversion of the simulation relation, the left side must be a function application:

$$\ell^c \vdash N_1'\ M' \le \mathsf{app}\,N_1\,M\,A\,B\,\ell \Leftarrow C$$

We know $\ell^c \vdash N_1' \le N_1 \Leftarrow (A \xrightarrow{\ell^c \vee \ell} B)_\ell$ and $\ell^c \vdash M' \le M \Leftarrow A$. By the induction hypothesis, there exists $N_3, \mu_3, N_2', \mu_2'$ such that $N_2 \mid \mu_2 \mid PC \longrightarrow^* N_3 \mid \mu_3$, $N_1' \mid \mu_1' \mid \ell \longrightarrow^* N_2' \mid \mu_2'$, $\ell^c \vdash N_2' \le N_3 \Leftarrow (A \xrightarrow{\ell^c \vee \ell} B)_\ell$, and $\Sigma_2 \vdash \mu_2' \le \mu_3$ for some $\Sigma_2$. Choose $M_3 = plug\ N_3\ F = \mathsf{app}\,N_3\,M\,A\,B\,\ell$, $\mu_3 = \mu_3$, $M_2' = N_2'\ M'$, and $\mu_2' = \mu_2'$. We know $\mathsf{app}\,N_2\,M\,A\,B\,\ell \mid$

[7] In `stampₑ-security` and `stamp!ₑ-security` of `/src/LabelExpr/Security.agda`.

$\mu_2 \mid PC \longrightarrow^* \mathsf{app}\, N_3\, M\, A\, B\, \ell \mid \mu_3$, $N_1'\ M' \mid \mu_1' \mid \ell \longrightarrow^* N_2'\ M' \mid \mu_2'$, and $\Sigma_2 \vdash \mu_2' \leq \mu_3$. We still need to show $\ell^c \vdash N_2'\ M' \leq \mathsf{app}\, N_3\, M\, A\, B\, \ell \Leftarrow C$, which is proved by applying rule $\leq$*−app*.

**Case *prot-val*:** By inversion on the simulation relation, the left side must be a protection term. By Lemma 28, the body $V'$ must be a value:

$$g_1 \vdash \mathsf{prot}\ \ell'\ V' \leq \mathsf{prot}\, PC\, \ell\, V\, A \Leftarrow B \tag{13}$$

$$g_2 \vdash V' \leq V \Leftarrow A \tag{14}$$

$$\ell' \preccurlyeq \ell \tag{15}$$

$$\vdash PC \Leftarrow g_2 \tag{16}$$

We take a step by *prot-val* on the left side, we need to relate:

$$g_1 \vdash V' \curlyvee \ell' \leq \mathit{stamp}\ V\ \ell \Leftarrow B$$

which is proved by applying Lemma 29 on (14) (with Lemma 33 applied) and (15).

**Case *prot-ctx*:**

$$M \mid \mu_1 \mid PC_2 \longrightarrow N \mid \mu_2$$

By inversion on the simulation relation, the left side must be protection:

$$g_1 \vdash \mathsf{prot}\ \ell'\ M' \leq \mathsf{prot}\, PC_2\, \ell\, M\, A \Leftarrow B \tag{17}$$

$$g_2 \vdash M' \leq M \Leftarrow A \tag{18}$$

$$\ell' \preccurlyeq \ell \tag{19}$$

$$\vdash PC_2 \Leftarrow g_2 \tag{20}$$

By inversion on the typing of the protection term on the right:

$$|PC_1| \curlyvee \ell \preccurlyeq |PC_2| \tag{21}$$

where $PC_1$ is the original PC to reduce the protection term on the right. To get the induction hypothesis, we need to show $\ell'' \curlyvee \ell' \preccurlyeq |PC_2|$ where $\ell''$ is the PC to reduce the protection term on the left. We know $\ell'' \preccurlyeq |PC_1|$, thus:

$$\ell'' \curlyvee \ell' \preccurlyeq \ell' \curlyvee \ell \preccurlyeq |PC_1| \curlyvee \ell \preccurlyeq |PC_2|$$

The induction hypothesis shows that there exists $L, \mu_3, N', \mu_2'$ s.t $N \mid \mu_2 \mid PC_2 \longrightarrow^* L \mid \mu_3$, $M' \mid \mu_1' \mid \ell'' \curlyvee \ell' \longrightarrow^* N' \mid \mu_2'$, $g_2 \vdash N' \leq L \Leftarrow A$, and $\Sigma_2 \vdash \mu_2' \leq \mu_3$ for some $\Sigma_2$. We step the left side to $\mathsf{prot}\ \ell'\ N'$ and step the right side to $\mathsf{prot}\, PC_2\, \ell\, L\, A$ using the congruence of *prot-ctx*. We then relate the protection terms on both sides using rule $\leq$*−prot*.

**Case *cast*:** There are three sub-cases: *cast*, *cast-id*, and *cast-comp*.

**Sub-case *cast*:** We know $g^c \vdash M' \leq V_r\ \langle c_r,\ \bar{c}\rangle \Leftarrow B$. $V_r\ \langle c_r,\ \bar{c}\rangle$ is not a value, so $g^c \vdash M' \leq V_r \Leftarrow A$. We need to relate $g^c \vdash M' \leq V_r\ \left\langle c_r,\ \bar{d}\right\rangle \Leftarrow B$, which is directly proved by applying $\leq$*-cast*.

**Sub-case *cast-id*:** We know $g^c \vdash M' \leq V_r\ \langle \mathbf{id}(\iota),\ \mathbf{id}(g)\rangle \Leftarrow A$. Again, note that $V_r\ \langle \mathbf{id}(\iota),\ \mathbf{id}(g)\rangle$ is not a value, so $g^c \vdash M' \leq V_r \Leftarrow A$.

**Sub-case *cast-comp*:** We know $g^c \vdash M' \leq V_r\ \langle \boldsymbol{c}\rangle\ \langle \boldsymbol{d}\rangle \Leftarrow B$. By inversion using rule $\leq$*−cast*, $g^c \vdash M' \leq V_r\ \langle \boldsymbol{c}\rangle \Leftarrow A$; $\boldsymbol{c}$ is irreducible, by Lemma 28 $M'$ is a value. We need to show there exists $M$ such that $V_r\ \langle \boldsymbol{c} \mathbin{⨾} \boldsymbol{d}\rangle \mid \mu_1 \mid PC \longrightarrow^* M \mid \mu_1$ and $g^c \vdash M' \leq M \Leftarrow B$. Casing on $V_r$:

- **If $V_r = \$k$.** Suppose $\boldsymbol{c} \mathbin{⨾} \boldsymbol{d} = d_r,\ \bar{c} \mathbin{⨾} \bar{d}$ for some $d_r$. We take a step using *cast* by reducing the composed coercion sequence to its normal form: $\bar{c} \mathbin{⨾} \bar{d} \longrightarrow^+ \bar{c}_n$. By Lemma 2 (Composition models explicit flow), $|\bar{c}| \preccurlyeq |\bar{c}_n|$. By Lemma 25, we know $M' = (\$\ k)_{\ell'}$ and $\ell' \preccurlyeq |\bar{c}|$. By transitivity, $\ell' \preccurlyeq |\bar{c}_n|$. (1) If $\bar{c}_n$ is not an identity coercion, we relate $g^c \vdash (\$\, k)_{\ell'} \leq \$\, k\ \langle d_r, \bar{c}_n \rangle \Leftarrow B$ directly using rule $\leq$-*wrapped-const*. (2) If $\bar{c}_n = \mathbf{id}(\ell)$, we take one additional step by *cast-id*. We know $\ell' \preccurlyeq |\mathbf{id}(\ell)| = \ell$, thus $g^c \vdash (\$\, k)_{\ell'} \leq \$\, k \Leftarrow \iota_\ell$ by rule $\leq$-*const*.

- If $V_r = \lambda x.\, N$. Similar to the constant case above. The only difference is that we do not need to specially handle $\bar{c}_n = \mathbf{id}(\ell)$, because the function coercion $(\bar{d},\, \boldsymbol{c} \rightarrow \boldsymbol{d},\, \mathbf{id}(\ell))$ is already irreducible.
- If $V_r = \mathtt{addr}\, n$. Same as the lambda case above.

**Case $\beta$:** By inversion on the simulation relation, the left side must be a function application:

$$\ell^c \vdash L\ M \leq \mathtt{app}\,(\lambda x.\, N)\, V\, A\, B\, \ell \Leftarrow C$$

where $C = \mathit{stamp}\ B\ \ell$. We know that $L$ must be a $\lambda$ by rule $\leq$-*lam* and $M$ must be a value by Lemma 28:

$$\ell^c \vdash ((\lambda x.\, N')_{\ell'})\ V' \leq \mathtt{app}\,(\lambda x.\, N)\, V\, A\, B\, \ell \Leftarrow C \tag{22}$$

$$\ell' \preccurlyeq \ell \tag{23}$$

$$\ell^c \curlyvee \ell \vdash N' \leq N \Leftarrow B \tag{24}$$

$$\ell^c \vdash V' \leq V \Leftarrow A \tag{25}$$

We take one step by $\beta$ on the left side of the simulation relation:

$$((\lambda x.\, N')_{\ell'})\ V' \mid \mu' \mid \ell'' \longrightarrow \mathtt{prot}\ \ell'\ (N'[x := V']) \mid \mu'$$

Now we need to show

$$\ell^c \vdash \mathtt{prot}\ \ell'\ (N'[x := V']) \leq \mathtt{prot}\,(\mathit{stamp}\ PC\ \ell)\,\ell\,(N[x := V])\, B \Leftarrow C$$

Applying $\leq$-*prot* yields three sub-goals: (1) $\ell' \preccurlyeq \ell$ is directly proved by (23) (2) $\ell^c \curlyvee \ell \vdash N'[x := V'] \leq N[x := V] \Leftarrow B$ is proved by applying Lemma 24 on (24) and (25). (3) $\vdash \mathit{stamp}\ PC\ \ell \Leftarrow \ell^c \curlyvee \ell$ because stamping is well-typed.

**Case *app-cast*:**

$$(\mathit{stamp}\ PC_1\ \ell)\ \langle \bar{d} \rangle \longrightarrow^* PC_2 \tag{26}$$

$$V\ \langle \boldsymbol{c} \rangle \longrightarrow^* W \tag{27}$$

By inversion on the simulation relation, the left side must be a function application:

$$\ell^c \vdash L\ M \leq \mathtt{app}\,(\lambda x.\, N\ \langle \bar{d},\, \boldsymbol{c} \rightarrow \boldsymbol{d},\, \bar{c} \rangle)\, V\, C\, D\, \ell \Leftarrow E$$

where $E = \mathit{stamp}\ D\ \ell$, $\vdash \bar{d} : \ell^c \curlyvee \ell \Rightarrow g^c$, $\vdash \boldsymbol{c} : C \Rightarrow A$, and $\vdash \boldsymbol{d} : B \Rightarrow D$.
We know that $L$ must be a $\lambda$ by Lemma 27 and $M$ must be a value by Lemma 28:

$$\ell^c \vdash ((\lambda x.\, N')_{\ell'})\ V' \leq \mathtt{app}\,(\lambda x.\, N\ \langle \bar{d},\, \boldsymbol{c} \rightarrow \boldsymbol{d},\, \bar{c} \rangle)\, V\, C\, D\, \ell \Leftarrow E \tag{28}$$

$$\ell' \preccurlyeq |\bar{c}| \tag{29}$$

$$g^c \vdash N' \leq N \Leftarrow B \tag{30}$$

$$\ell^c \vdash V' \leq V \Leftarrow C \tag{31}$$

We take a step by $\beta$ on the left side of simulation relation:

$$((\lambda x.\, N')_{\ell'})\ V' \mid \mu' \mid \ell'' \longrightarrow \mathtt{prot}\ \ell'\ (N'[x := V']) \mid \mu'$$

Now we need to show

$$\ell^c \vdash \mathtt{prot}\ \ell'\ (N'[x := V']) \leq \mathtt{prot}\, PC_2\, \ell\, ((N[x := W])\ \langle \boldsymbol{d} \rangle)\, D \Leftarrow E$$

By rule $\leq$-*prot*, it is equivalent to showing:

$$\ell' \preccurlyeq \ell \tag{32}$$
$$g^c \vdash N'[x := V'] \leq (N[x := W]) \langle \boldsymbol{d} \rangle \Leftarrow D \tag{33}$$
$$\vdash PC_2 \Leftarrow g^c \tag{34}$$

We know $\vdash PC_2 \Leftarrow g^c$ because reducing label expressions preserves types. By (29) and $|\bar{c}| = \ell$ we prove (32). Apply Lemma 23 on (31) and (27), we get $\ell^c \vdash V' \leq W \Leftarrow A$. By Lemma 24, Lemma 33, and (30), $g^c \vdash N'[x := V'] \leq N[x := W] \Leftarrow B$. Apply rule $\leq$-*cast* and we prove (33).

**Case *app⋆-cast*:** Similar to *app-cast*. The only noteworthy difference is that on the right side, we use $|\bar{c}|$ instead of the $\ell$ from the syntax of the function application, both in the protection term and when stamping the PC.

**Cases *β-if-true* and *β-if-false*:** By inversion on the simulation relation, the left side must also be an if-conditional:

$$\ell^c \vdash \texttt{if}\ (\$\ \texttt{true})_{\ell'}\ M'\ N' \leq \texttt{if}\ \$\,\texttt{true}\,A\,\ell\,M\,N \Leftarrow B$$

We know:

$$\ell' \preccurlyeq \ell \tag{35}$$
$$\ell^c \curlyvee \ell \vdash M' \leq M \Leftarrow A \tag{36}$$
$$\ell^c \curlyvee \ell \vdash N' \leq N \Leftarrow A \tag{37}$$

Take one step on the left side. We need to relate $\ell^c \vdash \texttt{prot}\ \ell'\ M' \leq \texttt{prot}\,(\mathit{stamp}\ PC\ \ell)\,\ell\,M\,A \Leftarrow B$. The goal is directly proved by applying rule $\leq$-*prot*. The case for *β-if-false* is analogous.

**Cases *if-true-cast* and *if-false-cast*:** Note that the label partial order trivially holds because $\ell' \preccurlyeq \texttt{high}$ for any $\ell'$. The rest is analogous to *β-if-true* and *β-if-false*.

**Cases *if⋆-true-cast* and *if⋆-false-cast*:** By Lemma 25 we know $\ell' \preccurlyeq |\bar{c}|$. The rest is analogous to *β-if-true* and *β-if-false*.

**Case *ref*:** From the typing derivation (rule $\vdash$-*ref*), we also know:

$$\vdash PC \Leftarrow \ell^c \tag{38}$$
$$\ell^c \preccurlyeq \ell \tag{39}$$

which implies

$$|PC| \preccurlyeq \ell \tag{40}$$

We know $\ell' \preccurlyeq |PC| \preccurlyeq \ell$, where $\ell'$ is the PC that the left side is reduced with. We take a step by *ref?-ok* on the left side by choosing the same fresh address $n$. We need to show:

$$\ell^c \vdash (\texttt{addr}\ n_\ell)_{\texttt{low}} \leq \texttt{addr}\,n \Leftarrow (\texttt{Ref}\ T_\ell)_{\texttt{low}} \tag{41}$$
$$(\Sigma_1, \ell \mapsto n \mapsto T) \vdash (\mu', \ell \mapsto n \mapsto (V' \curlyvee \ell)) \leq (\mu, \ell \mapsto n \mapsto V) \tag{42}$$

(41) is proved directly by $\leq$-*addr*. To prove (42) we need to show:

$$\texttt{low} \vdash V' \curlyvee \ell \leq V \Leftarrow T_\ell \tag{43}$$

By inversion on the simulation relation and Lemma 33, we know $\texttt{low} \vdash V' \leq V \Leftarrow T_\ell$. From the definition of stamping we know $\mathit{stamp}\ V\ \ell = V$ if $V \Leftarrow T_\ell$; (43) is thus proved by Lemma 29.

**Case *ref?*:**

$$PC_1 \langle \star \Rightarrow^{p} \ell \rangle \longrightarrow^{*} PC_2$$

Reducing label expressions preserves types, so $\vdash PC_2 \Leftarrow \ell$. We know $|PC_2| = \ell$ and $\ell' \preccurlyeq |PC_1|$, where $\ell'$ is the PC that the left side is reduced with. Casting models explicit flow (Lemma 30), so $|PC_1| \preccurlyeq |PC_2|$. We thus have $\ell' \preccurlyeq |PC_1| \preccurlyeq |PC_2| = \ell$. Take one step by *ref?-ok* on the left side by choosing the same fresh address $n$. The rest of the proof follows that of *ref*.

**Case *deref*:**

$$\mu(\hat{\ell}, n) = V$$

By $\Sigma \vdash \mu' \leq \mu$ and Definition 22, there exists $V'$ s.t $\mu'(\hat{\ell}, n) = V'$ and $\texttt{low} \vdash V' \leq V \Leftarrow T_{\hat{\ell}}$ where $T = \Sigma(\hat{\ell}, n)$. By inversion on the simulation relation, the left side must be a dereference:

$$g^c \vdash\ !\ (\texttt{addr}\ n_{\hat{\ell}})_{\ell'} \leq\ !\ (\texttt{addr}\ n)\ T_{\hat{\ell}}\ \ell \Leftarrow B$$

where $\ell' \preccurlyeq \ell$. Take one step on the left using rule *deref*, we need to relate:

$$g^c \vdash \texttt{prot}\ \ell'\ V' \leq \texttt{prot}\ \texttt{high}\ \ell\ V\ T_{\hat{\ell}} \Leftarrow B$$

which is directly proved by rule $\leq$*-prot* and Lemma 33.

**Case *deref★-cast*:**

$$\mu(\hat{\ell}, n) = V$$

By $\Sigma \vdash \mu' \leq \mu$ and Definition 22, there exists $V'$ s.t $\mu'(\hat{\ell}, n) = V'$ and $\texttt{low} \vdash V' \leq V \Leftarrow S_{\hat{\ell}}$ where $S = \Sigma(\hat{\ell}, n)$. By inversion on the simulation relation, the left side must be a dereference:

$$g^c \vdash\ !\ (\texttt{addr}\ n_{\hat{\ell}})_{\ell} \leq\ !\star\ (\texttt{addr}\ n\ \langle \textbf{Ref}\ \boldsymbol{c}\ \boldsymbol{d},\ \bar{c} \rangle)\ T \Leftarrow T_{\star}$$

where $\vdash \boldsymbol{c} : T_{\star} \Rightarrow S_{\hat{\ell}}, \vdash \boldsymbol{d} : S_{\hat{\ell}} \Rightarrow T_{\star}$. By Lemma 26, $\ell \preccurlyeq |\bar{c}|$. We take a step on the left side using *deref*. We need to relate:

$$g^c \vdash \texttt{prot}\ \ell\ V' \leq \texttt{prot}\ \texttt{high}\ |\bar{c}|\ (V\ \langle \boldsymbol{d} \rangle)\ T_{\star} \Leftarrow B$$

which is proved directly by applying Lemma 33, $\leq$*-prot*, and then $\leq$*-cast*.

**Case *deref-cast*:** Analogous to *deref★-cast*.

**Case *β-assign*:** By inversion on the simulation relation, the left side is also an assignment:

$$\ell^c \vdash (\texttt{addr}\ n_{\hat{\ell}})_{\ell'} :=^{?} V' \leq \texttt{assign}\ (\texttt{addr}\ n)\ V\ T\ \hat{\ell}\ \ell \Leftarrow \texttt{Unit}_{\texttt{low}}$$

where $\ell' \preccurlyeq \ell$. From the typing derivation we know $\ell^c \curlyvee \ell \preccurlyeq \hat{\ell}, \vdash PC \Leftarrow \ell^c$. We know $\ell'' \preccurlyeq |PC| = \ell^c$ where $\ell''$ is the PC on the left, thus $\ell'' \curlyvee \ell' \preccurlyeq \hat{\ell}$. We take one step on the left side using *assign?-ok*. The units on both sides relate straightforwardly. We need to relate:

$$\Sigma \vdash [\hat{\ell} \mapsto n \mapsto (V' \curlyvee \hat{\ell})]\mu' \leq [\hat{\ell} \mapsto n \mapsto V]\mu$$

Given $\Sigma(\hat{\ell}, n) = T$, we need to show:

$$\texttt{low} \vdash V' \curlyvee \hat{\ell} \leq V \Leftarrow T_{\hat{\ell}}$$

which can be proved similar to (43) because $\vdash V \Leftarrow T_{\hat{\ell}}$.

**Case *assign-cast*:**

$$V\ \langle \boldsymbol{c} \rangle \longrightarrow^{*} W$$

where $\vdash \boldsymbol{c} : T_{\hat{\ell}_2} \Rightarrow S_{\hat{\ell}_1}$. By Lemma 32, $\hat{\ell}_2 \preccurlyeq \hat{\ell}_1$. From the typing derivation, $\vdash PC \Leftarrow \ell^c$ and $\ell^c \curlyvee \ell \preccurlyeq \hat{\ell}_2$. By inversion on the simulation relation, the left side is also an assignment:

$$\ell^c \vdash (\texttt{addr}\ n_{\hat{\ell}_1})_{\ell'} :=^{?} V' \leq \texttt{assign}\ (\texttt{addr}\ n\ \langle \textbf{Ref}\ \boldsymbol{c}\ \boldsymbol{d},\ \bar{c} \rangle)\ V\ T\ \hat{\ell}_2\ \ell \Leftarrow \texttt{Unit}_{\texttt{low}}$$

We know $\ell' \preccurlyeq |\bar{c}| = \ell$. $\ell'' \preccurlyeq |PC| = \ell^c$ where $\ell''$ is the PC that the left side is reduced with. Thus we have $\ell'' \curlyvee \ell' \preccurlyeq \hat{\ell}_2 \preccurlyeq \hat{\ell}_1$. The check succeeds so we take one step on the left side by *assign?-ok*. The units on both sides relate straightforwardly. We need to relate:

$$\Sigma \vdash [\hat{\ell}_1 \mapsto n \mapsto (V' \curlyvee \hat{\ell}_1)]\mu' \le [\hat{\ell}_1 \mapsto n \mapsto W]\mu$$

Need to show:

$$\texttt{low} \vdash V' \curlyvee \hat{\ell}_1 \le W \Leftarrow S_{\hat{\ell}_1}$$

which can be proved similar to (43).

**Case *assign?-cast*:**

$$(stamp!\ PC_1\ |\bar{c}|)\ \langle \star \Rightarrow^p \hat{\ell} \rangle \longrightarrow^* PC_2 \tag{44}$$

$$V\ \langle \boldsymbol{c} \rangle \longrightarrow^* W \tag{45}$$

By inversion on the simulation relation, the left side is also an assignment:

$$g^c \vdash (\texttt{addr}\ n_{\hat{\ell}})_\ell :=^? V' \le \texttt{assign?}^p\ (\texttt{addr}\ n\ \langle \textbf{Ref}\ \boldsymbol{c}\ \boldsymbol{d},\ \bar{c} \rangle)\ V\ T\ g \Leftarrow \texttt{Unit}_{\texttt{low}}$$

where $\vdash \boldsymbol{c} : T_g \Rightarrow S_{\hat{\ell}}, \vdash \boldsymbol{d} : S_{\hat{\ell}} \Rightarrow T_g$. By Lemma 26, $\ell \preccurlyeq |\bar{c}|$. Stamping models implicit flow (Lemma 31), so $|stamp\ PC_1\ |\bar{c}|| = |PC_1| \curlyvee |\bar{c}|$; casting models explicit flow (Lemma 30), so $|PC_1| \curlyvee |\bar{c}| \preccurlyeq |PC_2|$. Reduction of label expressions preserves types, so $\vdash PC_2 \Leftarrow \hat{\ell}$. Thus $|PC_2| = \hat{\ell}$, $|PC_1| \curlyvee |\bar{c}| \preccurlyeq \hat{\ell}$. We know $\ell' \preccurlyeq |PC_1|$ where $\ell'$ is the PC the left side reduces with. Thus $\ell' \curlyvee \ell \preccurlyeq \hat{\ell}$; the check succeeds so we take one step on the left side by *assign?-ok*. The units on both sides relate straightforwardly. We need to relate:

$$\Sigma \vdash [\hat{\ell} \mapsto n \mapsto (V' \curlyvee \hat{\ell})]\mu' \le [\hat{\ell} \mapsto n \mapsto W]\mu$$

Need to show:

$$\texttt{low} \vdash V' \curlyvee \hat{\ell} \le W \Leftarrow S_{\hat{\ell}}$$

which again can be proved similar to (43).

□

Lemma 35 (Multi-step simulation). *Suppose $M$ is a well-typed $\lambda^c_{\texttt{IFC}}$ term $\Gamma; \Sigma; g^c; |PC| \vdash M \Leftarrow A$, $PC$ is a well-typed label expression: $\vdash PC \Leftarrow g^c$, and $\mu_1$ is a well-typed heap: $\Sigma \vdash \mu_1$. Suppose $g^c \vdash M' \le M \Leftarrow A$, $\Sigma \vdash \mu'_1 \le \mu_1$ ,and $\ell \preccurlyeq |PC|$. If $M \mid \mu_1 \mid PC \longrightarrow^* V \mid \mu_2$, then there exists $V', \mu'_2$ such that $M' \mid \mu'_1 \mid \ell \longrightarrow^* V' \mid \mu'_2$ and $g^c \vdash V' \le V \Leftarrow A$.*

Proof. By induction on multi-step reduction $M \mid \mu_1 \mid PC \longrightarrow^* V \mid \mu_2$.

**Zero step** If $M$ is already a value, choose $V'$ to be $M'$. By Lemma 28, $M'$ is also a value. The values are in sync because $g^c \vdash M' \le M \Leftarrow A$.

**One or more steps**

$$M \mid \mu_1 \mid PC \longrightarrow N \mid \mu_3 \tag{46}$$

$$N \mid \mu_3 \mid PC \longrightarrow^* V \mid \mu_2 \tag{47}$$

Apply Lemma 34 and then use the induction hypothesis.

□

Lemma (Noninterference for $\lambda^c_{\texttt{IFC}}$). *If $M$ is well-typed: $(x{:}\texttt{Bool}_{\texttt{high}}); \emptyset; \texttt{low}; \texttt{low} \vdash M \Leftarrow \texttt{Bool}_{\texttt{low}}$ and*

$$M[x := \$\,b_1] \mid \emptyset \mid \texttt{low} \longrightarrow^* V_1 \mid \mu_1 \quad \text{and} \quad M[x := \$\,b_2] \mid \emptyset \mid \texttt{low} \longrightarrow^* V_2 \mid \mu_2$$

*then $V_1 = V_2$.*

$\boxed{\epsilon\ M\ A = M'}$

$$
\begin{aligned}
\epsilon\ x\ - &= x \\
\epsilon\ (\$\, k)\ (\iota_\ell) &= (\$\, k)_\ell \\
\epsilon\ (\lambda x.\, N)\ ((A \xrightarrow{-} B)_\ell) &= (\lambda x.\, \epsilon\ N\ B)_\ell \\
\epsilon\ (\mathsf{addr}\ n)\ (\mathsf{Ref}\ (T_{\hat{\ell}})_\ell) &= (\mathsf{addr}\ n_{\hat{\ell}})_\ell \\
\epsilon\ (\mathsf{app}\ M\ N\ A\ B\ \ell)\ - &= (\epsilon\ M\ (A \xrightarrow{\star} B)_\ell)\ (\epsilon\ N\ A) \\
\epsilon\ (\mathsf{app}{\star}\ M\ N\ A\ T)\ - &= (\epsilon\ M\ (A \xrightarrow{\star} T_\star)_\star)\ (\epsilon\ N\ A) \\
\epsilon\ (\mathsf{if}\ L\ A\ \ell\ M\ N)\ - &= \mathsf{if}\ (\epsilon\ L\ \mathsf{Bool}_\ell)\ (\epsilon\ M\ A)\ (\epsilon\ N\ A) \\
\epsilon\ (\mathsf{if}{\star}\ L\ T\ M\ N)\ - &= \mathsf{if}\ (\epsilon\ L\ \mathsf{Bool}_\star)\ (\epsilon\ M\ T_\star)\ (\epsilon\ N\ T_\star) \\
\epsilon\ (\mathsf{ref}\ \ell\ M)\ (\mathsf{Ref}\ T_\ell)_{\mathsf{low}} &= \mathsf{ref}^?\ \ell\ (\epsilon\ M\ T_\ell) \\
\epsilon\ (\mathsf{ref?}^p\ \ell\ M)\ (\mathsf{Ref}\ T_\ell)_{\mathsf{low}} &= \mathsf{ref}^?\ \ell\ (\epsilon\ M\ T_\ell) \\
\epsilon\ (!\ M\ A\ \ell)\ - &= !\ (\epsilon\ M\ (\mathsf{Ref}\ A)_\ell) \\
\epsilon\ (!{\star}\ M\ T)\ - &= !\ (\epsilon\ M\ (\mathsf{Ref}\ T_\star)_\star) \\
\epsilon\ (\mathsf{assign}\ L\ M\ T\ \hat{\ell}\ \ell)\ - &= (\epsilon\ L\ (\mathsf{Ref}\ T_{\hat{\ell}})_\ell)\ :=^?\ (\epsilon\ M\ T_{\hat{\ell}}) \\
\epsilon\ (\mathsf{assign?}^p\ L\ M\ T\ \hat{g})\ - &= (\epsilon\ L\ (\mathsf{Ref}\ T_{\hat{g}})_\star)\ :=^?\ (\epsilon\ M\ T_{\hat{g}}) \\
\epsilon\ (N\ \langle c \rangle)\ - &= \epsilon\ N\ A \quad \text{, where } \vdash c : A \Rightarrow B \\
\epsilon\ (\mathsf{prot}\ PC\ \ell\ M\ A)\ - &= \mathsf{prot}\ \ell\ (\epsilon\ M\ A)
\end{aligned}
$$

Fig. 24. Erasure from $\lambda^c_{\mathsf{IFC}}$ to $\lambda_{\mathsf{SEC}}$

Proof. From the definition of $\epsilon$ (Figure 24), we know $\mathsf{low} \vdash \epsilon(M)[x := (\$\, b_i)_{\mathsf{high}}] \leq M[x := \$\, b_i] \Leftarrow \mathsf{Bool}_{\mathsf{low}}$. By Lemma 35, there exists $V_i', \mu_i'$ s.t $\epsilon(M)[x := (\$\, b_i)_{\mathsf{high}}] \mid \emptyset \mid \mathsf{low} \longrightarrow^* V_i' \mid \mu_i'$ and $\mathsf{low} \vdash V_i' \leq V_i \Leftarrow \mathsf{Bool}_{\mathsf{low}}$. The simulation relation must be of form $\mathsf{low} \vdash (\$\, a_i)_{\mathsf{low}} \leq \$\, a_i \Leftarrow \mathsf{Bool}_{\mathsf{low}}$ where $a_i$ is the output boolean. By Lemma 21, $a_1 = a_2$, thus $V_1 = \$\, a_1 = \$\, a_2 = V_2$. □

Theorem (Noninterference for $\lambda^\star_{\mathsf{IFC}}$). *Suppose a $\lambda^\star_{\mathsf{IFC}}$ program $M$ is well-typed:* $(x{:}\mathsf{Bool}_{\mathsf{high}}); \mathsf{low} \vdash M : \mathsf{Bool}_{\mathsf{low}}$. *If for any boolean inputs $b_1, b_2$*

$$(C\ M)[x := \$\, b_1] \mid \emptyset \mid \mathsf{low} \longrightarrow^* V_1 \mid \mu_1 \quad \text{and} \quad (C\ M)[x := \$\, b_2] \mid \emptyset \mid \mathsf{low} \longrightarrow^* V_2 \mid \mu_2$$

*then the resulting values $V_1 = V_2$.*

Proof. By Theorem 15 (Compilation preserves types), $(x{:}\mathsf{Bool}_{\mathsf{high}}); \emptyset; \mathsf{low}; \mathsf{low} \vdash M' \Leftarrow \mathsf{Bool}_{\mathsf{low}}$. By Lemma 14 (Noninterference for $\lambda^c_{\mathsf{IFC}}$), $V_1 = V_2$. □